\documentclass[english,prd,eqsecnum,nofootinbib,floatfix,preprintnumbers]{revtex4}
\usepackage[T1]{fontenc}
\usepackage[latin9]{inputenc}
\usepackage{amsmath}
\usepackage{amssymb}
\usepackage{graphicx,epsfig}
\usepackage{hyperref}

\usepackage{subfig}

\makeatletter
\@ifundefined{textcolor}{}
{%
 \definecolor{BLACK}{gray}{0}
 \definecolor{WHITE}{gray}{1}
 \definecolor{RED}{rgb}{1,0,0}
 \definecolor{GREEN}{rgb}{0,1,0}
 \definecolor{BLUE}{rgb}{0,0,1}
 \definecolor{CYAN}{cmyk}{1,0,0,0}
 \definecolor{MAGENTA}{cmyk}{0,1,0,0}
 \definecolor{YELLOW}{cmyk}{0,0,1,0}
 }

\usepackage{amsfonts}
\providecommand{\U}[1]{\protect\rule{.1in}{.1in}}
\newcommand{\BOX}{\hbox {$\sqcap$ \kern -1em $\sqcup$}}\newcommand{\be}{\begin{equation}}\newcommand{\ee}{\end{equation}}\newcommand{\ba}{\begin{eqnarray}}\newcommand{\ea}{\end{eqnarray}}\newcommand{\ban}{\begin{eqnarray*}}\newcommand{\bea}{\begin{eqnarray}}\newcommand{\eea}{\end{eqnarray}}\newcommand{\ean}{\end{eqnarray*}}\newcommand{\barr}{\begin{array}}\newcommand{\earr}{\end{array}}

\makeatother

\usepackage{babel}

\makeatother

\usepackage{babel}

\makeatother

\usepackage{babel}

\makeatother

\usepackage{babel}

\begin{document}
\preprint{CERN-PH-TH/2010-284}
\preprint{KCL-PH-TH/2010-27}
\preprint{IFIC/10-48}
\vspace*{0.1cm}

\title{Implications of a Stochastic Microscopic Finsler Cosmology}
\author{Nick E.\ Mavromatos}
\affiliation{King's College London, Department of Physics, Strand, London WC2R~2LS, UK\\ CERN, Theory Division, CH-1211  Geneva 23, Switzerland}

\author{Vasiliki A.\ Mitsou}
\affiliation{Instituto de F\'{i}sica Corpuscular (IFIC), CSIC -- Universitat de Val\`{e}ncia, \\
Parc Cient\'{i}fic, Apartado de Correos 22085, E-46071 Valencia, Spain }

\author{Sarben Sarkar and Ariadne Vergou}
\affiliation{King's College London, University of London, Department of Physics, Strand,
London WC2R2LS, UK}

\begin{abstract}
Within the context of supersymmetric space-time (D-particle) foam in string/brane-theory,
we discuss a Finsler-induced Cosmology and its implications for (thermal) Dark Matter abundances.
This constitutes a truly microscopic model of dynamical space-time, where Finsler geometries arise naturally.
The D-particle foam model involves point-like brane defects (D-particles),
which provide the topologically non-trivial foamy structures of space-time.
The D-particles can capture and emit stringy matter and this leads
to a recoil of D-particles. It is indicated how one effect of such
a recoil of D-particles is a back-reaction on the space-time metric
of Finsler type which is stochastic. We show that such a type of stochastic
space-time foam can lead to acceptable cosmologies at late epochs of the Universe, due to
the non-trivial properties of the supersymmetric (BPS like)
D-particle defects, which are such so as not to affect significantly the Hubble expansion.
The restrictions placed on the free parameters
of the Finsler type metric are obtained from solving the Boltzmann
equation in this background for relic abundances of a Lightest Supersymmetric
Particle (LSP) dark matter candidate. It is demonstrated that the
D-foam acts as a source for particle production in the Boltzmann equation,
thereby leading to enhanced thermal LSP relic abundances relative
to those in the Standard $\Lambda$CDM Cosmology. For D-particle masses of order TeV, such effects may be relevant for dark matter searches at colliders. The latter constraints complement those coming from high energy gamma-ray astronomy on the induced vacuum refractive index that D-foam models entail. We also comment briefly on the production mechanisms of such TeV-mass stringy defects at colliders, which,  in view of the current LHC experimental searches, will impose further constraints on their couplings.

\end{abstract}

\maketitle

\date{\today}

\section{Introduction}

There has been recent interest in physics beyond the Standard Model which can incorporate Lorentz symmetry violations~\cite{kostelecky} in a Finsler geometry setting~\cite{finsler,finsler2,finsler2b}. The Finsler metrics are functions not only of the space-time coordinates but also of the tangent vectors (momenta) at points of the curved manifolds. Physical examples of Finsler metrics have readily arisen in analogue condensed matter systems~\cite{finsler3} and so recent attempts to base Cosmology on Finsler have been necessarily purely phenomenological~\cite{finslercosmo}. Hence the fundamental viability, from a physical point of view, cannot be tested in such studies.

In a recent Letter~\cite{ariadneplb} we have proposed a scheme, which involved back reaction of space-time defects within a supersymmetric string/brane framework. The back reaction has the form of an induced Finsler metric~\cite{finsler2}, but of a stochastic kind. In this longer article, we shall give details and address the implications of the induced stochastic Finsler geometry on the Hubble expansion of the Universe and the Dark Matter (thermal) relic abundance. Just as candidates for Dark Matter have to be sought from Physics beyond the Standard Model, we need to have a framework which goes beyond local effective Lagrangians, to incorporate the Finsler dynamical background. Furthermore, the localised space-time defects (D-particles~\cite{dfoam,westmuckett,emnnewuncert,li}) that we consider, are generic in string/brane theory. These localised defects, although point-like, nevertheless cannot be just considered as another form of particle excitation, as we shall discuss below. We give arguments which show that, owing to the presence of D-particles and D-particle/string bound states in the  vicinity of the brane Universe, a viable and plausible Cosmology, with small but measurable in principle implications for the Dark matter relic abundances, exists.
Together with the results of our earlier Letter~\cite{ariadneplb}, this represents the first attempt to construct a microscopic Finsler Cosmology, based on our current understanding of Gravitational Physics beyond the Standard Model, and in particular within a string/brane framework.

The nature of the Dark sector of our Universe constitutes one of the
major unresolved puzzles of modern physics. Indeed, according to observations
over the past twelve years, 96\% of our Universe energy budget consists
of unknown entities: 23\% is Dark Matter (DM) and 73\% is Dark Energy
(DE), a mysterious form of ground state energy. DE is believed to
be responsible for the current-era acceleration of the Universe. These
numbers have been obtained by best-fit analyses of a plethora of astrophysical
data to the so-called Standard Cosmological Model ($\Lambda$CDM),
which is a Friedmann-Robertson-Walker (FRW) cosmology, involving cold
DM as the dominant DM species, and a positive cosmological constant
$\Lambda>0$; the data range from direct observations of the Universe
acceleration, using type-Ia supernovae~\cite{snIa}, to cosmic microwave
background~\cite{cmb,7yrwmap}, baryon oscillation~\cite{bao} and weak
lensing data~\cite{lensing}. It should be stressed that the afore-mentioned
energy budget depends crucially on the theoretical model for the Universe
considered.

An interesting, and not commonly discussed, class of cosmological
models that may lead to modifications of the Dark sector, involves
space-time with a ``foamy'' structure at microscopic (Planckian
or string) scales~\cite{wheeler}, due to quantum gravitational interactions.
In the past, for a variety of reasons, such models (differing in the
details of the constructs of space-time foam) have been considered
by many authors. They exhibit a profusion of features that can be
falsified experimentally and their predictions range from light-cone
fluctuations caused by stochastic metric fluctuations~\cite{ford},
to macroscopic Lorentz symmetry violations~\cite{finsler,lorentz}. As stated earlier, in this
note we would like to present a first study examining the contribution
of such space-time metric stochastic fluctuations on the Dark sector
and in particular on the DM sector of the Universe. We shall focus
on a particular class of stochastic space-time foam models, inspired
by certain types of string theory. These involve localized space-time
defects (D0-branes or D-particles)~\cite{dfoam} which are either
allowed background configurations in (supersymmetric) type IA string
models~\cite{dfoam,westmuckett,emnnewuncert} or arise effectively
from suitable compactifications of higher dimensional branes (e.g.\
D3-branes wrapped up in appropriate three cycles in the context of
type~IIB strings~\cite{li}). Observers on the brane detect a foam-like
structure due to the crossing of the brane by D-particles. In this
higher-dimensional geometry, only gravitational fields are allowed
to propagate in the bulk; all other particle excitations, including
DM candidates, are assumed to be described by open strings with their
ends attached to the brane world. The brane is assumed to have three
large spatial dimensions, and, depending on the model of string
theory considered, it may have a number of compactified extra dimensions.

Dynamical D-particles should \emph{not} be viewed as material excitations
of the vacuum but rather as \emph{vacuum structures}. This contrasts
with attempts to represent such D-particles as ordinary superheavy
DM excitations from the vacuum~\cite{D0matter}, owing to the completely
localized nature of the D-particles in the extra dimensions. In our
construction, closest in spirit to weak coupling string theory, they
are just \emph{vacuum defects}. In fact it can be shown that the gravitational
interactions among such (BPS) D-particles are cancelled by appropriate
gauged repulsive forces induced on them from other branes in our supersymmetric
models of D-foam~\cite{dfoam,westmuckett}. Hence such a collection
of D-particles in the bulk does not affect the Hubble expansion on
the D3-brane worlds, and so their concentration cannot be restricted
by considerations on overclosure of the Universe within our models.
However important constraints on the density of defects in D-foam models can
still be imposed by astrophysical experiments on the arrival times
of high energy cosmic photons~\cite{arrival}. According to these
string-foam models, there should be an effective \emph{refractive}
index in vacuo, such that higher energy photons would be delayed more,
since they would cause stronger disturbance on the background space-time~\cite{dfoam,emnnewuncert,li}.
The interaction of material open strings (such as photons) with the
D-particles leads to an induced distortion of space-time described
by a metric, which depends on both the coordinate and momentum transfer
of the photon during its scattering with the defect and so has similarities
to a Finsler metric~\cite{finsler}. This is a topologically non-trivial
process, involving the creation of a non-local intermediate string
state, oscillating from zero length to a maximum one, according to
a time-space stringy uncertainty~\cite{sussk1,emnnewuncert}. This
causes a time delay for the photon emerging after capture by the defect,
proportional to the incident energy of the photon.

The purpose of this article is to analyze, in the same spirit, the
modification of the estimate of the DM budget of the Universe as compared
to the $\Lambda$CDM model due to the quantum fluctuations of the
D-particles.%
\footnote{ The D-foam contributions to the Dark Energy budget
have been discussed already in~\cite{westmuckett,emnnewuncert}, and found to be
phenomenologically acceptable in magnitude, and
almost constant for the late eras of the Universe; the contribution may
be guaranteed by tuning the adiabatic velocity of the slowly moving
D-brane world in the bulk. The existing supersymmetries in the construction
of~\cite{westmuckett} guarantee a vanishing vacuum energy when no
relative motion between D-branes and D-particles is present. This non-trivial input
to DE comes from stretched strings between the bulk D-particles
and the brane world, in directions perpendicular to the (uncompactified)
brane dimensions. Any motion of D-particles parallel to the brane
world, including their recoil during interaction with matter/radiation
on the brane, does not lead to vacuum energy contributions. %
} In fact, as we shall argue, the propagation of massive DM particles
on such space-times, induces a back-reaction, which in turn has consequences
for the amount of thermal DM relics of these particles; this, in turn,
impacts on astroparticle tests of particle physics models incorporating
supersymmetry (SUSY), that provide currently one of the leading DM candidate
species, the neutralino. Its thermal abundance, calculated within
the simplest SUSY models (minimal supersymmetric model embedded
in minimal supergravity~\cite{msugra}), is heavily restricted by
cosmic microwave background data; hence the available parameter space
for these simplest supersymmetric models may vanish in the near future
on incorporating also data from collider experiments such as the LHC
at CERN~\cite{neutralino}. These constraints depend strongly on
details of theoretical models. In the presence, for instance, of extended
supergravity models with time-dependent dilaton-$\phi$ sources~\cite{elmn},
which characterize certain (non-equilibrium) string cosmology models~\cite{cosmo},
the calculated amount of thermal neutralino relic abundance can be
smaller than the one calculated within the $\Lambda$CDM-minimal supergravity
cosmology. Such dilaton models allow more scope for SUSY,
which can thus survive the otherwise stringent tests at the LHC~\cite{dutta}.
In our work we will find that the effects of the D-foam on the thermal
relic abundances oppose those from the dilaton models. These effects
become bigger for models allowing low string mass scales.

For clarity we will first outline the basic reason behind such modifications
in the DM thermal abundances~\cite{ariadneplb}. Non-equilibrium cosmology models are
associated with space-time distortions, due to either the presence
of time-dependent dilaton sources (cf.\ string cosmologies involving supercritical (SSC) dilaton
quintessence~\cite{elmn}), or the induced back-reaction
of the DM particles onto the space-time itself. Boltzmann equations
are used to determine the thermal DM species cosmic abundances. The
effect of space-time distortions appears as extra contributions to
a source $\Gamma(t)$ for particle production (at cosmic Robertson-Walker
time $t$) on the right-hand side of the appropriate Boltzmann equation.
In general (for a Universe with three spatial dimensions (i.e.\ a 3-brane),
with no disappearance of particles into the bulk), the apposite Boltzmann
equation for $n(t)$, the density of the DM species, reads: \begin{equation}
\frac{d\, n}{dt}+3H\, n=\Gamma(t)\, n+\mathcal{C}[n],\label{boltzmann}\end{equation}
 where $H=\frac{\frac{da}{dt}}{a}\equiv\frac{\dot{a}}{a}$ is the
Hubble ratio, $a$ being the scale factor in the Robertson-Walker
metric, and $\mathcal{C}[n]$ is the standard collision term describing
the deviations from thermal equilibrium of the species population. In keeping with standard cosmologies, this has the form~\cite{kolb}
\begin{equation}
\mathcal{C}[n]=-\langle\tilde{\sigma}\, v\rangle\left(n^{2}-n_{{\rm eq}}^{2}\right),\end{equation}
 with $n_{{\rm eq}}$ a thermal equilibrium density of a heavy dark matter species, $\tilde{\sigma}$
the total cross-section evaluated in the background metric, and $v$
the Mueller velocity.

The cosmological implications of the D-particle foam model are not restricted to
Dark Matter abundances. It is also natural to explore the possible r\^{o}le of D-particle foam~\cite{dfoam,westmuckett} on particle statistics,
extending and completing the ideas outlined in~\cite{tsallissarkar}.
The implication in the early universe of the statistical description
of particles has been investigated~\cite{pessah} within the context
of non-extensive statistics pioneered by Tsallis~\cite{tsallis}.
It is interesting, owing to the pervasive nature of space-time foam within our current physical
frameworks, to compare its implications for statistics
with the paradigm of non-extensive statistics.

The structure of the article is as follows: in section~\ref{sec:dfoam}
we review the model of stringy space-time foam (D-particle) on which
we base our analysis. We pay particular attention in discussing the
induced modifications to the space-time metric as a result of the
interactions of matter (and radiation) particle excitations with the
D-particle defects of the foam. As already mentioned, such metrics
depend on the momentum variables of the particle, in addition to any
coordinate dependence, and hence the induced metric is of Finsler
type~\cite{finsler}.  We also discuss briefly non-trivial issues associated with
the breaking or better obstruction of target-space supersymmetry that should characterize any phenomenologically realistic
model.
In section~\ref{sec:statistics} we analyze the implications
of D-foam on particle statistics. The reasons why the statistical
distributions of particles change in the presence of the foam is the
modification of the energy dispersion relations of particle probes,
as a result of the stochastic metric background. This affects the
relevant distributions. An interesting feature of our model is that
the foam effects are different between fermion and boson excitations.
In section~\ref{sec:cosmology} we discuss cosmological aspects of the D-foam,
and in particular its effects on the Hubble expansion of the Universe and  the induced modifications to the Boltzmann equation
that determines the (thermal) Dark Matter relic abundances. The
modifications to the Boltzmann equation are due to the Finsler~\cite{finsler}
nature of the foam-induced space-time metric. We find that the presence
of the foam acts as a source for particle production and affects the
form of the Boltzmann equation in such a way as to lead to an increase
of the DM thermal relic abundance in comparison to the foam-free Standard
Cosmology model. Moreover, contrary to conventional string/brane approaches, we argue that the D-particles in the models of D-foam
discussed here behave as a dark energy fluid rather than dark matter.
Some elementary phenomenological considerations,
concerning falsification of foam models involving low-string scales
via WMAP constraints, are presented in section~\ref{sec:wmap}.
We also comment briefly on the production mechanisms of TeV-mass-scale D-particles at colliders,
and argue that current experimental searches at the LHC can impose further constraints on the effective couplings
of these defects to Standard Model particles. The strength of such couplings, however, as well as the production mechanism itself, are highly string-model-dependent, and thus we reserve a detailed analysis along these lines for a future work. Finally, in section~\ref{sec:conclu}, we give our conclusions and outlook. Technical aspects of the work are presented in two appendices.

\section{Stochastic D-foam Basics \label{sec:dfoam}}

\begin{figure}[ht]
\centering
\includegraphics[width=7.5cm]{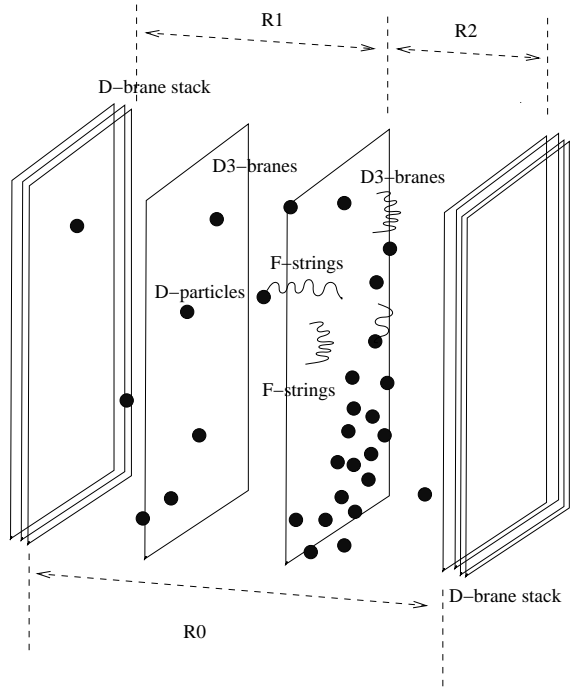}
\caption{Schematic
representation of a generic D-particle space-time foam model. The
model of ref.~\cite{westmuckett}, which acts as a prototype of a D-foam, involves two stacks of D8-branes,
each stack being attached to an orientifold plane. Owing to their special reflective properties, the latter provide
a natural compactification of the bulk dimension. The bulk is punctured by D0-branes (D-particles), which are allowed in the type~IA string theory of~\cite{westmuckett}. The  presence of a D-brane
is essential due to gauge flux conservation, since an isolated
D-particle cannot exist. Open strings live on the brane world, representing Standard Model Matter and
they can interact in a topologically non-trivial way with the D-particle defects in the foam. Recoil of the
D-particle during such interactions creates appropriate distortion in the space-time geometry, which depend on the momenta of the incident string states, and thus are of a generalized Finsler type.}%
\label{fig:recoil}%
\end{figure}

A class of model stochastic space-time foam has been suggested which
is based on a gravitational foam consisting of real (rather than virtual)
space-time defects in higher dimensional space-time, consistent with
the viewpoint that our world is a brane hyperspace embedded in a bulk
space-time~\cite{dfoam,westmuckett}. In general the construction
of a model involves a number of parallel brane worlds with three large
spatial dimensions, the required number being determined by target
space SUSY. These brane worlds move in a bulk space-time
containing a gas of point-like bulk branes, called D-particles, which
are stringy space-time solitonic defects (cf.\ fig.~\ref{fig:recoil}).
One of these branes is the observable universe. On this brane the
D-particles will appear as space-time defects. Typically open strings
interact with D-particles and satisfy Dirichlet boundary conditions
when attached to them. Closed and open strings may be cut by D-particles.
D-particles are allowed in certain string theories such as bosonic
and type~IIA. Here we will consider them to be present in string
theories of phenomenological interest~\cite{emnnewuncert} since,
even when elementary D-particles cannot exist consistently, as is
the case of type~IIB string models,
there can be effective D-particles formed by the compactification
of higher dimensional D-branes~\cite{li}. Moreover D-particles are
non-perturbative constructions since their masses are inversely proportional
to the the string coupling ${g_{s}}$. The study of D-brane dynamics
has been made possible by Polchinski's realization that such solitonic
string backgrounds can be described in a conformally invariant way
in terms of world sheets with boundaries~\cite{polch2}. On these
boundaries Dirichlet boundary conditions for the collective target-space
coordinates of the soliton are imposed~\cite{coll}. When low energy
matter given by a closed string propagating in a $\left(d+1\right)$-dimensional
space-time collides with a very massive D-particle (D0-brane) embedded
in this space-time, the D-particle recoils as a result~\cite{kogan}.

In this article, we shall consider the simple case of bosonic
stringy matter coupling to D-particles and so we can only discuss
matters of principle and ignore issues of stability. However we should
note that an open string model needs to incorporate for completeness,
higher dimensional D-branes such as the D3-brane. This is due to the
vectorial charge carried by the string owing to the Kalb-Ramond field.
Higher dimensional D-branes (unlike D-particles) can carry the charge
from the endpoints of open strings that are attached to them. For
a closed bosonic string model the inclusion of such D-branes is not
imperative. The requirement of higher dimensional branes is for consistency,
but is, otherwise, not pertinent to our
analysis. The current state of phenomenological modelling
of the interactions of D-particle foam with stringy matter will be
briefly summarized now. Since there are no rigid bodies in general
relativity the recoil fluctuations of the brane and their effective
stochastic back-reaction on space-time cannot be neglected. D-particle
recoil in the ``tree approximation'', i.e.\ in lowest order in the
string coupling $g_{s}$, corresponds to the punctured disc or Riemann
sphere approximation in open or closed string theory and induces a
non-trivial space-time metric.

\subsection{Elements of Logarithmic Conformal Field Theory of D-particle Recoil and Space-Time Implications \label{sec:lcft}}

In a flat target-space time, the recoil conformal field theory has
been developed in~\cite{kogan,szabo}. The pertinent vertex operator
for our purposes reads: \begin{equation}
V=\int_{\partial\Sigma}u_{i}X^{0}\Theta_{\varepsilon}(X^{0})\partial_{n}X^{i}\label{recvec}\end{equation}
 with \begin{eqnarray}
\Theta_{\varepsilon}\left(t\right) & = & \frac{1}{2\pi i}\int_{-\infty}^{\infty}\frac{dq}{q-i\varepsilon}e^{iqt}.\label{heaviside}\end{eqnarray}
 Above, $u_{i}$ is the recoil velocity of the D-particle, related
to the momenta $k^{(1)}\left(k^{(2)}\right)$ of the propagating string
state before (after) the recoil via: \begin{equation}
u_{i}=g_{s}\frac{k_{i}^{(1)}-k_{i}^{(2)}}{M_{s}}\equiv g_{s}\frac{\Delta k_{i}}{M_{s}}\label{defu}\end{equation}
 as a result of momentum conservation during the scattering event.
In the above relation $\Delta k_{i}$ is the momentum transfer during
a collision, $g_{s}<1$ is the (weak) string coupling, and $M_{s}$
is a string mass scale. $M_{D}=M_{s}/g_{s}$ is the D-particle mass,
which may be identified with the Planck mass $M_{P}$. These relations
have been calculated for non-relativistic D0-branes where $u_{i}$
is small.

The operator (\ref{recvec}) has non-trivial anomalous dimension $-\varepsilon^{2}/2<0$,
and as such the theory requires Liouville dressing to restore the
conformal symmetry~\cite{dfoam}. It can be shown that the theory
is supercritical (once the underlying string theory lives in its critical
target-space dimension), and the Liouville dressing results in the
operator (\ref{recvec}) being replaced by the world-sheet bulk operator~\cite{dfoam}:
\begin{equation}
\int_{\Sigma}e^{\alpha\varphi}u_{i}\left(\Theta_{\varepsilon}\left(t\right)\partial_{\alpha}t\partial^{\alpha}X^{i}+t\,\delta_{\varepsilon}(t)\partial_{\alpha}t\partial^{\alpha}t+t\partial^{2}X^{i}\right),\end{equation}
 where $\alpha=\frac{\varepsilon}{\sqrt{2}}$ is the Liouville anomalous
dimension. Upon considering times $t$ relatively long after the scattering,
and using the world-sheet equations of motion in flat target space
time, we observe that only the first term remains. Moreover, upon
identifying the target-time $t$ with the Liouville zero mode $\varphi_{0}/\sqrt{2}$,
we obtain for the target-space time metric~\cite{dfoam,mav2} \begin{equation}
g_{ij}=\delta_{ij},\, g_{00}=-1,\, g_{0i}=\frac{1}{2}u_{i}e^{\varepsilon\,\frac{\varphi_{0}}{\sqrt{2}}-\varepsilon\, t}=\frac{1}{2}u_{i}~,\qquad t>0,\label{recoil}\end{equation}
 where the suffix $0$ denotes temporal (Liouville) components. Notice
that the world-sheet $\sigma$-model time $t$ is assumed of Euclidean
signature, for reasons of convergence of the world-sheet path integral.
It is in this signature that the logarithmic conformal algebra of
recoil is rigorously valid~\cite{kogan,gravanis}. However, the target-space
time signature is Minkowskian, due to the supercriticality of the
Liouville string, after the identification of the Liouville mode with
the target time ($\varphi=\sqrt{2}t$): \begin{equation}
ds^{2}=-d\varphi^{2}+dt^{2}+u_{i}dtdX^{i}+(dX^{j})^{2}=-dt^{2}+u_{i}dtdX^{i}+(dX^{j})^{2}.\end{equation}
 The fact that the induced deformation $g_{0i}$ depends on the energy
content ($u_{i}$, cf.\ (\ref{defu})) of the low energy particle,
implies~\cite{dfoam} that the space-time metric in this case is
akin to a Finsler metric~\cite{finsler}. Such a feature does not
arise in other approaches to space-time foam and will be important
in the formulation of our microscopic model. In a stochastic D-particle
recoil model the velocity $u_{i}$ is taken to be \emph{random} (owing
to a lack of any particular prior knowledge). Specifically it is taken~\cite{Dparticle}
to have a distribution which is Gaussian with zero mean and variance
$\sigma$. The \emph{zero mean} of the distribution preserves \emph{Lorentz
symmetry on average}, which is a desirable feature for a (string theory)
model, both conceptually and phenomenologically. However, the stochastic
recoil velocity fluctuations break the symmetry (but the effects are
suppressed since the respective variances are assumed to be inversely
proportional to the effective Planck mass) and, on cosmological scales,
macroscopic effects of such a breaking may arise. The form of this
stochastic induced metric will serve as the starting point of our
approach to non-conventional statistics~\cite{tsallissarkar}.

Embedding the space-time D-foam in a Friedmann-Robertson-Walker
expanding Universe with scale factor $a(t)$, requires some discussion.
First of all we need an understanding of the recoil motion of D-particles,
and thus of the associated Finsler-distortions of the neighboring
space-time. For late eras, of particular relevance for us, the FRW
Universe obeys a power-law expansion, this issue has been discussed
in detail in~\cite{gravanis}. In this case the deformation (\ref{recoil})
can be found by considering the geodesic motion of the D-particles.

For completeness of our discussion let us review briefly the derivation
of the induced metric deformations due to recoil in such a FRW universe
background. We shall work with space-times of the form \begin{equation}
ds^{2}=-dt^{2}+a(t)^{2}(dX^{i})^{2}~\quad~{\rm with~scale~factor}\quad a(t)=a_{0}t^{p},\qquad p<1,\label{rwmetric}\end{equation}
 characterizing late eras of the Universe (radiation and matter dominance)
of interest to us here. The pertinent geodesic equations in this case
read: \begin{eqnarray}
\ddot{t}+pt^{2p-1}(\dot{y}^{i})^{2} & = & 0,\nonumber \\
\ddot{y}+2\frac{p}{t}(\dot{y}^{i})\dot{t} & = & 0,\end{eqnarray}
 where the dot denotes differentiation with respect to the proper
time $\tau$ of the D-particle.

For a single D-particle defect in the expanding Universe, for which
the impact with a matter string state occurred at the time moment
$t_{0}$, one has the initial conditions $y^{i}(t_{0})=0$ (for simplicity),
and $dy^{i}/dt(t_{0})\equiv u^{i}$, with $u_{i}$ the recoil velocity
(\ref{defu}). In this case one easily finds that, for relatively
(but not cosmologically) long times $t\gg t_{0}$ after the event,
the solution is: \begin{equation}
y^{i}(t)=\frac{u^{i}}{1-2p}\left(t^{1-2p}t_{0}^{2p}-t_{0}\right)+{\mathcal O}(t^{1-4p}),\qquad t\gg t_{0}.\label{pathexpre}\end{equation}
 To leading order in $t$, therefore, the appropriate vertex operator
describing the recoil of the $D$-particle in the FRW universe (\ref{rwmetric})
is: \begin{equation}
V=\int_{\partial\Sigma}a_{0}^{2}\frac{u^{i}}{1-2p}
\Theta(t-t_{0})\left(tt_{0}^{2p}-t_{0}t^{2p}\right)\partial_{n}X^{i},\label{path2}
\end{equation}
 where $\Theta(t-t_{0})$ is the Heaviside step function, expressing
an instantaneous action \emph{(impulse)} on the D-particle at
$t=t_{0}$.

To find the induced metric distortions due to the D-particle recoil
(\ref{path2}), we first pass into the T-dual Neumann picture for
the (spatial coordinate) $X^{i}$ $\sigma$-model field (assuming
T-duality to be an exact symmetry of string theory) of D-particle
recoil and rewrite (\ref{path2}) as a world-sheet bulk operator~\cite{gravanis}:
\begin{eqnarray}
V=\int_{\partial\Sigma}g_{ij}y^{j}(t)\partial_{\tau}X^{i}=\int_{\Sigma}\partial_{\alpha}
\left(y_{i}(t)\partial^{\alpha}X^{i}\right)=\int_{\Sigma}
\left(\dot{y}_{i}(t)\partial_{\alpha}t\partial^{\alpha}X^{i}+y_{i}\partial^{2}X^{i}\right),\label{bulkop}
\end{eqnarray}
 where the dot denotes derivative with respect to the target time
$t$, $\alpha$ is a world-sheet index, and $\partial^{2}$ is the
world-sheet Laplacian. This deformation is non-conformal, and one
should dress it with a Liouville mode, which should then be identified
with the target time. The discussion parallels the flat space-time
case outlined above and will not be repeated here. Notice that it
is the covariant vector $y_{i}$ which appears in the formula, and
incorporates the metric $g_{ij}$, since $y_{i}=g_{ij}y^{j}$.

To determine the background geometry, for the string motion, it suffices
to use the classical motion of the string, described by the world-sheet
equations: $\partial^{2}X^{i}+{\Gamma^{i}}_{\mu\nu}\partial_{\alpha}X^{\mu}\partial^{\alpha}X^{\nu}=0$,
where $\mu,\nu$ are space-time indices, $\alpha=1,2$ is a world-sheet
index, $\partial^{2}$ is the Laplacian on the world sheet, and $i$
is a target spatial index. The relevant Christoffel symbol in our
FRW background case (\ref{rwmetric}), is $\Gamma_{0i}^{i}$, and thus
the operator (\ref{bulkop}) becomes: $\int_{\Sigma}\left({\dot{y}}_{i}-2y_{i}(t){\Gamma^{i}}_{0i}\right)\partial_{\alpha}t\partial^{\alpha}X^{i}$,
from which we can read off an induced non-diagonal component for the
space time metric \begin{eqnarray}
2g_{0i}=\dot{y}_{i}-2y_{i}(t)\Gamma_{ti}^{i}.\label{indmetr}\end{eqnarray}
 In the FRW background, the \emph{covariant} path $y_{i}(t)=g_{ij}y^{j}(t)$
is described by: \begin{eqnarray}
y_{i}(t)=\frac{u_{i}\, a_{0}^{2}}{1-2p}\left(tt_{0}^{2p}-t_{0}t^{2p}\right)\end{eqnarray}
 (with $a_{0}$ the current value of the scale factor), which gives
\begin{equation}
2g_{0i}=a^{2}(t_{0})u_{i}~.\label{rwdistortion}\end{equation}
 This yields for the metric line element (after the identification
of the Liouville mode with time): \begin{equation}
ds^{2}=-dt^{2}+u_{i}a^{2}(t_{0})dtdX^{i}+a^{2}(t)(dX^{i})^{2},\qquad{\rm for}\quad t>t_{0}.\label{fixedpoint}\end{equation}
 As expected, this space-time has precisely the form corresponding
to a Galilean-boosted frame (the rest frame of the D-particle), with
the boost occurring suddenly at time $t=t_{0}$.

Such metrics can be generalized to describe the distortions of the
space-time as a consequence of collisions of both open and closed
strings with D-particle defects. Open strings are of particular importance
to us, since they describe matter excitations of brane worlds, which
can propagate in bulk space-times punctured by D-particles. As the
brane world moves in the bulk, the D-particles cross it, and from
the point of view of an observer on the brane they appear as space-time
defects which {}``flash on and off''. ``Foamy'' space-time
features on the brane result as a consequence (cf.\ fig.~\ref{fig:recoil}).
When considering a collection of uniformly distributed D-particles,
there are moments $t_{0}$, at which matter strings collide with defects.
In (\ref{fixedpoint}) the time $t$ is taken to be after the moment
of impact but not much after on a cosmological scale. In a coarse-grained
approximation, for non-clustering D-particle populations, one may
assume that $a(t_{0})\simeq a(t)$ to a good approximation. From now
on it will be assumed that matter with momentum $\vec{p}$ and energy
$E$ induces the following space-time distortions (on our brane world,
cf.\ fig.~\ref{fig:recoil})) as a result of its interaction with
the D-foam: \begin{equation}
ds^{2}\simeq-dt^{2}+u_{i}a^{2}(t)\left(dtdX^{i}+(dX^{i})^{2}\right).\label{approx}\end{equation}
 Notice that in this formalism the time is already of Minkowskian
signature, as a result of the Liouville dressing.

The reader should notice that the recoil velocity of the point-like D-particles has been taken to have non-zero components only on the brane world, where the matter excitation resides. This is an approximation, which proves sufficient for the phenomenology of our model. Moreover, for the case of D-particles represented by wrapped-up D3  branes in type IIB strings~\cite{li}, the situation is complicated; the defects from such constructions entail additional degrees of freedom. For simplicity we shall ignore these complications, and concentrate on \textit{effectively} point-like defects, whose recoil induces a back-reaction on space-time represented by a metric of the coarse-grained form (\ref{approx}).

In phase space, for a D3-brane world, the function $u_{i}$, (\ref{defu}),
involving a momentum transfer, $\Delta k_{i}$, can be modelled by
a local operator using the following parametrization~\cite{Dparticle}:
\begin{equation}
u_{i}=g_{s}\frac{\Delta k_{i}}{2M_{s}}=r_{i}k_{i}\,~,\,{\rm no~sum}\, i=1,2,3,\label{defu2}\end{equation}
 where the (dimensionful) variables $r_{i},i=1,2,3$, appearing above,
are related to the fraction of momentum that is transferred at a collision
with a D-particle in each spatial direction $i$. These parameters
are taken as gaussian normal random variables with a range $-\infty$
to $+\infty$ and defining moments \begin{equation}
\langle r_{i}\rangle=0,\label{moment1}\end{equation}
 \begin{equation}
\langle r_{i}r_{j}\rangle=0,\quad\textrm{if }\, i\neq j\label{moment2}\end{equation}
 and \begin{equation}
\sigma_{i}^{2}=\langle r_{i}^{2}\rangle-\langle r_{i}\rangle^{2}=\langle r_{i}^{2}\rangle\neq 0.\label{moment3}\end{equation}
 A \emph{homogeneous} foam situation will be assumed for cosmological eras after freeze-out of species, the situation of interest to us. It is, of course, understood that one may encounter inhomogeneities in more general models of D-foam, due to variations in the population of the D-particle defects at various epochs of the  evolution of the Universe. However, such inhomogeneous contributions would be severely constrained by cosmological data at late eras of the Universe, and, in view of the weak effect of the foam, will be ignored in our analysis, to a first approximation.

An \emph{isotropic} foam situation would require $r_{i}=r$, for all $i=1,2,3$.
In such a case the variances $\langle r_{i}^{2}\rangle=\sigma_{0}^{2}$ along all
spatial directions. In what follows we keep our discussion more general,
and treat $r_{i}$ as differing along the various spatial directions.
These differences are small, and could be possibly responsible for
small-scale anisotropies in the large scale structure of the observable Universe,
which, to a large degree, is isotropic and homogeneous.

\subsection{Supersymmetry breaking and implications for D particles \label{sec:susy}}

An important issue~\cite{emnnewuncert} concerning our analysis pertains to effects of SUSY breaking in a phenomenologically realistic cosmological model of D-foam cosmology~\cite{westmuckett,li}. The important constituents in our construction are D-particles. The word particle in the latter is somewhat misleading. BPS D-branes in general ( and D0-branes or D particles in particular) are described within a conformal field theory as hypersurfaces on which open strings end. They can also be described as classical solitons (or defects) of an appropriate supergravity. For BPS D-particles (which break half the space-time SUSY) gravitational attractions are canceled by gauge repulsions (i.e. a no force condition holds). In this way, their gravitational effects in the expansion of the brane Universe, along the longitudinal (brane) directions, are suppressed. Typically the conformal field theory description is appropriate for weak string coupling (relevant for us) whereas the soliton representation is appropriate for strong coupling. However, in  constructions of physical relevance, where many D-branes (through stacks, intersections and foam) are involved, a solitonic (i.e. defect) description is also possible at weak coupling. However SUSY is broken and the BPS condition will not hold. We will discuss the degree to which the no force condition still applies.

 Sen has shown that D-branes that break SUSY can be constructed in terms of branes and anti-branes (which can also be interpreted in terms of K theory)~\cite{sen}. In fact the D-brane spectrum of some orbifolds  of toroidal compactifications of type IIA/B superstring theory are compatible with the predictions of K theory. In general an orbifold of type IIA and IIB are generated by $g_{1}={\cal I}_n$ which denotes the reflection of $n$ co-ordinates or $g_2={\cal I}_n (-1)^{F_{L}}$ where ${F_{L}}$ is the fermion number of left moving particles. When $n$ is even the $g_{i}$ represent symmetries of type II theories. For $n=4$ a boundary state $|D0 \rangle$ for a stable non-BPS D-particle has been analysed at
 a perturbative level in the string coupling, at tree-string level in ref.~\cite{eyras} and at one-string-loop level in ref.~\cite{loops}. The interaction of two D-particles separated by a distance $r$ is determined by the amplitude
 $$\langle D0|{\cal D}_{a}|D0 \rangle$$
where ${\cal D}_{a}$ is the closed string propagator
$${\cal D}_{a}=\frac{\alpha}{4 \pi}\int_{|z|\leq 1}\frac{d^{2}z}{|z|^{2}}z^{L_{0}-a}{\bar{z}}^{\bar{L}_{0}-a}$$  with $a=\frac{1}{2}$ in the untwisted NS-NS sector and $a=0$ in the twisted R-R sector. The quantities $L_0$, $\bar{L}_0$ are the appropriate Virasoro operators.
A similar calculation can be done for two D-particles in relative motion with a speed $v$. This analysis results in the absence of a force between two non-BPS D-particles for short distances up to order $v^{2}$ (the next order being $O(v^{4})$).We should emphasize that the \emph{no-force} condition
among the \emph{non} BPS D-particles is found to occur at large
scales $r$, as compared with the string length $\ell_s = \sqrt{\alpha '}$, where
(target space) effective low-energy string action
methods are applicable. This feature is exclusive of a
\emph{critical-radius} orbifold compactification, and points to the fact that specific compactifications may indeed
preserve such conditions. However, the construction
of \cite{eyras} ignored higher-string loop effects. The incorporation
of such effects results in general in the destruction
of the no force condition~\cite{loops}, although under some circumstances
it may be made valid up to one loop. Non-perturbative effects from
string-loop resummation cannot be answered at present in this compactification framework. The above considerations will be used in section
\ref{sec:Hubble}, when we discuss the contributions of the D-foam on the expansion of the Universe and its vacuum energy budget.

  There are other compactification schemes. Currently it is believed that phenomenology requires a spontaneously broken SUSY reducing to a $N=1$ matter sector. This is hard to arrange in string theory model building. In models involving toroidal orbifold compactifications (as discussed above), attempts to obtain a realistic SUSY breaking scenario rely on non-perturbative effects such as gaugino condensation. The incorporation of D-branes altered the situation since they are sources of RR fluxes, permitting models with background fluxes. In our model we have a compact bulk dimension and fluxes cannot be turned on at will since they contribute to the vacuum energy. Orientifold planes are negative tension sources and are necessary to incorporate for consistency.For D-foam, we have however considered~\cite{Gravanis} a scenario of supersymmetry \emph{obstruction}~\cite{witten} rather than breaking, in the following sense: one may compactify the extra dimensions (relative to the three longitudinal directions) of the D8-brane world in  the model of \cite{westmuckett,emnnewuncert},( or the D7 brane worlds in the type IIB string model of \cite{li}), using certain ``magnetized'' manifolds (\emph{e.g}. tori)~\cite{bachas}. The resultant vacuum of the theory is still supersymmetric whereas the spectrum of excitations on the brane world (with three uncompactified longitudinal directions) is not. The bosonic and fermionic excitations couple differently to the ``magnetic'' fields in the extra compactified dimensions. Fermion-boson induced mass splittings are induced and can be made realistic~\cite{bachas} by appropriate choices of the ``magnetic'' field intensities -this is the analogue of the Zeeman effects in quantum mechanics. Thus the D-particles can be viewed as defects in a supersymmetric vacuum of a theory with a non-supersymmetric excitation spectrum; hence the no force condition would still be valid when there is no relative motion.  Local fluctuations of the recoil velocities of the foam, as a result of the interaction with stringy excitations, would break further the SUSY, as a result of the relative (slow) motion of the heavy D particle defects, but this phenomenon would not be sufficient to lead to significant gravitational effects of the D-particles on the brane worlds. Moreover, at late eras in the expansion of the Universe, the motion of the brane world in the bulk may be viewed as adiabatic and very slow, so again the relevant contribution to the breaking of the bulk supersymmetry would not be strong enough to create significant effects in the Brane Universe expansion along its longitudinal directions.

Before dealing with the stochastic fluctuations of the cosmological
space-time (\ref{approx}), we find it instructive to discuss aspects
of ``fuzzy'' particle statistics in the current era. The evolution of the scale
factor is ignored, i.e.\ the recoil of the D-particle is considered
in a Minkowski background. This is done in the next section, and will
help the reader to clarify the most important issues in our approach.

\section{Particle Statistics and the D-Foam Model \label{sec:statistics}}

At the macroscopic level there is some evidence~\cite{Dmatter} that
the velocity distribution of self-gravitating collisionless particles
follows non-extensive statistics~\cite{tsallis} and is consistent
with properties of dark matter halos in galaxies.
Here we will show that, at a microscopic level, our D-foam model shares
some features of non-extensive statistics for elementary particle
dark matter candidates. The usual formulation of non-extensive statistics
is based on the postulates of a generalized entropy $S_{q}$ and a
($q$) expectation value, $\langle q \rangle$, where $q$ is real. Explicitly, \begin{equation}
S_{q}=\frac{\sum_{i=1}^{w}(p_{i}-p_{i}^{q})}{q-1}\label{entropy}\end{equation}
 for a system with $w$ microstates having a probability $p_{i}$
of being in the $i$-th microstate. Moreover we still require $\sum_{i=1}^{w}p_{i}=1$.
As $q\rightarrow 1$ we recover the usual Shannon entropy. The $q$
expectation value $\langle O\rangle_{q}$ for an operator $O$ whose
expectation value in the $i$-th microstate is $O_{i}$, is defined
as \begin{equation}
\langle O\rangle_{q}=\sum_{i=1}^{w}O_{i}p_{i}^{q}.\label{expectn}\end{equation}
 These postulates lead to modified particle distributions which, in
the limit $q\rightarrow 1$, reduce to either Bose or Fermi particle
distributions. Via maximum entropy arguments the generalized particle
occupation numbers are given by \begin{equation}
\langle n_{j}\rangle_{q}=\frac{1}{[1+(q-1)\beta(\varepsilon_{j}-\mu)]^{1/(q-1)}+\xi},\label{genoccupn}\end{equation}
 where $\varepsilon_{j}$ is the energy of a single particle state
labelled by $j$ and $\xi=-1,\,1$ for Bose-Einstein and Fermi-Dirac
statistics respectively~\cite{tsallis}. The consequences of such a
$q$-deformation of the usual statistics for ``radiation''
and ``dust'' energy densities in the
early universe as well as for relic densities has been already discussed
when $q-1$ is very small~\cite{pessah}. Indeed this analysis used
the available experimental bounds to deduce the constraint $|q-1|<2.6\times 10^{-4}$.
In this approach $q$ remains a parameter which needs to be determined
outside the framework.

In the analysis ensuing from the space-time foam model, the paradigm
of D-particle recoil and induced metrics is allied to a conventional
statistical mechanics (with an unconventional metric background). The
parameter that we introduce will measure the ``fuzziness''
of space-time induced by the interaction of D-particles with stringy
matter. The resulting randomly fluctuating space-time on a flat, non-expanding
FRW background has the following form ((\ref{defu2})): \begin{equation}
g_{\mu\nu}=\left(\begin{array}{cccc}
-1 & r_{1}k_{1} & r_{2}k_{2} & r_{3}k_{3}\\
r_{1}k_{1} & 1 & 0 & 0\\
r_{2}k_{2} & 0 & 1 & 0\\
r_{3}k_{3} & 0 & 0 & 1\end{array}\right).\label{metric}\end{equation}
 The reader should recall that the (dimensionful) variables $r_{i},\, i=1,2,3$
are taken as gaussian normal random variables with a range $-\infty$
to $+\infty$ and defining moments as given by (\ref{moment1}), (\ref{moment2}),
(\ref{moment3}). Any potentially unphysical values of momenta introduced
by this approximate model are rendered insignificant by the smallness
of the variance of the $r_{i}$ (suppressed as inverse Planck mass ).

\subsection{Energy-momentum dispersion relations}

For the statistical mechanical analysis we will need the energy-momentum
dispersion relation in this unconventional metric. One way this can
be found is by analyzing a massive spinless scalar particle moving
in this background. In an arbitrary gravitational field $g_{\mu\nu}$
the massive Klein-Gordon equation for a field $\Phi$ reads as:

\begin{equation}
[g^{\mu\nu}\partial_{\mu}\partial_{\nu}-m^{2}]\Phi=0.\end{equation}

If we expand this for a general metric where all terms contribute,
we have: \begin{equation}
\left[g^{00}(\partial_{0})^{2}+2g^{0i}\partial_{0}\partial_{i}+\sum_{i}g^{ii}(\partial_{i})^{2}+2g_{i\neq j}^{ij}\partial_{i}\partial_{j}-m^{2}\right]\Phi=0,\end{equation}
 where Latin indices run over 1,2,3; in the interests of a streamlined
notation, the Einstein summation convention is adopted when we have
two identical indices but not when we have more than two. For plane-wave solutions \begin{equation}
\Phi(\vec{x},t)=\phi(\vec{k},\omega_{r})\exp(i(-\omega_{r}t+\vec{k}\cdot\vec{x})),\end{equation}
 $\omega_{r}$ satisfies \begin{equation}
g^{00}\omega_{r}^{2}-2g^{0j}k_{j}\omega_{r}+\sum_{i}g^{ii}k_{i}^{2}+2\left(g^{ij}k_{i}k_{j}\right)_{i\neq j}+m^{2}=0,\end{equation}

\noindent or equivalently,

\begin{equation}
\omega_{r}=\frac{g^{0j}k_{j}\pm\sqrt{\left(g^{0j}k_{j}\right)^{2}-g^{00}\left(g^{ij}k_{i}k_{j}+m^{2}\right)}}{g^{00}}.\label{energy1}\end{equation}
 This can be further simplified on noting the form of the contravariant
metric: \begin{equation}
g^{\mu\nu}=\left(\begin{array}{cccc}
-\frac{1}{1+B} & \frac{r_{1}k_{1}}{1+B} & \frac{r_{2}k_{2}}{1+B} & \frac{r_{3}k_{3}}{1+B}\\
\frac{r_{1}k_{1}}{1+B} & 1-\frac{r_{1}^{2}k_{1}^{2}}{1+B} & -\frac{r_{1}r_{2}k_{1}k_{2}}{1+B} & -\frac{r_{1}r_{3}k_{1}k_{3}}{1+B}\\
\frac{r_{2}k_{2}}{1+B} & -\frac{r_{1}r_{2}k_{1}k_{2}}{1+B} & 1-\frac{r_{2}^{2}k_{2}^{2}}{1+B} & -\frac{r_{2}r_{3}k_{2}k_{3}}{1+B}\\
\frac{r_{3}k_{3}}{1+B} & -\frac{r_{1}r_{3}k_{1}k_{3}}{1+B} & -\frac{r_{2}r_{3}k_{2}k_{3}}{1+B} & 1-\frac{r_{3}^{2}k_{3}^{2}}{1+B}\end{array}\right),\label{metric2}\end{equation}
 where $B\equiv \sum_{j=1}^3 r_{j}^{2}k_{j}^{2}$.

Choosing from (\ref{energy1}) the positive-energy solution (($-$)
sign for the metric signature that we have chosen) and keeping terms
up to $r_{i}^{2}$, we obtain the dispersion relation:

\begin{equation}
\omega_{r}=E-\sum_{j}k_{j}^{2}r_{j}+\frac{E}{2}\sum_{j}k_{j}^{2}r_{j}^{2},\label{energy2}\end{equation}

\noindent where we dropped cross terms of the form $r_{i}r_{j}$ for $i\neq j$
(these terms would in any case vanish in the end because of (\ref{moment2})), i.e.\ used that:

\begin{equation}
\left(\sum_{j}k_{j}^{2}r_{j}\right)^{2}=\sum_{j}k_{j}^{4}r_{j}^{2}\end{equation}

\noindent and $E$ is simply a notation and stands for the standard Minkowski energy: \begin{equation}
E=\sqrt{\sum_{j=1}^{3}k_{j}k^{j}+m^{2}}.\label{onshell}\end{equation}
 We remind the reader at this point that we use the standard Minkowski metric
to transform between contravariant and covariant vectors, i.e.\
\begin{equation}
k^{i}=k_{i},\quad k^{0}=\omega_{r}=-k_{0}~.\label{contrav}\end{equation}
\noindent
When the expansion of the Universe will be taken into account in subsequent sections, the full Friedmann-Robertson-Walker metric will be used.

\subsection{Distribution functions}

From the standard version of statistical mechanics, in equilibrium,
at a finite temperature $T$, we have for the particle distribution function (in units $\hbar = c = 1$) \begin{equation}
\langle n\rangle_{r}=\frac{1}{\exp(\beta(\omega_{r}-\mu))+\xi},\label{statmech}\end{equation}
 where $\xi=+1$ applies to fermions and $\xi=-1$ applies to bosons,
$\beta=\frac{1}{T}$ and $\omega_{r}$ is given in (\ref{energy2}).
Writing down the Taylor expansion of (\ref{statmech}) with respect
to the small parameters $r_{j}$ and keeping up to terms $r_{j}^{2}$
yields:
\begin{eqnarray}\label{dist}
\langle n\rangle_{r} & = & \frac{1}{\exp\left(\beta\left(E-\mu\right)\right)+\xi}\nonumber \\
 &  & -\frac{\exp\left(\beta\left(E-\mu\right)\right)}{\left(\exp\left(\beta\left(E-\mu\right)\right)+\xi\right)^{2}}
 \left[\beta\left(-2\sum_{j}k_{j}^{2}r_{j}+\frac{E}{2}\sum_{j}k_{j}^{2}r_{j}^{2}\right)+2\beta^{2}\left(\sum_{j}k_{j}^{2}r_{j}\right)^{2}\right]\nonumber \\
 &  & +\frac{\exp\left(2\beta\left(E-\mu\right)\right)}
 {\left(\exp\left(\beta\left(E-\mu\right)\right)+\xi\right)^{3}}4\beta^{2}\left(\sum_{j}k_{j}^{2}r_{j}\right)^{2}
 +{\mathcal O}\left(r^{3}\right).\end{eqnarray}

In our case, since $r$ was introduced as a random variable, we will
need to perform an ensemble average on $\langle n\rangle_{r}$. This averaging
entails integration over the statistical parameters $r_{i}$ through
integrals of the form: \begin{equation}
\frac{1}{\sigma_{j}\sqrt{2\pi}}\int\limits _{-\infty}^{\infty}dr_{j}\langle n\rangle_{r}\,\exp\left(-\frac{r_{j}^{2}}{2\sigma_{j}^{2}}\right),\label{int}\end{equation}
where $\sigma_{j}$ is defined as:
\begin{equation}
\left\langle r_{j}^{2}\right\rangle =\sigma_{j}^{2}.\end{equation}

 In our case, the variance of the recoil velocity depends on the string
coupling $g_{s}$, which itself may fluctuate. Now we have: $g_{s}^{-2}=e^{-2\left\langle \phi\right\rangle }$,
where $\left\langle \phi\right\rangle $ is the vacuum expectation
value of the dilaton field $\phi$. In a stochastic space-time, the
D-foam~\cite{dfoam}, there are in general non-trivial fluctuations
of the dilaton field, and, therefore, of the string coupling itself.
Hence, our knowledge of such fluctuations is very limited since it
depends on properties of string vacua. In our analysis we simply assume
that $\sigma_{j}$ are also normally distributed with $\left\langle \sigma_{j}^{2}\right\rangle =\sigma_{0j}^{2}$.
At this point we should also note that $\sigma_{0j}$ are taken to
be different along the various spatial directions, for the sake of
generality. However the observed isotropy of the Universe in large
scales imposes such potential anisotropies to be quite small, consistently
with our ansatz of smallness of the $\sigma_{j}$.

On performing both averages on $\langle n\rangle_{r}$, the result, denoted by
$\langle\langle n\rangle\rangle$, on second order in the statistical parameters $\sigma_{0j}$
is given by: \begin{eqnarray}
\langle\langle n\rangle\rangle & \simeq & \frac{1}{\exp(\beta(E-\mu))+\xi}\nonumber \\
 &  & -\frac{\exp(\beta(E-\mu))}{(\exp(\beta(E-\mu))+\xi)^{2}}\left(\beta\frac{E}{2}\sum_{j}k_{j}^{2}\sigma_{0j}^{2}+2\beta^{2}\sum_{j}k_{j}^{4}\sigma_{0j}^{2}\right)\nonumber \\
 &  & +\frac{\exp\left(2\beta\left(E-\mu\right)\right)}
 {\left(\exp\left(\beta\left(E-\mu\right)\right)+
 \xi\right)^{3}}4\beta^{2}\sum_{j}k_{j}^{4}\sigma_{0j}^{2}.\label{dist2}\end{eqnarray}
 We will adopt the notation $\langle\langle\ldots\rangle\rangle$ to generically denote
the above double average.

One can see that (\ref{dist2}) has the same general form
as the distribution functions appearing in~\cite{pessah}, where Tsallis
statistics was applied. Indeed, in that case, the corresponding expanded
formula for the number density distribution function is: \begin{equation}
\langle n\rangle_{q}=\frac{1}{e^{\beta\left(\epsilon-\mu\right)}+\xi}+\frac{q-1}{2}\frac{\left(\beta\left(\epsilon-\mu\right)\right)^{2}e^{\beta\left(\epsilon-\mu\right)}}{\left(e^{\beta\left(\epsilon-\mu\right)}+\xi\right)^{2}},\label{tsallispessah}\end{equation}
 where $q$ is the non-extensive parameter accounting for the non-additivity
of Tsallis entropies.

Comparing with our case here, we find that there are
two major differences. The first is that, unlike in~\cite{pessah}, the
quantity $E$ appearing in the first term of (\ref{dist2}) is \emph{not}
the actual \emph{energy} of the particle, except in a Minkowski background
(cf.\ (\ref{onshell})). Therefore, in our case, by setting
the stochastic effects to zero (which, in the Tsallis cosmologies~\cite{pessah},
corresponds to setting $q=1$ in (\ref{tsallispessah})), leads us
back to the standard result by eliminating the last two terms in (\ref{dist2});
the metric (\ref{metric}), of course, is also reduced to the Minkowski
form, so that the actual energy gets the standard form (\ref{onshell}).
The second difference is that the extra term in (\ref{tsallispessah})
depends only on the particle's energy, whereas, in our case, there
is an explicit dependence on the particle's 3-momentum as well, owing
to the induced Finsler-like metric distortion.

The similarity of our
approach with that of~\cite{pessah}, lies in the fact that the corrections
to the number density distribution have broadly similar structure;
the second term is a \emph{perturbatively} \emph{small} deviation
from the standard case. In the Tsallis cosmologies, this is a consequence
of the (assumed) smallness of the quantity $q-1$, while in our case,
this is guaranteed by the smallness of the statistical parameters
$\sigma_{0j}$, describing the stochastic fluctuations of the D-foam.

In the above sense, our D-particle foam may be viewed as providing
a microscopic framework for modified particle statistics in the Early
Universe, in a similar (but not identical) spirit to the Tsallis cosmological
models of~\cite{pessah}. At this stage we would also like to recall
the approach of~\cite{tsallissarkar}, where a further \emph{non-extensivity}
in the framework of Tsallis, has been discussed, as being
due to quantum fluctuations of the string coupling $g_{s}$. Now $g_{s}=e^{\langle\phi\rangle}$,
where $\phi$ is the dilaton field, and $\langle\dots\rangle$ denotes
a vacuum expectation value in target space-time. By including all possible
quantum space-time metric and topology fluctuations in a fully (yet
unknown) quantum gravity framework, $g_{s}$ becomes stochastic. Such effects are ignored,
as sub-leading, in our present approach. In principle they
should, at least formally, be included in a truly non-perturbative
formalism of the very Early Universe in the presence of space-time
foam.


\subsection{Number and energy densities \label{sectc}}

On applying the basic formulae:
\begin{equation}\label{no}
\\n=\frac{g}{(2\pi)^3}\int{\langle\langle n \rangle\rangle d^3k}
\end{equation}
and
\begin{equation}\label{dens}
\rho=\frac{g}{(2\pi)^3}\int{ \langle\langle n \omega_r \rangle\rangle d^3k },
\end{equation}
with $\langle\langle n \rangle\rangle $ given in (\ref{dist2}) and $\langle\langle n \omega_r \rangle\rangle$ calculated in Appendix A,
we can derive the number density $n$ and energy density $\rho$ for
both fermions and bosons in typical eras for the evolution of the
universe. The integrals in relations (\ref{no}) and (\ref{dens})
are with respect to $k^{j}$, which in the non-expanding case satisfies (\ref{contrav}).
In our work here we restrict ourselves to non-degenerate species, so the chemical potential contributions will be ignored, since  $T \gg \mu$, with $T$ the temperature.

As our primary interest is the effect of space-time foam on heavy, non-relativistic Dark matter particles, we will start with this case.
After some straightforward calculations, whose details are given in Appendix A, we obtain for the density of heavy non-relativistic particles, for which $m^2 \gg k^2$:
\begin{eqnarray}\label{nonrel2}
n&=&\frac{g}{8\pi^3}\int{\langle\langle n \rangle\rangle d^3k}\nonumber \\
&=&g\left(\frac{m}{2\pi\beta}\right)^\frac{3}{2}e^{-(m-\mu)\beta}
+\frac{\sqrt{2}g}{5\pi^{3/2}}\left(m^{7/2}\beta^{-3/2}\right)\bar{\sigma}_0^2\left(\frac{15}{2}e^{-(m-\mu)\beta}-\frac{45}{16\sqrt{2}}\xi e^{-2(m-\mu)\beta}+\frac{15}{27\sqrt{3}}\xi^2e^{-3(m-\mu)\beta}\right)\nonumber\\
&&-\frac{g}{12\pi^{3/2}}\left(\beta^{-2}+m^2\right)\bar{\sigma}_0^2\left(\frac{m}{\beta}\right)^{3/2}\left(3e^{-(m-\mu)\beta}-\frac{3}{4\sqrt{2}}\xi e^{-2(m-\mu)\beta}\right),
\end{eqnarray}
where $\bar{\sigma}_0^2=\sum_i\sigma_i^2$ and $\xi=+1$ ($\xi=-1$) applies to fermions (bosons).

As one can see from (\ref{nonrel2}), the first term on the right-hand side represents the standard result for the non-relativistic number density of fermions and bosons, whereas the other terms, proportional to $\bar{\sigma}_0^2$,
describe the foam-induced corrections. It is interesting to note at this stage, that the D-particle foam seems to affect differently fermions from bosons, as manifested by the different $\xi$ dependence of the foam-correction terms in the distribution functions. This also holds for the relativistic case, which is included  below for completeness, and the details of its derivation appears in Appendix A. For the number density we get:
\begin{equation}
n_b=\frac{g_bT_b^3}{2\pi^2} \Gamma(3)\zeta(3)+\frac{g}{8\pi^3}T_b^5\bar{\sigma}_0^2\zeta(5)\left(-\frac{2\pi}{3}\Gamma(6)+\frac{8\pi}{5}\Gamma(7)\right)
\end{equation}
for bosons, and

\begin{equation}
n_f=\frac{g_fT_f^3}{2\pi^2}\frac{3}{4}\Gamma(3)\zeta(3)+\frac{g}{8\pi^3}T_f^5\bar{\sigma}_0^2\left(\frac{2\pi}{3}\Gamma(6)\left[-2\eta(6)+\eta(5)\right]+\frac{8\pi}{5}\Gamma(7)\left[6\eta(7)-6\eta(6)+\eta(5)\right]\right)
\end{equation}
for fermions, whereas the corresponding relativistic energy densities are found to be:

\begin{equation}\label{reldens1}
\rho_{b}=\frac{g_b}{2\pi^2}\Gamma(4)\zeta(4)T_b^4+\frac{g_b}{\pi^2}\bar{\sigma}_0^2T_b^6\left(\frac{1}{12}\Gamma(6)\zeta(6)-\frac{17}{60}\Gamma(7)\zeta(6)+\frac{1}{5}\Gamma(8)\zeta(6)\right)=\frac{\pi^2}{30}g_bT_b^4+g_b\frac{2\pi^4}{189}\bar{\sigma}_0^2T_b^6,
\end{equation}
\noindent for bosons and
\begin{eqnarray}\label{reldens2}
\rho_{f}&=&\frac{g_f}{2\pi^2}\Gamma(4)\eta(4)T_f^4+\frac{g_f}{\pi^2}\bar{\sigma}_0^2T_f^6\left(\frac{1}{12}\Gamma(6)\eta(6)-\frac{17}{30}\Gamma(7)\eta(7)+\frac{17}{60}\Gamma(7)\eta(6)+\frac{6}{5}\Gamma(8)\eta(8)-\frac{6}{5}\Gamma(8)\eta(7)+\frac{1}{5}\Gamma(8)\eta(6)\right)\nonumber\\
&=&\left(\frac{7}{8}\right)\frac{\pi^2}{30}g_fT_f^4+\frac{g_f}{\pi^2}\bar{\sigma}_0^2T_f^6\left(\frac{18941\pi^6}{15120}-\frac{50841}{8}\zeta(7)+\frac{127\pi^8}{200}\right)=\left(\frac{7}{8}\right)\frac{\pi^2}{30}g_fT_f^4+g_f\frac{793.32}{\pi^2}\bar{\sigma}_0^2T_f^6,
\end{eqnarray}
\noindent for fermions.

These equilibrium distributions will be important later on when
estimating the number of degrees of freedom of relativistic species.
Changes to distribution functions induced by the foam can have cosmological
implications, such as modifications to the freeze-out point of species and their (thermal) relic abundances,
which could lead to observable effects, depending on the size of the effects and the associated
string mass scale and coupling. In the next section we therefore proceed to discuss such cosmological aspects of the model.


\section{Cosmological Aspects of the model \label{sec:cosmology}}

 The full rigorous cosmology of the D-foam is a difficult and rather complicated subject yet~\cite{emnnewuncert} to be analyzed in a complete way. It depends on the underlying microscopic string theory model, which is not presented here~\cite{dfoam,westmuckett}. Below, we shall only concentrate on one aspect, which however is interesting for particle phenomenology, namely the effects of the non extensive statistics of the D-foam on thermal dark matter relics. However, before embarking on a detailed discussion of the effects of the interaction of the Dark matter with the D-foam on the thermal relic abundance of the former, it is instructive to discuss first briefly the effects of the D-foam fluid on the energy budget of the Universe \emph{per se}. In this way the special r\^ole of the D-particle defects as supersymmetric vacuum defects, and not ordinary dark matter contributors, will hopefully be elucidated. We do so in the next subsection.

 \subsection{D-particles, the energy Budget of the Universe  and the Hubble Expansion Rate \label{sec:Hubble}}

Our focus, for the interaction of D-particles with cold dark matter,
has been on the induced stochastic Finsler-like metric fluctuations
that are induced. Although D-particle is a generally useful term it
is of course a D$0$-brane and represents strictly a space-time defect.
Consequently it needs to be dealt with using the methods of string
theory when considering its contribution to the energy of the system.
We have noted that, if there is relative bulk motion, SUSY is broken
and there are non-trivial forces among the D-particles as well as
between the D-particles and the brane world or orientifolds~\cite{westmuckett}.
The resulting non-zero contribution to the energy is proportional
to $v^{2}$ for \emph{transverse} relative motion of branes with different
dimensionalities, and to $v^{4}$ for branes of the same dimensionality
(there is \emph{no} contribution to the energy of a p-brane-world
from motion of other branes in directions \emph{parallel} to its longitudinal
directions). There is also a dependence on the relative distance of
the various branes. In particular, the interaction of a single D-particle,
that lies far away from the D3 brane (D8-compactified) world, and
moves adiabatically with a small velocity $v_{\perp}$ in a direction
transverse to the brane, results in the following potential~%
\footnote{For brevity, in what follows we ignore potential contributions induced
by compactification of the D8 brane worlds to D3 branes, stating only
the expressions for the induced potential on the uncompactified brane
world as a result of a stretched string between the latter and the
D-particle - the compactitication does not affect our arguments on
the negative energy contributions to the brane vacuum energy.%
}~\cite{westmuckett} \begin{eqnarray}
\mathcal{V}_{D0-D8}^{long}=+\frac{r\,(v_{\perp}^{{\rm long}})^{2}}{8\pi\alpha^{\prime}}~,~r\gg\sqrt{\alpha^{\prime}}.\label{long-1}\end{eqnarray}
 On the other hand, a D-particle close to the D3-brane (compactified
D8), at a distance $r'\ll\sqrt{\alpha'}$, moving adiabatically in
the perpendicular direction with a velocity $v_{\perp}^{{\rm short}}$
will induce the following potential to it:
\be\label{pot1} \mathcal{V}_{D0-D8}^{short}=
- \frac{\pi\alpha^\prime (v_\perp^{\rm short})^2}{12{r'}^3}.
\ee
We now mention that the analysis~\cite{westmuckett} of this short distance behaviour also reveals a characteristic
minimum length
\begin{equation}\label{rmin}
r_{min}\sim\sqrt{v_\perp}l_{s}
\end{equation}
which for $v_\perp \ll1$ is much
smaller than the string length $l_{s}=\sqrt{\alpha^\prime}$, a feature common in theories involving D-branes~\cite{strongcoupl}. The short range force is an attractive one and
D-particles can make bound states with the D-brane through the exchange
of light strings. This process in perturbation theory has
divergences which lead also to  instabilities. These are cured by a non-perturbative
treatment of string theory, which reveals that the string coupling $\exp\left(\left\langle \phi\right\rangle \right)$
at these short distances $\left(r\ll l_{s}\right)$ is given by~\cite{strongcoupl}
 \begin{equation}
\exp\left(2\left\langle \phi\right\rangle \right)\sim g_{s}^{2}\left(1+\frac{g_{s}}{(r/\sqrt{\alpha^\prime})^{7}}\right)^{3/2}
\label{stringcoupl}
\end{equation}
where $g_{s}$ is the string coupling on the brane for interactions
between stringy matter. Clearly this coupling is enhanced and t
associated mass of the bound D-particle $\frac{M_{s}}{g_s\, \exp\left(\left\langle \phi\right\rangle \right)}$
is much reduced, compared to the mass ($M_s/g_s$) of the D-particles attached on the brane world and co-moving with it in the bulk direction~\footnote{That the mass of such D-particles is $M_s/g_s$ can be determined by considering their recoil during scattering with fundamental strings having their ends attached on the brane, and moving in directions parallel to the brane's longitudinal ones~\cite{kogan,szabo}.}.
Hence, there will be two components of the D-particle
contribution to the energy on the brane: one due to a D-particle density
on the brane (a form of dust, in view of their no force conditions, discussed in section \ref{sec:susy}) and the other due to the D-particle
bound states which will contribute a negative interaction energy on
the brane. Hence the presence of both contributions to the energy
density allows a larger density of D-particles on the brane without
overclosing the universe.

Let us explain this point in more details.  To this end, we first assume that the bulk space expands cosmologically, with a scale factor $a(t)$, only along the directions that are parallel to the longitudinal dimensions of our D3 brane world (appropriate compactification down to three spatial dimensions is assumed from now on for the model of \cite{westmuckett}, without further details; this by itself is an important topic for investigation that depends on the underlying models, but hopefully the main features of our analysis, regarding the contributions of the D-foam to the Hubble expansion, will not be affected significantly by such details).
Let also $n_f \equiv \frac{\mathcal{N}_{f}}{a^3(t)\,l_{s}^{3}}$ be the
three-space-dimensional density of D-particles on the D3-brane world (with $\mathcal{N}_f$ the corresponding number of (non-interacting) D-particles on the brane and  $a^3(t)$ the proper volume of an expanding Universe for a co-moving observer) and $n_b \equiv \frac{\mathcal{N}_{b}}{a^3(t)\,l_{s}^{3}}$
the corresponding density of bound D-particles (with $\mathcal{N}_b$ the corresponding number of bound D-particles, assumed non interacting among themselves).  In view of their dust state,
the D-particles bound on our brane world, and comoving with it in the bulk, will contribute a term in the energy density of the brane Universe of the form
\begin{equation}\label{mass}
\rho_{\rm mass} \sim  \frac{M_s}{g_s}\, n_f~.
\end{equation}
Since, as mentioned above, the bound state masses are much smaller, suppressed by powers of $v_\perp$ (for distances $r \sim r_{min}$ given by Eq.~(\ref{rmin})), due to the non-perturbative effects on the string coupling (\ref{stringcoupl}), their mass-induced contribution to the ground state energy on the brane world will be subdominant, leaving only (\ref{pot1}) as the dominant contribution from each one of them. Assuming, for simplicity, that the effects of the nearby bulk D-particles are effectively described by a model where these defects are  mainly concentrated on a thin D3 brane parallel to our brane world, at bulk distances $r_{min}$ away, we obtain the dominant (negative) contributions to the vacuum energy density of the brane world due to the nearby bulk D-particles (with $\ell_s = 1/M_s$):
\begin{equation}\label{pot1b}
\rho_{\rm nearby~D0}^{\rm short} = n_b  \, \mathcal{V}_{D0-D8}^{short}(r \sim r_{min}) \sim - n_b  \,\frac{\pi}{12\, \ell_s} \sqrt{v_\perp} ~.
\end{equation}
Then the cancellation
of the mass effect (\ref{mass}) of the D-particles on the brane world
and those due to the short-distance potential $V^{{\rm short}}$ (\ref{pot1b})  gives
\begin{equation}\label{popul}
{n}_{b}\sim\frac{12\, {n}_{f}}{\pi g_{s}\sqrt{v_\perp}}~.
\end{equation}

We will now use some estimates from a phenomenological study~\cite{emninfl}
of the order of magnitude of $v_\perp$. In general, for a distribution of nearby D-particles in the bulk, this denotes the average
velocity of the distribution. To simplify things, for the (qualitative) purposes of this section, we may make the reasonable assumption that the bulk D-particles are almost stationary in the bulk, and only our D-brane world moves slowly in the bulk dimension of the models of D-foam presented in \cite{westmuckett}. This makes $v_\perp$ the velocity of the brane motion, which has been bounded phenomenologically in \cite{emninfl} by the requirement that the models agree with the slow-roll conditions for inflation (defined appropriately for this particular class of brane cosmology models  in ref.~\cite{emninfl}, where we refer the interested reader for details).
The relevant phenomenological bound is
\begin{equation}\label{maxv}
v_\perp^{2}\leq1.48\times10^{-5}g_{s}^{-1}
\end{equation}
and, on account of (\ref{popul}),  this leads to
\begin{equation}\label{popul2}
{n}_{b}\sim\frac{60}{g_{s}^{\frac{3}{4}}} {n}_{f}~.
\end{equation}

Another important estimate we need to make for our cosmological considerations in this paper concerns the density of defects, in both the bulk space and on the brane Universe. As we discuss briefly in section \ref{sec:wmap}, constraints on the density of D-foam are obtained by comparing the (subluminal) refractive effects of the D-particle foam vacuum in relation to time lags
in the arrival of cosmic $\gamma-$rays depending linearly on energy (as reported
by MAGIC~\cite{MAGIC2} and Fermi~\cite{fermi} collaborations).
In such studies it was found~\cite{emnnewuncert,mavroreview2010} that, for a uniform D-foam density, at least for late epochs of the Universe's history, an order of magnitude  ${n}_{f}\sim10^{-3}-10^{-2}$ was sufficient
to make the D-foam explanation of the effect viable, in particular concerning the bounds imposed by the Fermi satellite~\cite{fermi} (for more discussion see section \ref{sec:wmap} and references therein). Since $g_{s}\sim O\left(1\right)$,  we
see that ${n}_{b}\sim 60\,  {n}_{f} \sim 6 \times (10^{-2}-10^{-1}) < 1$
and we have the required (for consistency of our perturbative treatment) \emph{dilute}
limit of D-particle foam. The situation in a semi-realistic model, such as that of \cite{dfoam},
is complicated since there are stacks of D-branes, orientifolds and
brane intersections (leading to a complicated interplay of the effects
of $V^{{\rm short}}$ ((\ref{pot1})) and $V^{{\rm long}}$ ((\ref{long-1}))). However, we will consider
the implications of the broad possibilities of D-particles on the
brane and bound states. This argument has shown that ${n}_{f} \sim O\left(10^{-3}- 10^{-2}\right)$ does
\emph{not} need to \emph{overclose} the Universe.

The next issue we shall be concerned with is the contributions of D-foam to the Hubble expansion of the Universe, namely to the
equation of state of the D-foam fluid. As already mentioned in the introduction section of this article, in view of their point-like structure in space, D-particles may be naively thought of as ordinary (massive, of mass $M_s/g_s$) excitations of the vacuum, thus playing the r\^ole of dark matter~\cite{D0matter}. However, this is \emph{not} the case in the D-foam models~\cite{dfoam}, as becomes evident from the afore-mentioned properties and will hopefully become clearer from the following arguments.
If $n_{{\rm D3-foam}}(t) = n_f(t)$ denotes the number density of the D-particle
effects on a D3-brane, at any given instance of cosmic time, $t$,
then according to earlier arguments, we would have the following cosmic
time evolution: \begin{equation}
\frac{d}{dt}n_{{\rm D3-foam}}(t)+3Hn_{{\rm D3-foam}}(t)=\frac{1}{V}\frac{d\mathcal{N}}{dt}\label{evoldens}\end{equation}
 where $V\sim a^{3}(t)$ is the proper higher-dimensional (bulk) volume, expanding only along the D3-brane world longitudinal dimensions;
from the point of view of an observer on the three-brane Universe,$\frac{d}{dt}\mathcal{N}$
is the effective rate of \emph{bulk} D-particles crossing the brane
as the latter moves in the bulk space (\emph{cf.} fig.~\ref{fig:recoil}).
For eras of the Universe long after a catastrophic cosmic event, such
as collision of two brane Universes in the arrangement of fig.~\ref{fig:recoil},
the brane motion in the bulk is adiabatic to a good approximation.
Secondly we assume that in such eras a steady state situation arises
the effects of the source term in (\ref{evoldens}) compensate those
of the cosmic dilution on the effective population of D-particles
in the expanding D$3-$ brane Universe. Hence $n_{{\rm D3-foam}}(t)$,
in (\ref{evoldens}) is approximately \emph{constant} in time, i.e.
\begin{equation}
\frac{d}{dt}n_{{\rm D3-foam}}(t)\simeq0~.\label{consdens}\end{equation}
 In such a case, the D-foam would behave as a \emph{dark energy} (cosmological
constant-like) contribution to the energy budget rather than dark
matter, thereby leading to different considerations than in the work
of \cite{D0matter}, where, as mentioned earlier, the D-particles are viewed as dark matter
due to their localized nature in space. As the D$3$-brane moves in the bulk,
such phenomena are obviously cosmological era dependent, given the
likely inhomogeneities in the bulk population of D-particles.

To summarise,
the negative contributions to the Brane Universe's vacuum energy induced
by averaging (\ref{pot1}) over appropriate populations of bound states
may then cancel out (or suppress significantly, on average) the positive energy contributions of the D-particles trapped
on our world, due to their mass $M_s/g_s$.
According to the above picture, the effective Hubble rate $H(z)$
of the expansion of the D3 brane Universe, is \emph{not} affected in the
eras where (\ref{consdens}) is valid by the presence of D-foam, thereby
following the rule \begin{equation}
(H(z)/H(z=0))^{2}=\frac{8\pi G_{N}}{3}\left(\Omega_{M}(1+z)^{3}+\Omega_{{\rm DE}}+\dots\right)~,\end{equation}
 where $z$ is the cosmic red-shift, $\Omega_{i}$ are various energy
densities (in units of the critical density), and the $\dots$ denote
terms due to the brane Universes, such as dark radiation , which we
do not discuss here explicitly. The quantity $\Omega_{M}$ contains
dust contributions, including field-theoretic contributions to Dark
matter (\emph{e.g}. supersymmetric matter contribution coming from
the effective (broken) supersymmetric field theories of the excitations
on the D3 brane Universe). On the other hand, $\Omega_{{\rm DE}}$
corresponds to dark energy/cosmological-constant-like contributions,
with equation of state approaching $w\to-1$ for redshifts $z\to0$
(today), which remain constant in time and dominate over matter for
more than 1.9 billion years, according to observations.

Although a detailed and complete microscopic bulk model, with realistic
particle phenomenology and Cosmology, is far from being available
in this framework at present, nevertheless, we consider the above
picture as plausible, and this will be our main assumption in what
follows. For a more detailed, but definitely incomplete, microscopic
picture, where one takes into account also the bulk distributions
of D-particles lying far away from the brane world (\emph{cf}. Eq.~(\ref{long-1})),
we refer the reader to the second paper in \cite{emnnewuncert}. The
possibility of approximately constant D-foam contributions to the
vacuum energy (\ref{consdens}) was also raised there.

Based on the above assumptions and plausibility arguments, we shall
assume in what follows that the dominant effects of the D-foam on
the energy budget of the Cosmos concern primarily the (ordinary) Dark Matter
content of the brane Universe. These are in the form of the distortion
of the neighbouring space-time during the propagation of (electrically
neutral, weakly interacting) matter in regions of space-time with
populations of D-particle defects; this leads to an effective background
space-time of the form (\ref{approx}), felt by the Dark Matter particles.
This space-time fluctuates in the framework of stochastic D-foam according
to (\ref{defu2})), (\ref{moment1}), (\ref{moment2}), (\ref{moment3}).
Such space-time distortions depend on the momentum transfer between
the defect and the matter string (i.e. akin to Finsler metrics), and
as such affect the relic abundance from the Boltzmann equation for
the evolution of Dark Matter. We now proceed to discuss this effect
in some detail, in the context of thermal relics that we restrict
our attention upon in this article.

\subsection{Effective number of degrees of freedom}\label{effsection}

We now come to the announced detailed discussion on the effects of the D-foam on thermal Dark Matter abundances, as a non-trivial consequence of the distortion of space-time (\ref{approx}) on account of the non-trivial interaction of the pertinent matter string excitations, representing Dark Matter relic particles,  with the D-particle defects.
To this end, we shall first need to calculate the effective number of degrees of freedom participating in the thermal history of the D-foam universe. The reader is invited to compare our findings with the approach of~\cite{pessah}, where conventional cosmology but with Tsallis statistics~\cite{tsallis} has been considered. If one assumes that the dominant energy contribution comes from the relativistic species (e.g.\ the radiation dominated era (r.d.e.)), then the total energy density can be written in the very compact form:
\begin{equation}\label{rel1}
\rho_T=\frac{\pi^2}{30}g_\textit{eff}T^4
\end {equation}
\noindent where $g_\textit{eff}$ is called the effective number of degrees of freedom and in the standard approaches is given by:
\begin{equation}\label{gfeff}
\\ g_\textit{eff}= \sum_{i}{g_{i,b}\left(\frac{T_{i,b}}{T}\right)^4} + \frac{7}{8}\sum_{j}{g_{j,f}\left(\frac{T_{j,f}}{T}^4\right)}.
\end{equation}
where $T_{i,b}$ ($T_{i,f}$) denotes the temperature of the thermal distribution
satisfied by bosonic (fermionic) species $i$; in principle, $T_{i,b}\neq T$,
where $T$ is the temperature of the photon gas.
The quantity (\ref{gfeff}) is very important in cosmology because it allows us to count the species that are relativistic at a given era. Obviously it is a time (temperature) dependent quantity since, as the Universe expands (and freezes), less and less species can be effectively massless, i.e.\ highly relativistic. For instance, at very early eras, when $T>175~{\rm GeV}$, all the present species were relativistic yielding a very big effective number of degrees of freedom: $g_\textit{eff}\approx 106.8$. After the $e^-$ annihilation,  at $T>500~{\rm keV}$, only the three light neutrinos (restricted to one helicity state) and their anti-neutrinos and the photon remain massless, giving: $g_\textit{eff}\approx 3.37$ (for more details on this see~\cite{kolb}).

Within the context of our model, the effective number of degrees of freedom is expected to be modified. To obtain this correction, let us
substitute in the left hand side of (\ref{rel1})  the relativistic energy density for fermions and bosons that we derived in the previous section (expressions (\ref{reldens1}), (\ref{reldens2})). Then we have:
\begin{equation}\label{eff}
g_\textit{eff}'=g_\textit{eff}+\frac{30}{\pi^2}\bar{\sigma}_0^2\left(\frac{2\pi^4}{189}\sum_{i}{g_{i,b}T_{i,b}^2\left(\frac{T_{i,b}}{T}\right)^4 }+\frac{793.92}{\pi^2}\sum_{j}{g_{j,f}T_{j,f}^2\left(\frac{T_{j,f}}{T}\right)^4 }\right).
\end{equation}
Now the cosmological evolution equations, for instance in a r.d.e, will still be given by the standard relations:
\begin{equation}\label{hubble}
H=1.66(g_\textit{eff}')^\frac{1}{2}\frac{T^2}{m_{pl}}
\end{equation}
and
\begin{equation}\label{time}
t=0.30(g_\textit{eff}')^{-\frac{1}{2}}\frac{m_{pl}}{T^2},
\end{equation}
but with the effective number of degrees of freedom having the corrected form (\ref{eff}), that carries the D-foam effects. Hence, it can be seen  that all the additional effects stemming from the D-particles recoil, can be entirely subsumed in a  modification of the effective number of degrees of freedom as presented by (\ref{eff}).

\subsection{Boltzmann equation}
In order to account for the expansion of the universe by the factor $a(t)$ we should rewrite the distorted metric (\ref{approx}) in the following form:
\begin{equation}
g_{\mu\nu}=\left(\begin{array}{cccc}
-1 & a^{2}(t)r_{1}k_{1} & a(t)^{2}r_{2}k_{2} & a^{2}(t)r_{3}k_{3}\\
a^{2}(t)r_{1}k_{1} & a^{2}(t) & 0 & 0\\
a^{2}(t)r_{2}k_{2} & 0 & a^{2}(t) & 0\\
a^{2}(t)r_{3}k_{3} & 0 & 0 & a^{2}(t)\end{array}\right)\label{metric3}.\end{equation}

Hence the relations between covariant and contravariant momenta now become:
\begin{equation}
k_{0}=-k^{0},\, k_{i}=a^{2}(t)k^{i},\, i=1,2,3,
\end{equation}where in our notation the scale factor $a\left(t\right)$ is dimensionless and the recoil effect has not been taken into account at this point. We also have the general relations:
\begin{equation}
k^{\mu}=m\frac{dx^{\mu}}{d\tau},\mu=0,1,2,3,
\end{equation}
where $\tau$ is an affine parameter (the proper time in our case)
and $m$ is the mass of the particles under investigation.

We should  point out this perturbation is different in spirit (and mathematically) from metric perturbations in cosmological perturbation theory. Our considerations of quantum gravity phenomenology suggest that an effective theory cannot be just based on a 4-d pseudo-Riemmanian manifold $M$ but rather the tangent bundle $TM$. Diffeomorphisms would have to be appropriate to $TM$ and hence issues of our metric being diffeomorphically equivalent to those used in cosmological perturbation theory cannot materialize. To illustrate this, in the conformal Newtonian (longitudinal) gauge, a line element in cosmological perturbation theory can be written as
\begin{equation}
ds^{2}=a^{2}(\tau)[-(1+2\Psi)d\tau^{2}+(1-2\Phi)\delta_{ij}dx^{i}dx^{j}]\label{metric4}
\end{equation}
where $\Psi$ and  $\Phi$ are functions on $M$. It is interesting to see how similar we can make the metric in (\ref{metric3})
to that in (\ref{metric4}).

To this end, the reader should recall that the metric (\ref{metric3}) has been derived by a Liouville-dressing procedure in the world-sheet conformal field theory describing the propagation of a string excitation in a background space time, as reviewed briefly in section~\ref{sec:lcft}. It can be diagonalized by means of an appropriate T duality transformation on the target-space coordinates/$\sigma$-model fields, as follows~\cite{mavroreview2010}: Ignoring for the moment the cosmological expansion, \emph{i.e.} setting $a(t)$=constant, we observe that the vertex operator (\ref{recvec}) for the description of the recoil of a D-particle can be rewritten, using the two dimensional version of Stokes theorem, to give~\cite{mavroreview2010}
\begin{equation}\label{Tdual}
     \int_{\Sigma}d^{2}z u_{i}\varepsilon_{\alpha\beta}\partial^{\beta}t[\Theta_{\varepsilon}(t)+t\delta_{\varepsilon}(t)]\partial^{\alpha}X^{i}~,
\end{equation}
where $\alpha, \beta =1,2$ are two-dimensional world-sheet indices.
For times long after the impact of the open-string state with the D-particle, $t > 0$, this can be interpreted as  an open string propagating in a target-space antisymmetric background $B_{\mu\nu}$ with
$$
B_{0i}\sim u_{i} \; {\rm and} \; B_{ij}=0~.
$$
On making a canonical T duality transformation of the co-ordinates, the presence of the $B$ field leads to the following mixed boundary condition on the boundary $\partial \Sigma$ of the world-sheet $\Sigma$:
$$
g_{\mu\nu}\partial_{n}X^{\nu}+B_{\mu\nu}\partial_{\tau}X^{\nu}|_{\partial \Sigma}=0
$$
where $\Sigma$ is homeomorphic to a disc.
The induced open-string effective target space-time metric is obtained from calculating the world sheet propagator $\langle X^{\mu}(z,\overline{z})X^{\nu}(0,0)\rangle$. The resulting metric is~\cite{mavroreview2010}
\begin{eqnarray}\label{elfield}
  g_{\mu\nu} &=& (1-\overrightarrow{u}^{2})\eta_{\mu\nu},\;  \mu,\nu=0,1 \nonumber \\
   g_{\mu\nu} &=& \eta_{\mu\nu}, \; \mu,\nu={\rm all \ other \ values}~,
\end{eqnarray}
where we have considered a frame in which the matter particle has motion only in the spatial $X^{1}$ direction. Also, it should be noted, the metric implies the existence of a critical recoil velocity (viz. the speed of light \emph{in vacuo}), which stems from Lorentz invariance of the underlying string theory~\cite{mavroreview2010}.

On embedding such diagonal metrics in Cosmology, one should multiply the spatial parts by the appropriate Robertson-Walker scale factor $a(t)$. However, the reader should recall that the dependence of the metric elements (\ref{elfield}) on the recoil velocity implies in turn a momentum-transfer dependence of the metric. Then, compared to (\ref{metric4}), although in diagonal form, the space part of the cosmological extension of the metric (\ref{elfield}) has a velocity dependent piece, quite different from metric perturbations used in cosmology, that depend only on coordinates. Also, the time part has a similar form. Consequently, the Finslerian nature~\cite{finsler} of our metrics differentiate them significantly from usual metric perturbations.

Because of the above differences, it is not possible to put the back-reacted metric (\ref{metric3}) into the longitudinal gauge of
scalar metric perturbations by an ordinary (\emph{i.e}. only a space-time coordinate dependent) diffeomorphism. Consequently, the metric due to the clumping of ordinary particles \emph{cannot} be put into the form of the metric due to back-reaction from D-foam. As we have argued previously, the r\^ole of the D-particle defects is quite different from ordinary dark matter. As discussed above, in view of their argued no-force condition, which may be maintained in our supersymmetry obstruction scenarios
characterizing models of D-foam, these defects constitute a dark fluid, contributing to the dark energy
of the vacuum~\cite{emnnewuncert}. The fluid affects the propagation of ordinary particle excitations on the brane world through the above-mentioned Finsler-type metric distortions (\ref{metric3}), and via this the relic abundance of ordinary dark matter particles. This will be discussed in detail below.

The momentum dependence
of the (Finsler) metric implies a modification of the geodesic equation for
$\frac{d^{2}x^{\mu}}{d\tau^{2}}$ along with the expected modification of the Christoffel symbols.
Details are provided in Appendix B.

However the distribution function of a particle species $f$ is specified by the microscopic properties of the system and, in this sense, it would be convenient to define a \emph{local}, in space-time, momentum in an expanding universe~\cite{bernstein}:
\begin{equation}\label{bernstein}
\overline{k}^{i}\equiv a(t) k^{i}~,\qquad i=1,2,3~.
\end{equation}
In terms of these re-scaled momenta, the energy-momentum dispersion relation for a (dark matter) particle of mass $m$ in the spatially flat Robertson-Walker space-time background  we are working on, assumes an effectively ``Minkowski-space-time'' form, $ k^\mu k^\nu g_{\mu\nu} = -\omega^2 + a^2(t) k^i k^j \delta_{ij} = -\omega^2 + \overline{k}^i \overline{k}^j \delta_{ij} = -m^2$, and hence it is essential to define the phase-space densities as functions of the coordinates $(x^i, t)$ and the local momenta $\overline{k}^i$, $f(x^i, t, \overline{k}^i)$. This is particularly important in keeping the correct scaling properties of the Dark Matter density with the scale factor, as we shall see below.  In an isotropic Robertson Walker background, the momenta are assumed to be on-shell, so that the phase-space density depends only on the amplitude $|\vec{\overline{k}}|$, and thus the energy $\omega$, of the Dark Matter particle. However,
in our case, the small anisotropies that characterize the foam fluctuations should be taken into account, and this is performed by assuming
a dependence on the individual components $\overline{k}_i$ as well. Then, the Boltzmann equation will be written in
terms of the one-particle distribution function $f\left(x^{\mu},\overline{k}^{i}\right)$ and explicit dependence on the energy $\left(k^0=\overline{k}^0\right)$  can be omitted due to the on shell condition:
\begin{equation}\label{onshell2}
g_{\mu\nu}k^{\mu}k^{\nu}=-m^{2}.
\end{equation}
For such a distribution function,  Liouville operator takes the form:
\begin{equation}
\hat{L}[f]=k^{\mu}\frac{\partial f}{\partial x^{\mu}}+m \sum _{i}\frac{\partial f}{\partial \overline{k}^{i}}\frac{d\overline{k}^{i}}{d\tau},
\end{equation}
and after applying the isotropy condition: $\frac{\partial f}{\partial x^{i}}=0$ for $i=1,2,3,$ it simplifies to:
\begin{equation}
\hat{L}[f]=\overline{k}^{0}\frac{\partial f}{\partial t}+m \sum_{i}\frac{\partial f}{\partial \overline{k}^{i}}\frac{d\overline{k}^{i}}{d\tau}.
\end{equation}Noting that:
\begin{equation}
\frac{d\overline{k}^{i}}{d\tau}=a(t)\frac{dk^{i}}{d\tau}+\dot{a}(t)\frac{dt}{d\tau}k^{i}=a(t)\frac{dk^{i}}{d\tau}+\dot{a}(t)\frac{\overline{k}^{0}}{m}k^{i}
\end{equation}
and using the notation $H=\frac{\dot{a}}{a}$, we get the final expression:
\begin{equation}\label{liouville}
\hat{L}[f]=\overline{k}^{0}\frac{\partial f}{\partial t}+m^2a(t)\sum_{i}\frac{d^2x^{i}}{d\tau^2}\frac{\partial f}{\partial \overline{k}^{i}}+H\overline{k}^{0}\sum_{i}\overline{k}^{i}\frac{\partial f}{\partial \overline{k}^{i}}.
\end{equation}

Substituting (\ref{geod2}) in the above, with $k^{0}$ being everywhere replaced by $\overline{k}^{0}$, and dividing throughout by $\overline{k}^{0}$, we obtain:
\begin{eqnarray}\label{final1}
 &  & \frac{\hat{L}[f]}{\overline{k}^{0}}=\frac{\partial f}{\partial t}-H\sum_{i}\overline{k}^{i}\frac{\partial f}{\partial\overline{k}^{i}}-2Ha^2(t)\overline{k}^{0}\sum_{i}r_{i}\overline{k}^{i}\frac{\partial f}{\partial\overline{k}^{i}}+8Ha^4(t)\sum_{i}r_{i}^{2}\left(\overline{k}^{i}\right)^{3}\frac{\partial f}{\partial\overline{k}^{i}}\nonumber \\
 &  & +\frac{2}{\overline{k}^{0}}Ha^2(t)\sum_{j}\left(\overline{k}^{j}\right)^2\sum_{i}r_{i}\overline{k}^{i}\frac{\partial f}{\partial\overline{k}^{i}}+4Ha^4(t)\left(\overline{k}^{0}\right)^{2}\sum_{i}r_{i}^{2}\overline{k}^{i}\frac{\partial f}{\partial\overline{k}^{i}}-4a^{4}(t)H\sum_{j}\left(\overline{k}^{j}\right)^2\sum_{i}r_{i}^{2}\overline{k}^{i}\frac{\partial f}{\partial\overline{k}^{i}}\end{eqnarray}
 Applying at this point the mass shell condition (\ref{onshell2}), yields the energy dispersion relation:
\begin{equation}
\overline{k}^{0}=a^{2}(t)\sum_{i}\left(\overline{k}^{i}\right)^{2}r_{i}+\sqrt{\sum_{i}\left(\overline{k}^{i}\right)^{2}+m^{2}}\left[1+\frac{a^{4}(t)\left(\sum_{i}\left(\overline{k}^{i}\right)^{2}r_{i}\right)^{2}}{\sum_{i}\left(\overline{k}^{i}\right)^{2}+m^{2}}\right]^{\frac{1}{2}}.
\end{equation}
Further analysis involves the approximation of heavy dark matter, which we consider in this work, since the phenomenologically dominant species are the heavy ones that in general are expected to cause the most significant distortions in the space time background,
$m^{2}\gg \sum_{i} \overline{k}^{i\:2}$, in which case
\begin{equation}
\overline{k}^{0}\sim m\left[1+\frac{a^{4}(t)}{2m^{2}}\sum_{i}\left(\overline{k}^{i}\right)^{4}r_{i}^2+\frac{a^{2}(t)}{m}\sum_{i}\left(\overline{k}^{i}\right)^{2}r_{i}\right],
\end{equation}
or  expanding also  up to second order and dropping cross terms,
\begin{equation}
\frac{1}{\overline{k}^{0}} \sim \frac{1}{m}\left[1+\frac{a^{4}(t)}{2m^{2}}\sum_{i} r_{i}^2\left(\overline{k}^{i}\right)^{4}-\frac{a^2(t)}{m}\sum_{i} r_{i}\left(\overline{k}^{i}\right)^{2}+\cdots\right].
\end{equation}
With these in hand, (\ref{final1}) gives:
\begin{eqnarray}
\frac{\hat{L}[f]}{\overline{k}^{0}} &=&\frac{\partial f}{\partial t}-H\sum_{i}\overline{k}^{i}\frac{\partial f}{\partial\overline{k}^{i}}-2Hma^2(t)\sum_{i}r_i\overline{k}^{i}\frac{\partial f}{\partial\overline{k}^{i}}+\frac{2}{m}Ha^2(t)\sum_{j}\left(\overline{k}^{j}\right)^2\sum_{i}r_i\overline{k}^{i}\frac{\partial f}{\partial\overline{k}^{i}}+6Ha^4(t)\sum_{i}r_i^2\left(\overline{k}^{i}\right)^3\frac{\partial f}{\partial\overline{k}^{i}}\nonumber\\
&&-\frac{2}{m^{2}}Ha^{4}(t)\sum_{j}\left(\overline{k}^{j}\right)^2\sum_{i}r_{i}^{2}\left(\overline{k}^{i}\right)^{3}\frac{\partial f}{\partial\overline{k}^{i}}
+4Hm^{2}a^4(t)\sum_{i}r_{i}^{2}\overline{k}^{i}\frac{\partial f}{\partial\overline{k}^{i}}-4Ha^4(t)\sum_{j}\left(\overline{k}^{j}\right)^2\sum_{i}r_{i}^{2}\overline{k}^{i}\frac{\partial f}{\partial\overline{k}^{i}}\nonumber\\
\end{eqnarray}
Now we can average over the ensembles for the random variables $r_{i}$. It is not trivial why this can be done at the equation level. The reason is because we assume that the
time scale of D-particle scatterings is much shorter
than $H^{-1}.$ Remember also that terms $r_ir_j$ that would have zero contribution for $i\neq j$, have already been ignored, in previous steps of our analysis. Then, the Boltzmann equation (with the binary collision
term taken also into account) is written as:

\begin{eqnarray}\label{boltz}
 &  & \frac{\partial f}{\partial t}-H\sum_{i}\overline{k}^{i}\frac{\partial f}{\partial\overline{k}^{i}}+6Ha^4(t)\sum_{i}\sigma_{0i}^{2}\left(\overline{k}^{i}\right)^{3}\frac{\partial f}{\partial\overline{k}^{i}}-\frac{2}{m^{2}}Ha^{4}(t)\sum_{j}\left(\overline{k}^{j}\right)^2\sum_{i}\sigma_{0i}^{2}\left(\overline{k}^{i}\right)^{3}\frac{\partial f}{\partial\overline{k}^{i}}\nonumber\\
 &  &+4Hm^{2}a^4(t)\sum_{i}\sigma_{0i}^{2}\overline{k}^{i}\frac{\partial f}{\partial\overline{k}^{i}}-4Ha^4(t)\sum_{j}\left(\overline{k}^{j}\right)^2\sum_{i}\sigma_{0i}^{2}\overline{k}^{i}\frac{\partial f}{\partial\overline{k}^{i}},
  =\frac{C[f]}{\overline{k}^{0}},\end{eqnarray}
\noindent where the fourth and sixth terms can be neglected as very small compared to the other terms, always under the assumption of superheavy dark matter: $\left(\frac{\overline{k}_{i}}{m}\right)^{2}\ll 1$.
Since the number density $n\left(t\right)$ is defined as:
\begin{equation}
n\left(t\right)\equiv\frac{g}{\left(2\pi\right)^{3}}\int d^{3}\overline{k}\, f\left(t,\overline{k}^{i}\right),
\end{equation}
\noindent with $d^{3}\overline{k}\equiv d\overline{k}^{1}d\overline{k}^{2}d\overline{k}^{3}$, the following simple identities can be derived:
\begin{eqnarray}\label{identity}
\frac{g}{\left(2\pi\right)^{3}}\int d^{3}\overline{k}\,\frac{\partial}{\partial t}f\left(t,\overline{k}^{i}\right)&=&\frac{dn}{dt},\nonumber\\
\frac{g}{\left(2\pi\right)^{3}}\left(\sum_{i} \int d^{3}\overline{k}\,\overline{k}^{i}\frac{\partial f}{\partial \overline{k}^{i}}\right)&=&-3n,\nonumber\\
\frac{g}{\left(2\pi\right)^{3}}\sum_{i}\sigma_{0i}^{2}\int d^{3}\overline{k}\,\overline{k}^{i}\frac{\partial f}{\partial\overline{k}^{i}}&=&-\left(\sigma_{01}^{2}+\sigma_{02}^{2}+\sigma_{03}^{2}\right)n\nonumber\\
\frac{g}{\left(2\pi\right)^{3}}\sum_{i}\sigma_{0i}^{2}\int d^{3}\overline{k}\,\left(\overline{k}^{i}\right)^{3}\frac{\partial f}
{\partial\overline{k}^{i}}&=&-3\frac{g}{\left(2\pi\right)^{3}}\sum_{i}\sigma_{0i}^{2}\int d^{3}\overline{k}\,\left(\overline{k}^{i}\right)^{2}f~.
\end{eqnarray}

We now define an average temperature $T$ through (see~\cite{Wu}):
 \begin{equation}
\frac{g}{\left(2\pi\right)^{3}}\int d^{3}\overline{k}\: \left(\overline{k}^{i}\right)^2 f\equiv Tmn\label{temperature}\end{equation}
\noindent and, as a result, the last identity in (\ref{identity}) becomes:
\begin{equation}
\frac{g}{\left(2\pi\right)^{3}}\sum_{i}\sigma_{0i}^{2}\int d^{3}\overline{k}\,\left(\overline{k}^{i}\right)^{3}\frac{\partial f}{\partial\overline{k}^{i}}=-\left(\sigma_{01}^2+\sigma_{02}^2+\sigma_{03}^2\right)3Tmn.
\end{equation}

Therefore integrating (\ref{boltz}) with respect to $d^3\overline{k}$ yields~\cite{ariadneplb}:
 \begin{equation}\label{boltz2}
\frac{dn}{dt}+3Hn=\Gamma(t)n+\frac{g}{(2\pi)^{3}}\int{d^{3}\overline{k}\frac{C[f]}{\overline{k}_0}}~,
\end{equation}
where we have managed to incorporate all the foam-induced correction terms in a time-dependent
source term given by:
 \begin{equation}\label{source}
\Gamma(t)=2Ha^4(t)m\left(\sigma_{01}^2+\sigma_{02}^2+\sigma_{03}^2\right)\left[9T+2m\right].\end{equation}
The reader is reminded at this point that the above result has been derived under the approximation of superheavy dark matter species $\chi$.

We will elaborate a little on the contributions of the foam to the interaction terms $C\left[f\right]$
in the Boltzmann equation (\ref{boltz2}). We assume for simplicity that the dominant scattering
process affecting the abundance of $\chi$ is $1+2\longleftrightarrow 3+4$,
where $1$ and $2$ are $\chi$ particles and $3$ and $4$ denote light
(Standard Model) particles. The latter are assumed to be strongly
coupled to the cosmic plasma and hence in equilibrium. Following standard
treatments we have
\begin{eqnarray}\label{collision}
\frac{1}{(2\pi)^{3}}\int{d^{3}\overline{k}\frac{C[f]}{\overline{k}_{0}}} &\equiv&\int\frac{d^{3}\overline{k}^{1}}{(2\pi)^{3}2\omega_{1}}\int\frac{d^{3}\overline{k}^{2}}{(2\pi)^{3}2\omega_{2}}\int\frac{d^{3}\overline{k}^{3}}{(2\pi)^{3}2\omega_{3}}\int\frac{d^{3}\overline{k}^{4}}{(2\pi)^{3}2\omega_{4}}\nonumber\\
&&\times(2\pi)^{4}\delta^{3}\left(\overline{k}^{1}+\overline{k}^{2}-\overline{k}^{3}-\overline{k}^{4}\right)\delta\left(\omega_{1}+\omega_{2}-\omega_{3}-\omega_{4}\right)\left|\mathcal{M}\right|^{2}\nonumber\\
&&\times\left[f_{3}f_{4}\left(1\pm f_{1}\right)\left(1\pm f_{2}\right)-f_{1}f_{2}\left(1\pm f_{3}\right)\left(1\pm f_{4}\right)\right],\end{eqnarray}
where $(\pm)$ denotes Bose enhancement $(+)$ and Pauli blocking
$(-)$; $f_{i}$ is the distribution function associated
with species $i$; $\mathcal{M}$ is the scattering amplitude; and $\omega_{i}$
is the foam dressed energy for particle $i$. For $T<\omega-\mu$
the quantum statistics of particles can be replaced by the Maxwell-Boltzmann
distribution, i.e.\ $f_{i}\rightarrow\exp\beta\left(\mu_{i}-\omega_{i}\right)$
and clearly the effects of foam in $f_{i}$ will only occur through the modified $\omega_{i}\left(=\overline{k}_{i}^{0}\right)$. D-foam effects will also appear in $\mathcal{M}$.
For neutralinos (which are spin-$\frac{1}{2}$ Majorana fermions in
MSSM and are linear combinations of the bino and wino) the tree level
amplitude involves a weak vector boson propagator. Vector bosons have
masses close to $100$~GeV, i.e.\ of the order of WIMPs and so (for
simplicity working in the unitary gauge) the foam effects at the tree
level are governed by the vector boson propagator; this leads to a
suppression of the effect of the foam relative to that in the source
by the square of the WIMP mass. From the form
of the Boltzmann factor $f$, the thermal averaging has a similarly suppressed foam contribution.

Finally, it is important to note that equation (\ref{boltz2}) yields the standard scaling for ordinary massive matter in the absence of D-particles (term $3Hn$), because of the use of the local momenta formalism~\cite{bernstein}.
From (\ref{boltz2}) we can also see  that stochastically fluctuating  Finsler metrics, induced by the D-particles foam model, are equivalent  to particle-production source terms in the Boltzmann equations. As a consequence of these extra terms,  relic abundances of dark matter candidates are expected to be modified, more or less significantly, according to the free parameters of our model. To understand this modification, the reader is referred to~\cite{elmn},  where the generic effect of source terms on the Boltzmann equation is investigated, but where the source is generated by  off-shell / dilaton terms. Another difference from~\cite{elmn} is that, in our case, the foam-induced
source effects come with a positive sign, therefore are expected to increase today's dark matter relic abundances, when compared to the standard ones and as a result,  leave even less room for supersymmetry at colliders~\cite{elmn}.
However, the corrections will in general be quite small since they will always scale as $\sigma^2_{0i}$, which are perturbative free parameters in our analysis depending on the string scale, as will be explained at the end of the article.

\subsection{Solutions to the Boltzmann Equation and Foam-Modified Thermal Dark Matter Relic Abundances}

 In (\ref{boltz2}), from standard arguments, the  collision term  has the form: $-\left\langle \sigma \upsilon\right\rangle\left(n^{2}-n_{\rm eq}^{2}\right)$~\cite{kolb}. Then the Boltzmann equation is re-written as:
 \begin{equation}\label{bz}
 \dot{n}+3Hn=\tilde{\Gamma}(t)n-\left\langle \sigma \upsilon\right\rangle\left(n^2-n_{\rm eq}^2\right),
 \end{equation}
with $n_{\rm eq}$ being the (thermal) equilibrium distribution, $\left\langle \sigma \upsilon\right\rangle$
the thermally averaged annihilation cross section  summed over all
contributing channels and $\upsilon$ the Mueller velocity. Our source, from now on, will be denoted by $\tilde{\Gamma}(t)$, not to be confused with the particles interaction rate $\Gamma(t)$.

An analytical solution of (\ref{bz}) is difficult to be given generally since the collision part is era dependent. In standard approaches~\cite{kolb}, the total annihilation cross section is assumed to depend on temperature through:
\begin{equation}\label{cross}\left\langle \sigma \upsilon\right\rangle=\sigma(x)=c_0 x^{-n},
\end{equation}
where $x=\frac{m}{T}$, $c_0$ is an arbitrary constant (discussed in more detail later) and $n$ is related to the channel of annihilation: $n=0$ for $s$-wave and $n=1$ for $p$-wave annihilators. Since these are usually the dominant types of annihilation, we will restrict our attention on these two possible values for the integer $n$.

It proves convenient to use instead of the particles' number $n(x)$, the parameter number per entropy density $Y(x)=\frac{n(x)}{s}$. Under this re-parametrization and along with using the conservation of the entropy per comoving volume ($sa^3=\text{const}\Rightarrow \frac{ds}{dt}=-3Hs$), (\ref{bz}) is written in the form:
\begin{equation}
\frac{dY}{dt}=\tilde{\Gamma}(t)Y-\left\langle \sigma \upsilon\right\rangle s\left(Y^2-Y_{\rm eq}^2\right).
\end{equation}

Also, since the interaction term depends explicitly on temperature rather than time, it is useful to take $x$ as the independent evolution variable. Applying (\ref{time}) then, yields:
\begin{equation}\label{bz2}
\frac{dY}{dx}=\frac{x\tilde{\Gamma}(x)}{H_m}Y-\frac{\left\langle \sigma \upsilon\right\rangle u_0}{H_m}\frac{1}{x^2}\left(Y^2-Y_{\rm eq}^2\right),
\end{equation}
\noindent where we have set $u_0\equiv \frac{2\pi^2}{45}h~m^3$, with $h$ denoting the entropy degrees of freedom and $H_m\equiv Hx^2= 1.67g_\textit{eff}'^{1/2}\frac{m^2}{M_{}pl}$, in a r.d.e, with $H$ being the Hubble rate.

Before the freeze-out ($x<x_f$), particles are in thermal equilibrium, therefore
\begin{equation}
Y(x)\approx Y_{\rm eq}(x)
\end{equation}
with the equilibrium density being:
 \begin{equation}\label{eq}
Y_{\rm eq}(x)=n_{eq}(x)/s={u_0}^{-1}x^{3} n_{\rm eq}(x)
\end{equation}
and $n_{\rm eq}(x)$ is given by (\ref{nonrel2}).

After the freeze-out and at relatively late times ($x\gg x_f$), $Y$ tracks $Y_{\rm eq}$ very poorly, therefore assuming that $Y\gg Y_{\rm eq}$, one can safely neglect $Y_{\rm eq}$ from (\ref{bz2}) obtaining:
\begin{equation}\label{bz4}
\frac{d}{dx}\frac{1}{Y}+\frac{x\tilde{\Gamma}(x)}{H_{m_\chi}}\frac{1}{Y}=\frac{\left\langle \sigma \upsilon\right\rangle u_0}{H_{m_\chi}}\frac{1}{x^2},
\end{equation}
\noindent where we have also divided both sides by $1/Y^2$ and  have restricted the discussion on species $\chi$ to superheavy dark matter candidates of mass $m_{\chi}$.
Now (\ref{bz4}) is in the form of a Bernoulli equation ($y'+P(x)y=Q(x)y^n$, for $n=0$) and as such can be exactly solved.  The general solution to (\ref{bz4}) is then:
\begin{equation}\label{sol1}
\frac{1}{Y(x)}=\exp\left(-\int_{a}^{x}{\frac{x'\tilde{\Gamma}(x')}{H_{m_\chi}}dx'}\right)\left\{\int_{b}^{x}{\exp\left(\int_{a}^{s}{\frac{x\tilde{\Gamma}(x)}{H_{m_{\chi}}}dx}\right)\frac{\sigma(s) u_0}{H_{m_{\chi}} s^2}ds}\right\}
\end{equation}
with $a$ and $b$ two arbitrary constants and $s$ just an integration dummy variable, not to be confused with the entropy density.  Choosing $a=x_f$, keeping $b$ general and applying (\ref{sol1}) on the freeze-out point $x_f$ as well (this is only approximately correct since the solution (\ref{sol1}) is formally valid for $x>x_f$), we can obtain today's solution:

\begin{equation}\label{sol2}
\frac{1}{Y(x_0)}-\exp\left(-\int_{x_f}^{x_0}{\frac{x\tilde{\Gamma}(x)}{H_{m_\chi}}dx}\right)\frac{1}{Y(x_f)}=\exp\left(-\int_{x_f}^{x_0}{\frac{x\tilde{\Gamma}(x)}{H_{m_\chi}}dx}\right)\left\{\int_{x_f}^{x_0}{\exp\left(\int_{x_f}^{s}{\frac{x\tilde{\Gamma}(x)}{H_{m_\chi}}dx}\right)\frac{\sigma(s) u_0}{H_{m_\chi} s^2}ds}\right\},
\end{equation}
where obviously $x_0$ denotes today.
Expanding the exponentials with respect to the small parameters $\sigma_{i}^{2}$ that are hidden in the source term $\tilde{\Gamma}(x)$ and keeping only corrections up to order $\sigma_i^2$, that is terms linear in $\tilde{\Gamma}$, one obtains:
\begin{eqnarray}\label{sol3}
&&\frac{1}{Y(x_0)}-\frac{1}{Y(x_f)}=\nonumber\\
&&-\left(\int_{x_f}^{x_0}{\frac{x\tilde{\Gamma}(x)}{H_{m_\chi}}dx}\right)\frac{1}{Y(x_f)}+\int_{x_f}^{x_0}{\frac{\sigma(s) u_0}{H_{m_\chi} s^2}ds}\left(1-\int_{x_f}^{x_0}{\frac{x\tilde{\Gamma}(x)}{H_{m_\chi}}dx}\right)+\int_{x_f}^{x_0}{\left(\int_{x_f}^{s}{\frac{x\tilde{\Gamma}(x)}{H_{m_\chi}}dx}\right)\frac{\sigma(s) u_0}{H_{m_\chi} s^2}ds}.
\end{eqnarray}
 The function $\sigma(s)$ should in general also depend on the statistical parameters $\sigma_i$ of our model through $c_0$. However, at this point this is not taken into account since formally the only effect it would have is to add extra undetermined constant terms in our analysis. An important point to be remarked here is that the freeze-out point $x_f$, is also affected by the source term (\ref{source}). To get the shift to $x_f$ we apply the freeze-out criterion~\cite{kolb}:
 \begin{equation}
 Y\left(x_f\right)-Y_{\rm eq}\left(x_f\right)\approx d~Y_{\rm eq}\left(x_f\right),
 \end{equation}
 \noindent where $d$ is a phenomenological constant.
 Using (\ref{sol1}) and (\ref{eq}), we get:
 \begin{equation}\label{freeze-out}
 \int_{b}^{x}{\exp\left(\int_{x_f}^{s}{\frac{x\tilde{\Gamma}(x)}{H_{m_{\chi}}}dx}\right)\frac{\sigma(s) u_0}{H_{m_{\chi}} s^2}ds}=(d+1)^{-1}u_0 x_f^{-3} n_{\rm eq}^{-1}(x_f),
 \end{equation}
 \noindent with $n_{\rm eq}(x_f)$ given by (\ref{nonrel2}).

Denoting the left hand side of (\ref{freeze-out}) by I,  a very rough estimation of the freeze-out temperature comes by assuming that:
$I\approx I^{(0)}$
where by $I^{(0)}$ we mean the source-free part.

Then, choosing $b\equiv x_{in}$, where $x_{in}$ stands for some  initial time, e.g.\  after just after the exit from inflation and expanding the exponential with respect to the source $\tilde{\Gamma}(t)$, we get:
\begin{eqnarray}
&I&=-\frac{A}{n+1}\left(\frac{1}{x_f^{n+1}}-\frac{1}{x_{in}^{n+1}}\right)+\frac{AB}{3\left(2-n\right)}\left(x_f^{2-n}-x_{in}^{2-n}\right)+\frac{AB                                                   }{3\left(n+1\right)}x_f^3\left(x_f^{-n-1}-x_{in}^{-n-1}\right)\nonumber\\
&& \approx -\frac{A}{n+1}\left(\frac{1}{x_f^{n+1}}-\frac{1}{x_{in}^{n+1}}\right)-\frac{AB}{3\left(n+1\right)}x_f^{2-n}\left(\frac{x_{in}}{x_f}\right)^{-n-1}\nonumber\\
&&= -\frac{A}{n+1}\left(\frac{1}{x_f^{n+1}}-\frac{1}{x_{in}^{n+1}}\right)-\frac{AB}{3\left(n+1\right)}x_f^{(0)^{2-n}}\left(\frac{x_{in}}{x_f^{(0)}}\right)^{-n-1},
\end{eqnarray}
since  $x_{in}\ll x_{f}$ and  $B$ is already of order $\sigma^2$ (see below). We also get
\begin{equation}
I^{(0)}=-\frac{A}{n+1}\left(\frac{1}{x_f^{(0)^{n+1}}}-\frac{1}{x_{in}^{n+1}}\right),
\end{equation}
where $A\equiv \frac{c_0u_0}{H_{m_{\chi}}}$ and $B\equiv 18\frac{m_{\chi}^2}{x_0^4}\bar{\sigma}_0^2$.
Applying our assumption then and always to linear  order in $B$,  yields:
\begin{equation}
x_f \approx x_f^{(0)}\left(1+\frac{B}{3\left(n+1\right)}x_f^{(0)^{3}}\left(\frac{x_f^{(0)}}{x_{in}}\right)^{n+1}\right),
\end{equation}
 or equivalently
\begin{equation}\label{shift}
x_f \approx x_f^{(0)}+\frac{6}{n+1}\left(\frac{x_f^{(0)}}{x_0}\right)^4\left(\frac{x_f^{(0)}}{x_{in}}\right)^{n+1}m^2\bar{\sigma}_0^2.
\end{equation}
We can see from (\ref{shift}) that the freeze-out point is positively shifted due to the foam meaning that dark matter species are expected to have decoupled later in the presence of the stochastic background. However what is important is that the correction to $x_f$, similarly to~\cite{elmn}, only scales linearly to the source term, therefore as $\sigma_i^2$.  And since in any case these parameters are very small, also the shift on the freeze-out temperature will be small, therefore one can retain the assumptions/approximations of Standard Cosmology~\cite{kolb,elmn}. In this sense, all terms involving $1/Y(x_f)$ can be neglected from  equation (\ref{sol3}).  For convenience, we additionally can set  $\int_{x_f}^{x_0}{\frac{\sigma(s)}{ s^2}ds}\equiv J$, as in standard approaches and $\frac{\sigma(s)}{s^2}\equiv J(s)$, where by $x_f$. We note at this point  that from now  by $x_f$, wherever it appears, we will refer to the modified freeze-out point. Then, equation (\ref{sol3}) can be written under the compact form:
\begin{equation}\label{sol4}
Y(x_0)^{-1}=\left(\frac{\pi}{45}\right)^{1/2}m_\chi M_{pl}~g_{\textit{eff},f}'^{-1/2}h'J\left(1-\int_{x_f}^{x_0}{\frac{x\tilde{\Gamma}(x)}{H_{m_\chi}}dx}+\frac{1}{J}\int_{x_f}^{x_0}{J(s)\left(\int_{x_f}^{s}{\frac{x\tilde{\Gamma}(x)}{H_{m_\chi}}dx}\right)}ds\right),
\end{equation}
where $g_{\textit{eff},f}'$ and $h'$ represent the effective number of degrees of freedom and the entropy degrees of freedom at the freeze-out point $x_f$, respectively.
Obviously, the term $\left(\frac{\pi}{45}\right)^{1/2}m_\chi M_{pl}~g_{\textit{eff},f}'^{-1/2}h'J$ corresponds to the standard result~\cite{kolb} whereas the other terms in the prefactor are our foam-induced corrections. However, one should bear in mind that since both the freeze-out point and the effective number of degrees of freedom carry foam corrections,  what we refer to as standard result is only approximately the standard one. We also note, that solving \emph{exactly} the differential equation~(\ref{sol4}) gave us an extra correction term (third term in the prefactor in~(\ref{sol4})) when compared to~\cite{elmn}, where similarly a source term in the Boltzmann equation was considered.

The energy density of the species $\chi$ will then be:
\begin{equation}
\rho=Y(x_0)s~m_\chi=\left(1+\int_{x_f}^{x_0}{\frac{x\tilde{\Gamma}(x)}{H_{m_\chi}}dx}-\frac{1}{J}\int_{x_f}^{x_0}{J(s)\left(\int_{x_f}^{s}{\frac{x\tilde{\Gamma}(x)}{H_{m_\chi}}dx}\right)}ds\right)\left(\frac{4\pi^3}{45}\right)^{1/2}\left(\frac{T}{T_{\gamma}}\right)^3\frac{T_{\gamma}^3}{M_{pl}}\frac{g_{\textit{eff},f}'}{J}
\end{equation}
or equivalently to~\cite{elmn}:
\begin{equation}\label{relic}
\frac{\Omega_{\chi}h_{0}^{2}}{(\Omega_{\chi}h_{0}^{2})_{\text{no~source}}}= \left(\frac{g_{\textit{eff},f}'}{g_{\textit{eff},f}}\right)^{\frac{1}{2}}
\left(1+\int_{x_f}^{x_0}{\frac{x\tilde{\Gamma}(x)}{H_{m_\chi}}dx}-\frac{1}{J}\int_{x_f}^{x_0}{J(s)\left(\int_{x_f}^{s}{\frac{x\tilde{\Gamma}(x)}{H_{m_\chi}}dx}\right)}ds\right).
\end{equation}

It is important to note that the formalism presented so far is true for an \emph{any} source term present in the Boltzmann equation, only under the restriction that it is perturbatively small. Substituting in then our specific source term (\ref{source})
and using relation (\ref{cross}) for a general $n$, the integrals of the first factor on the right hand side of (\ref{relic}) can be explicitly calculated yielding:
\begin{eqnarray}\label{ef1}
&&\left\{1+\int_{x_f}^{x_0}{\frac{x\tilde{\Gamma}(x)}{H_{m_\chi}}dx}-\frac{1}{J}\int_{x_f}^{x_0}{J(s)\left(\int_{x_f}^{s}{\frac{x\tilde{\Gamma}(x)}{H_{m_\chi}}dx}\right)}ds\right\}=\nonumber\\
&&\left\{1+x_0^{-4}m_\chi^2\bar{\sigma}_0^2\left[6x_0^3-6x_f^3+x_0^4-x_f^4-\frac{c_0}{J}\left(-\left(6x_f^3+x_f^4\right)\int_{x_f}^{x_0}\frac{1}{s^{2+n}}ds+6\int_{x_f}^{x_0}\frac{1}{s^{-1+n}}ds+\int_{x_f}^{x_0}\frac{1}{s^{-2+n}}ds\right)\right]\right\},\nonumber\\
\end{eqnarray}
with $n=0,1$.

Upon noticing that in all realistic cosmologies~\cite{spanos} $\frac{x_f}{x_0}\ll 1$,
 \begin{equation}
 J=\frac{c_0}{n+1}\left(\frac{1}{x_f^{n+1}}-\frac{1}{x_0^{n+1}}\right)\approx \frac{c_0}{n+1}\frac{1}{x_f^{n+1}}
 \end{equation}
 and therefore for both values of $n$, the following approximations can be made:
\begin{eqnarray}
\frac{c_0x_0^{-4}}{J}\left(6x_f^3+x_f^4\right)\int_{x_f}^{x_0}\frac{1}{s^{2+n}}ds&\approx& 6x_0^{-1}\left(\frac{x_f}{x_0}\right)^3+\left(\frac{x_f}{x_0}\right)^4-6x_0^{-1}\left(\frac{x_f}{x_0}\right)^{4+n}-\left(\frac{x_f}{x_0}\right)^{5+n}\ll 1,\nonumber\\
\frac{c_0x_0^{-4}}{J}\int_{x_f}^{x_0}\frac{1}{s^{-2+n}}ds&\approx& \frac{n+1}{3-n}\left[\left(\frac{x_f}{x_0}\right)^{n+1}-\frac{x_f}{x_0}\right]\ll 1,\nonumber\\
\frac{c_0x_0^{-4}}{J}\int_{x_f}^{x_0}\frac{1}{s^{-1+n}}ds&\approx& x_0^{-1}\left( \frac{n+1}{2-n}\right)\left[\left(\frac{x_f}{x_0}\right)^{n+1}-\left(\frac{x_f}{x_0}\right)^3\right]\ll 1.\nonumber\\
\end{eqnarray}
Hence, (\ref{relic}) reduces to:
\begin{eqnarray}
\frac{\Omega_{\chi}h_{0}^{2}}{(\Omega_{\chi}h_{0}^{2})_{\text{no~source}}}&=&
\left(\frac{g_{\textit{eff},f}'}{g_{\textit{eff},f}}\right)^{\frac{1}{2}}
\left\{1+m_\chi^2\bar{\sigma}_0^2\left(1+6x_0^{-1}-6\frac{x_f^3}{x_0^4}-\frac{6(n+1)}{2-n}\left[\frac{x_f^{n+1}}{x_0^{2+n}}-\frac{x_f^3}{x_0^4}\right]\right)\right\}\nonumber\\
&\approx& \left(\frac{g_{\textit{eff},f}'}{g_{\textit{eff},f}}\right)^{\frac{1}{2}}
\left\{1+m_\chi^2\bar{\sigma}_0^2\left(1+6x_0^{-1}\right)\right\},
\end{eqnarray}
since terms of the form $\left(\frac{x_f}{x_0}\right)^n$ are of subdominant contribution.
It is important to realize that no era assumption was made when calculating the integral $\int_{x_f}^{x_0}{\frac{x\tilde{\Gamma}(x)}{H_{m_\chi}}dx}$ in (\ref{relic}) although the form of the source term as written in (\ref{source}) assumes a r.d.e. In fact, one can easily check that the integrand does not require any specific form of the Hubble rate $H$ and thus is era independent.

However, totally in our calculation, it has been tacitly assumed that temperature scales as $T=\frac{T_0}{a(t)}$ throughout all the evolution of the Universe. This assumption is only \emph{approximately} true, since annihilating particles at different stages of the evolution of the universe deposited energy in the universe, slowing down its cooling. In order to get the corrected cooling law one should assume that the entropy, dominated by the relativistic species contribution, remains constant  (this is also an approximation since the presence of the source term may lead to entropy production~\cite{elmn}). Then  we get:  $a(t)\approx \left(\frac{g_{\textit{eff},0}}{g_\textit{eff}(t)}\right)^{1/3}\frac{a_0T_0}{T}$, where $g_\textit{eff}$ is an evolving with time (temperature) function  $g_\textit{eff}(t)$~\cite{kolb} and is such that today $g_{\textit{eff},0}\simeq 3.36$  (only photons and the light neutrinos contribute), while at freeze-out
$g_\textit{eff}\left(t_f\right)\simeq 106.75$ within the framework of the Standard Model. Therefore in principle, one should calculate  the integral in (\ref{relic}) numerically using this evolution of $g_\textit{eff}(t(T))$. For our  qualitative purposes in this work we have ignored such corrections which, at any rate, could not affect significantly the order of magnitude of foam contributions to DM relic abundances.

The last ingredient in our calculation involves the modification of the effective number of degrees of freedom, which has been presented in section \ref{effsection}. Applying (\ref{eff}) and counting carefully the relativistic species at an early era of the universe, when basically all species of the Standard Model were relativistic (the reader should recall at this stage that a typical freeze-out temperature is of the order of some $\rm GeV$ for typical dark matter candidates with masses in the range $m\approx 1~{\rm GeV}-10^4{\rm GeV}$), we find:
\begin{equation}
g_\textit{eff}'=106.75+22138\,\overline{\sigma_0}^2T^2
\end{equation}
(we used that: $g_{f}=g_\text{quarks}+g_\text{leptons}=6\times\left(2\times 2\times 3\right)+3\times\left(2\times 2\right)+3\times 2=90$ and $g_{b}=g_\text{gluons}+g_\text{EW}+g_\text{photon}+g_\text{Higgs}=8\times 2+ 3\times 3 + 2+1=28$).

Therefore the correction to the effective number of degrees of freedom will be roughly going as:

\begin{equation}
\left(\frac{g_\textit{eff}'}{g_\textit{eff}}\right)_{x_f}\approx 1+207.38\,\bar{\sigma}_0^2T_f^2
\end{equation}
and the modification to dark matter relic abundances is finally found to be~\cite{ariadneplb}:
\begin{equation}\label{final2}
\frac{\Omega_{\chi}h_{0}^{2}}{(\Omega_{\chi}h_{0}^{2})_{\text{no~source}}}=\left\{1+m_{\chi}^2\bar{\sigma}_0^2\left(6x_0^{-1}+1\right)\right\}\left\{1+103.69\,\bar{\sigma}_0^2m_{\chi}^2x_f^{-2}\right\}.
\end{equation}

The dominant contributions come from the terms $\bar{\sigma}_0^2m_\chi^2$ and $103.69\,\bar{\sigma}_0^2m_{\chi}^2x_f^{-2}$, since the freeze-out for, e.g.\ neutralino, dark matter candidates typically occurs at $x_f\approx 20$~\cite{spanos} (since any foam-induced freeze-out shift has been found to be negligible, we have retained the Standard Cosmology estimate). The other terms can be safely neglected from (\ref{final2}) since  $x_f \ll x_0$ in all phenomenologically realistic cosmological models.

Then, writing our statistical parameters $\sigma_{0i}$ as:
\begin{equation}
\sigma_{0i}=\frac{g_s}{M_s}\Delta_i,
\end{equation}
where $g_s$ is the string coupling, $M_s$ the string scale and $\Delta_i$ are dimensionless variances  $\frac{M_s}{g_s}\,\sqrt{\langle\langle r^2_i \rangle\rangle} \equiv \Delta_i$, $i=1,2,3$ that can be naturally up to ${\mathcal O}(1)$, the dominant contributions of the foam effects on DM relic abundances is of order:
\begin{equation}
\frac{\Omega_{\chi}h_{0}^{2}}{(\Omega_{\chi}h_{0}^{2})_{\text{no~source}}} \sim
1 + 1.259\,m_\chi^2\frac{g_s^2}{M_s^2} \sum_{i=1}^3 \Delta_i^2~.
\label{dmcontr}
\end{equation}
Then, we may use the standard WIMP estimate $(\Omega_{\chi})_{{\rm no~source}} \sim \frac{x_f T_0^3}{\rho_c M_P} \langle \sigma_A v \rangle^{-1}$, where $\rho_c$ is the critical density of the Universe, $T_0$ is the CMB temperature today, and the total annihilation cross section can be written on dimensional grounds as:
\begin{equation}\label{weakcross}
\sigma_A v \sim k \frac{g_{\rm weak}^4}{16\pi^2 m_\chi^2}~,
\end{equation}
for $s$-wave annihilators (cf.\ (\ref{cross})), that will restrict our qualitative considerations here. The weak interaction coupling constant $g_{\rm weak} \sim 0.65$, and the fudge factor $k$~\cite{spanos}, which in general parametrizes deviations from the above estimate, is assumed in the range $k \in (0.5 - 2)$. As well known, the WIMP miracle then implies
that, for a particle which is assumed to make up 100\% of DM, as is our case here, one must have masses in the range $m_\chi \sim 100 - 10^3~{\rm GeV}$. In view of (\ref{dmcontr}), then, we obtain (for $s$-wave annihilators):
\begin{equation}
\Omega_{\chi}h_{0}^{2} = h_{0}^{2} \,\frac{16 \pi^2 x_f T_0^3 \, m_\chi^2 }{\rho_c k \, g_{\rm weak}^4 \, M_P}\left(
1 + 1.259\,m_\chi^2 \frac{g_s^2}{M_s^2}\sum_{i=1}^3 \Delta_i^2\right)~.
\label{dmcontr2}
\end{equation}
 We shall make use of such estimates in our phenomenology analysis, in the next section.

It is understood that in estimating $\Omega_{\rm no~source}$ one may use the pertinent thermal abundances in the context of specific particle physics models of interesting dark matter candidates, such as supersymmetric or string-inspired models, with hidden sectors~\cite{spanos}. The Finsler foam corrections are \emph{universal} in this respect, since, as we have seen above, the only assumption is that electrically neutral candidates interact predominantly with the foam defects. The only difference in using models other than the standard WIMP ones lies on the fact that in such cases the weak interaction coupling constant $g_{\rm weak}$ should be replaced by  the appropriate coupling $g_X$, for instance describing the properties of the hidden sector of a string-inspired model~\cite{spanos}. The restriction is that the combination $m_X^2/g_X^4$, which enters the inverse total annihilation cross section of the dark matter candidate $X$, is fixed by the requirement of producing thermal relic in the ball park of WMAP observations~\cite{cmb,7yrwmap}.

The presence of the stochastic D-particle foam, results in an increase of the
thermal DM relic abundance, compared with the corresponding result in Standard Cosmology, in the absence of the foam. In our approach, the modifications are mainly due to the presence of a source-like term (\ref{source}) in the
pertinent Boltzmann equation (\ref{boltz2}), associated with particle production as a consequence of the D-particle foam.
This interpretation in terms of particle production is phenomenological. As is well known, the concept of particles itself is an ambiguous one in curved space-time. Particle production within the framework of kinetic theory can be incorporated as a viscous pressure. It has been shown that the inclusion of particle production, in a general relativistic background, leads to a kinetic equation of the form~\cite{triginer}
$$
p^{\mu}f_{,\mu}-{\Gamma ^{\mu}}_{\upsilon \nu}p^{\upsilon}p^{\nu}\frac{\partial f}{\partial p^{\mu}}=C[f]+H(x,p)
$$
(using standard notation). Here $H$ is the source term representing particle production (or decay) and $C$ collisional events. Moreover, in such a framework, it is assumed that $C$
 \begin{itemize}
 \item is a local function of $f$
 \item implies number and 4-momentum conservation and
 \item induces non-negative entropy production and vanishes if and only if $f$ has the form of local equilibrium distribution
 \end{itemize}
but otherwise its structure does not need to be specified. The particle number 4-flow, $N^{\mu}$ is defined as
$$
N^{\mu}=\int {\rm d}P p^{\mu} f(x,p)
$$
where ${\rm d}P= \delta (p_{\nu}p^{\nu}+m^{2})\sqrt{-g}dp^{0}dp^{1}dp^{2}dp^{3}$. $H$  is taken to have the form
$$
H(x,p)=\zeta(x,p) f^{0}(x,p)
$$
where $f^{0}$ is the equilibrium distribution; $\zeta(x,p)$  is parameterized in terms of 4-velocity and momentum as
$$
\zeta(x,p)=-\frac{u^{\nu}p_{\nu}}{\tau(x)}+\upsilon(x)
$$
and two arbitrary functions $\upsilon(x)$ and  $\tau(x)$. This form of $H$ is compatible with but more general than the source that we derived for D-foam. It can then be shown that~\cite{triginer}
$$
N^{\mu}_{;\mu}=\int H {\rm d}P.
$$
On writing,
$$
\Gamma=\frac{1}{n}\int H {\rm d}P
$$
(corresponding to our source notation), $n$ being the particle number density, the following relation~\cite{triginer}
$$
n\Gamma=-\frac{u^{\kappa}}{\tau}N_{\kappa}+\upsilon(x) n
$$
can be derived. In our case $\tau$ is large and so can be neglected. Hence a positive $\upsilon(x)$ corresponding to particle production corresponds to a positive $\Gamma$ in (\ref{source}). It is in this sense that we regard our source $\Gamma$ term in (\ref{source}) as corresponding to particle production.

In terms of particle-physics DM models, this implies that in addition to the $m_X^2/g_X^4$ there is another parameter,
the foam fluctuations $\bar{\sigma}_0^2$ which enters the game.
In our case, the foam corrections to the effective degrees of freedom, $\frac{g_\textit{eff}'}{g_\textit{eff}}$, contribute at most 26\% to the final result~(\ref{dmcontr2}).

Schematically therefore, we can say that the presence of the foam and the requirement of a WIMP-miracle-like situation, as implied by the cosmological data~\cite{cmb},
fixes combinations of parameters, which in order of magnitude look like:
\begin{equation}\label{combination}
\Omega_X \propto \frac{m_X^2}{g_X^4} \left[1 + \left|\mathcal{O}\left(m_X^2\frac{g_s^2}{M_s^2} \sum_{i=1}^3 \Delta_i^2\right)\right| \right]
\end{equation}
for cases where the particle $m_X$ of a model, including hidden sector ones, constitutes the dominant DM candidate.

Whether the D-foam effects on the DM relic abundances are observable depends on the value of the D-particle mass scale $M_s/g_s$, which in the modern version of string theory is a free parameter. In the next section we proceed to discuss briefly the phenomenology of the model, and compare it with some other related interesting cosmological models available to date.

\section{Comments on Phenomenology and Comparison with other Models}\label{sec:wmap}

The corrections due to D-foam are in general small, as expected, and they can only
be significant for low-string-scale models and heavy DM candidates.
For traditionally
high string scales ($M_{s}\gtrsim{\mathcal O}(10^{16}~{\rm GeV})$), in order for the
D-foam effects to be significant one needs superheavy DM, with masses
higher than $M_{s}/g_{s}$. However, the effects of such a superheavy
DM will be eroded by inflation~\cite{kolb}; moreover superheavy DM would not be
produced significantly during a reheating phase of the Universe after
its exit from the inflationary period.

For intermediate string scales~\cite{pioline},
where the quantity $M_{s}/g_{s}$ could be of order $10^{11}~{\rm GeV}$
(which is the order of the GZK cutoff of ultra-high energy cosmic
rays), there could be significant modifications in the relic abundances
of superheavy DM particles with masses of this order. Such superheavy
DM particles can be produced during reheating~\cite{riotto}, but
in view of our scenario above, their relic abundance will be modified
from the Standard Cosmology result. The presence of superheavy DM,
with increased relic abundances, might provide an explanation for
the production of at least part of the spectrum of the ultra high
energy cosmic rays, with energies of order $10^{20}~{\rm eV}$. Hence,
the effects of D-foam on such scenarios are worthy of investigating
further, especially in view of the fact that the density of D-particles
might be significantly higher at earlier eras of the Universe, leading
to stronger stochastic effects ${\mathcal O}(\Delta_{i}^{2})$.

\subsection{Low (TeV)-string-scale D-particles}

For low string scales, of order a few TeV, the effects of the D-foam
on thermal relic densities would be more significant. In fact depending
on the type of DM considered, the effect could be constrained or falsified
already by the WMAP data~\cite{cmb,7yrwmap}, since the induced increase of
thermal relic abundance leaves less room for supersymmetry in the
relevant parameter space. In certain cases, it may exceed the allowed
region set by WMAP.  Indeed, the DM relic abundance determined by the
WMAP seven-year-data mean (at 68\% C.L.)~\cite{7yrwmap}, is:
\begin{equation}\label{sevenyrdata}
\left(\Omega_{\chi} h_0^2\right)^{\rm WMAP,\Lambda CDM} = 0.1126 \pm 0.0036,
\end{equation}
thus putting the observational error at the ${\mathcal O}(10^{-3})$ level.

In our phenomenological analysis we have found it convenient to parameterize the foam-enhanced abundances (\ref{dmcontr}) as follows:
\begin{eqnarray}
\frac{\Omega_{\chi}h_{0}^{2}}{(\Omega_{\chi}h_{0}^{2})_{{\rm no~source}}} & = & 1 +  \sigma^2 \, m_\chi^2~, \nonumber \\
\sigma^2 \equiv   1.259\,\frac{g_s^2}{M_s^2} \sum_{i=1}^3 \Delta_i^2~ , &\quad &
(\Omega_{\chi}h_{0}^{2})_{{\rm no~source}} = (2.6\times 10^{-10}~{\rm GeV}^{-2})\frac{16\pi^2 m_\chi^2}{k \, g_{\rm weak}^4},
\label{dm2}
\end{eqnarray}
where the fudge factor $k$ parameterizes the deviation of the coupling from the weak interaction coupling
$g_{\rm weak}$ that characterizes the standard WIMP case~\cite{spanos} and the parameter $\sigma^2$ encodes the foam fluctuations.

\begin{figure}[htbp]
\begin{center}
\includegraphics[width=0.6\textwidth]{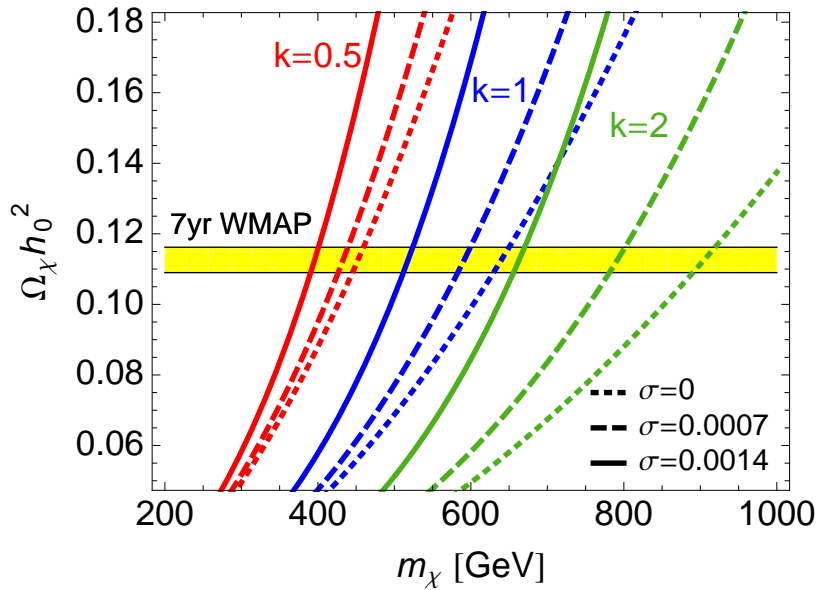}
\caption{Predicted values of $\Omega_{\chi}h^2$ as a function of the DM particle mass $m_{\chi}$ for various deviations from the weak coupling $g_\text{weak}$: $k=0.5$ (red), $k=1$ (blue) and $k=2$ (green). Different values of foam fluctuations $\sigma$ are drawn: $\sigma=0$ (dotted) corresponds to the no-source estimate; $\sigma=0.0007~{\rm GeV}^{-1}$ (dashed) and $\sigma=0.0014~{\rm GeV}^{-1}$ (solid) describe stochastic D-particle foam models. The yellow horizontal band represents the 7-year WMAP observational value~(\ref{sevenyrdata}) for $\Omega_{\chi}h^2$~\cite{7yrwmap}.}
\label{fig:omegaxh2}
\end{center}
\end{figure}

The impact of the space-time D-particle foam on the predicted DM relic abundance and the comparison of the latter with the latest WMAP measurements are depicted in fig.~\ref{fig:omegaxh2}. We observe that the parameter $\sigma$ enhances the thermal relic density $\Omega_{\chi}h_{0}^{2}$, decreasing thus the upper bound of the allowed DM masses compared to Standard Cosmology ($\sigma=0$). The $\Omega_{\chi}h_{0}^{2}$ curves are drawn for three values of the factor $k$, namely $k=0.5, 1, 2$, demonstrating the (large) dependence on the variation of the coupling constant.

\begin{figure}[htbp]
\begin{center}
\includegraphics[width=0.6\textwidth]{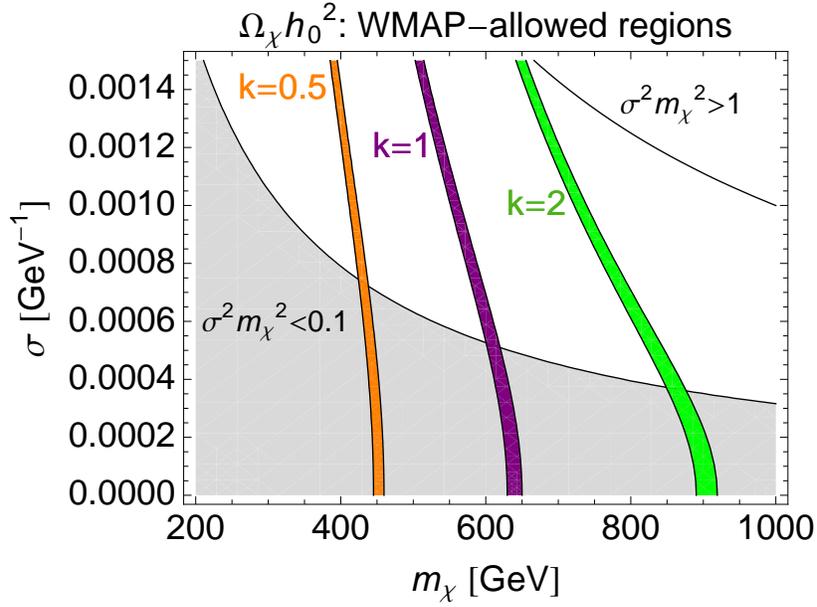}
\caption{Regions in the $(m_{\chi},\sigma)$ plane allowed by the latest WMAP observations~(\ref{sevenyrdata})~\cite{7yrwmap} for various deviations from the weak coupling $g_\text{weak}$: $k=0.5$ (orange), $k=1$ (purple) and $k=2$ (green). In the grey area, the foam fluctuations are small enough for the approximation~(\ref{dm2}) to be valid.}
\label{fig:omega_corridors}
\end{center}
\end{figure}

In fig.~\ref{fig:omegaxh2}, the range for the parameter $\sigma$ has been chosen so that $\sigma^2 m_\chi^2 \lesssim 0.1$, which is a necessary assumption for the validity of our perturbative approach on which (\ref{dmcontr}) and (\ref{dm2}) are based. This is evident in fig.~\ref{fig:omega_corridors}, where it is shown that this condition is met for the whole range of WIMP masses $m_{\chi}$. In the same figure, we also present the acceptable regions in the $(m_{\chi},\sigma)$ plane, i.e.\ those compatible with the ``WMAP corridor''  (\ref{sevenyrdata}), for various values of the parameter $k$ as defined in (\ref{dm2}). It is clear from the plot that there is a shift of the allowed range of $m_{\chi}$ towards lower values and a narrowing of this range with increasing parameter $\sigma$.

\begin{figure}[htbp]
\begin{center}
\includegraphics[width=0.6\textwidth]{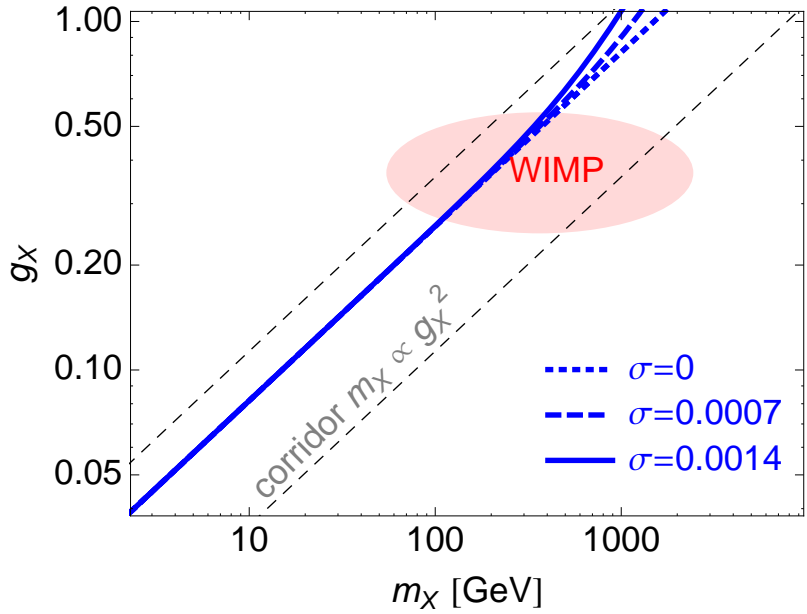}
\caption{Contours of $\Omega_X h^2=0.1126$~\cite{7yrwmap} in the $(m_X,g_X)$ plane for various values of foam fluctuations: $\sigma=0$ (dotted) corresponds to the standard, no-source estimate; $\sigma=0.0007~{\rm GeV}^{-1}$ (dashed) and $\sigma=0.0014~{\rm GeV}^{-1}$ (solid) describe stochastic D-particle foam models. The corridor $m_\text{weak}\equiv(m_X/g_X^2)g'^2\in(100~{\rm GeV},1~{\rm TeV})$ (between the thin dashed lines) shows the region where $m_X\propto g_X^2$. The ellipse marks the WIMP region, where $g_X\sim g'\sim g_\text{weak}$.}
\label{fig:gx_mx}
\end{center}
\end{figure}

For more general models, where the dark matter may even come from hidden sectors and it is unrelated to weak interactions, one may use for the DM relic abundance (cf.\ (\ref{combination}) and (\ref{dm2})):
\begin{equation}\label{generaldm}
\Omega_X h_{0}^{2} \propto \frac{m_X^2}{g_X^4}(1+\sigma^2 m_X^2).
\end{equation}
The effects of foam in such general models are considered in fig.~\ref{fig:gx_mx}, where we plot the WMAP-allowed contours~\cite{7yrwmap} of $\Omega_X h^2=0.1126$
in the $(m_X,g_X)$ plane for various characteristic values of the foam fluctuation parameter $\sigma$, including  the foamless case ($\sigma=0$), for comparison. The standard WIMP region, where the coupling $g_X\sim g'\sim g_\text{weak}$, is also indicated. We observe that, for the allowed values for the foam-fluctuations parameter, $0\lesssim\sigma\lesssim 0.0015$, the WMAP contour falls within the corridor $m_X \propto g_X^2$, which is a condition sufficient for a model to provide a viable DM candidate~\cite{spanos}.

We note, therefore, that it is a generic feature of the foam-induced Finsler backgrounds to tend to reduce the cosmologically allowed range of neutralino (or in general the DM candidate) masses, relative to the foamless cases, for fixed low string scales (translated to relatively low ---compared to Planck scale--- D-particle masses). For low string scales $({\mathcal O}(10~{\rm TeV}))$, D-foam models may lead to significant modifications of supersymmetry searches at colliders, especially in the context of neutralino DM models with the neutralino being higgsino- or wino-like, with masses up to ${\mathcal O}(1~\text{TeV})$.

We also note at this stage that in models where the neutralino DM is higgsino- or wino-like, there may be other reasons for an increased relic abundance, for instance slepton co-annihilation~\cite{slepton},
and in fact our effects of the foam are of comparable strength in
some of these cases. Hence, from our point of view, in such models, the addition of a low-scale D-foam fluctuating background will lead in general to further constraints on the cosmologically allowed DM masses, since the presence of the foam will lead to further enhancement of the DM thermal relic abundances.

\subsection{Comparison of D-foam with Other Cosmological Models}

Let us now compare the results on the effects of D-foam backgrounds on dark matter abundances obtained in the previous sections, with some other, but somewhat related, non-conventional cosmological models.
First of all, we shall consider the non-equilibrium string cosmologies of~\cite{cosmo}, involving
time-dependent dilatons $\phi (t)$. The low energy effective field theories are (super)gravity theories, with these non-trivial dilaton fields. As discussed in~\cite{elmn}, the presence of the latter, induces source terms in the Boltzmann equation (\ref{boltzmann}),
describing the thermal DM relic abundances in such Universes, of the form:
\begin{equation}
\Gamma_{\rm dil} (t) = \dot \phi,
\end{equation}
where the overdot denotes derivative with respect to the cosmic time in the Einstein frame, i.e.\ in a frame in which the four-dimensional Einstein term in the effective (super)gravity action assumes the canonical form.
A self-consistent solution of the cosmological equations of these models~\cite{cosmo} yields the result
$\dot \phi < 0$, which implies that, in contrast to our D-foam effect discussed here, the time-dependent dilaton acts as a sink rather than a source for particles. This was important in the foamless cases, since it could lead to
a dilution of the DM particle abundance by terms of ${\mathcal O}(1/10)$, as compared to the Standard Cosmology models in the absence of dilaton sources, while the baryon density remained unscathed.
This had profound phenomenological consequences for collider tests of supersymmetry~\cite{dutta}, given that in that case more room for supersymmetry was left by the WMAP astroparticle constraints~\cite{cmb,7yrwmap} on DM, and thus heavier partner masses are allowed as compared to the Standard Cosmology. Furthermore, other astrophysical observations such as type-Ia supernovae and galaxy ages set tight constraints~\cite{nonequil-observ} on the parameters of these models through their prediction for the dark energy contribution.

In the presence of D-foam, on the other hand, cosmologies with time dependent dilatons exhibit competing dilaton and foam effects, with the result that, for low string scales, the relevant contributions might cancel each other out. For instance, one might obtain a combined time-dependent-dilaton-foam source in the relevant Boltzmann equation of the form (cf.\ (\ref{source})), which in a schematic way would look like:
\begin{equation}
\Gamma_{\rm combined} \sim -|{\dot \phi }| + \frac{g_s^2}{M_s^2}\left|\mathcal{O}\left(Ha^4(t)m \Delta^2\left[9T+2m\right]\right)\right|,
\end{equation}
where $\Delta^2$ denote dimensionless foam fluctuations. As we have discussed previously, for low string scale the two (opposite) sign terms
might be of the same order, at least for some range of
the DM masses, thereby canceling effectively each other out  and thus modifying the conclusions of~\cite{dutta}.
Similarly to the dilaton cosmologies, though, the foam affects predominantly only the electrically neutral DM particles, and not the charged baryon density, for specifically stringy reasons outlined above, associated with either charge conservation (type-IA strings~\cite{dfoam,emnnewuncert}) or compactification details (type-IIB theory~\cite{li}).

Next we come to the models of~\cite{pessah}, involving non extensive  Tsallis statistics~\cite{tsallis} for particles.
As already discussed in section \ref{sec:statistics}, our D-foam models provide a microscopic origin of effects that mimic in some but not all respects, those of a non-extensive statistics a l\'a Tsallis. As in~\cite{pessah}, the distributions functions that determine the thermodynamics and statistical aspects of the Early Universe, differ slightly from the
Standard Cosmology ones, by perturbatively small terms proportional to the foam fluctuations, assumed small. However, as we have explicitly already mentioned, there are major differences from that approach, concerning not only the different functional form of the pertinent expressions between the two approaches, but also the fact that the energy appearing in our expressions is not the actual energy of the particle but rather its on-shell Minkowski counterpart, as well as, and more importantly, the dependence of the corrections on the momentum content of the particle, as a consequence of the Finsler nature of the space-time metric deformations, induced by the interaction of the DM particle with the foam defects.

Another important difference concerns the form of the Boltzmann equation, which
in the approach of~\cite{pessah} remains unmodified. The non-extensive statistics effects can only be seen in the equilibrium expressions of particle number densities. Indeed, in terms of the parameter $Y$, defined above, one has
for the Tsallis cosmologies~\cite{pessah}
\begin{equation}
\frac{dY_q}{dt}=-\langle \sigma u\rangle s_q\left(Y_q^2-Y^2_{q,eq}\right),
\end{equation}
\noindent where the modified $Y_{q,eq}$ can be calculated explicitly using the modified distribution functions. Thus, in the Tsallis cosmology cases~\cite{pessah}, the modifications on the DM thermal relic come mainly from the factor $\frac{g_\textit{eff}'}{g_\textit{eff}}$ as a result of the $q$-statistics deformation on the available degrees of freedom, which also affect the freeze-out points of the various species, by terms proportional to $(q-1)$, which in the approach of~\cite{pessah} is assumed perturbatively small.

 In contrast, in our D-foam model, as we have seen above,
in addition to a modification in the pertinent particle distribution functions,
which affect the effective degrees of freedom, the Boltzmann equation is also explicitly affected by the D-foam through a time dependent  source term $\Gamma (t)$ (\ref{boltzmann}), (\ref{source}), which, because of its positive sign, is expected to lead to particle production.
In fact, in our case,  the foam-induced modifications to the freeze-out point are negligible, at least for the weak foam cases we consider here.
On the other hand, for WIMP-based models, there is at most a 26\% correction to the available degrees of freedom from the foam contributions.
Thus, the main effects on the DM relic population come primarily from the foam-dependent source terms in the Boltzmann equation (\ref{boltzmann}), associated with particle production and only partly from the corrections to the available degrees of freedom. This is to be contrasted to the Tsallis entropy cosmology of~\cite{pessah}, where the non-extensive statistics effects are exclusively associated with modifications of the effective degrees of freedom of the system.

\subsection{Other effects of D-foam and complementary constraints}

There are other interesting effects of the D-foam model, which can provide complementary constraints on the parameters
of the model, by means of entirely different phenomenology. In particular, the non-trivial interactions of photons with the D-particle defects in the foam, result in a non-trivial sub-luminal refractive index~\cite{emnnewuncert,li}, with the anomalous effects being proportional to the first power of the ratio $E_{\rm obs}/M_{\rm eff, QG}$, where $E_{\rm obs}$ is the (observed) photon energy and $M_{\rm eff, QG}$ is the effective quantum gravity scale, given by
\begin{equation}
M_{\rm eff, QG} \equiv \frac{M_s}{n(z)}
\label{effqgscale}
\end{equation}
with $M_s$ the string scale. The function $n(z)$ is the density of D-particle defects in the foam, encountered by the photon as it traverses the cosmic distance from emission till observation. This quantity is essentially arbitrary in the context of the D-foam models, as it depends on microscopic details of the higher-dimensional bulk physics~\cite{dfoam,westmuckett}.
Phenomenologically, information on $n(z)$ (mainly constraints on upper bounds)  can be provided, for late epochs of the Universe, i.e.\ red-shifts $z \le \mathcal{O}(10)$, by studies of the arrival times of cosmic photons from distant celestial sources, such as Active Galactic Nuclei and/or Gamma Ray Bursts. The idea is simple: there is at present evidence for delayed arrival of high energy photons, as compared to their lower-energy counterparts, from a few cosmic sources, either AGN~\cite{MAGIC2} or GRB~\cite{fermi}. The interactions of photons with the D-particles in the foam, cause time delays of the more energetic photons by an amount~\cite{emnnewuncert,li}:
\begin{equation}
\label{totaldelay}
\Delta t_{\rm obs} = \int_0^z dz \frac{n(z)\, E_{\rm obs}}{M_s} \,\frac{(1 + z)}{H(z)}~,
\end{equation}
where $H (z)$ is the Hubble rate, which is a function that depends on the details of the underlying cosmological model. By assuming that the foam is the primary source of the delays (which, of course is an assumption, since astrophysical effects at the sources could also be responsible for (part of) the delays of the energetic photons), one can obtain restrictions on the density of defects and the cosmological parameters of the model, encoded in the functional form of $H(z)$. If at late eras of the Universe, $z \le 10$, the Standard Cosmological Cold-Dark Matter model with a cosmological constant $\Lambda > 0$
is assumed to be a good approximation of reality, as the current evidence suggests, then one can see~\cite{emnfits} that fits of all the available data on observed photons delays at present imply constraints on the density of D-particle defects $n(z)$, and in fact require that the latter is not constant but drops by almost two orders of magnitude as the red-shift climbs from $z=0.03$ (MAGIC photon delays~\cite{MAGIC2}), where $M_s/n(z)=\mathcal{O}(10^{18})$~GeV, to $z = \mathcal{O}(1)$ (GRB~090510 observations by Fermi), where $M_s/n(z=1) \ge 1.22 M_P$, with $M_P = 1.2 \times 10^{19}$ GeV the Planck mass.
In this way the ${\mathcal O}(1~\text{min})$ delays of the TeV photons from the AGN Mk501, observed by MAGIC~\cite{MAGIC2}, can be fitted with the same D-foam model as the ${\mathcal O}(0.5~\text{sec})$ delays of the 30~GeV photons from GRB~090510 (at $z \simeq 0.9$), observed by Fermi.

This is important information, since it implies that the density of defects in the bulk of the D-foam brane model of~\cite{dfoam,westmuckett} may be \emph{non-homogeneous} but increases as the time elapses, at least at late eras. Since
the density of defects on the brane world enters the expressions for the statistical averages over populations of D-particles, which in turn affect the induced stochastic fluctuations of the space time, and through them the cosmology of the model~\cite{westmuckett,emnnewuncert}, as we have discussed above, one should incorporate such a feature into the analysis.
This is left for future work. We stress once more, however, that we consider it as an interesting complementary test of the model, providing independent constraints on the model parameters, apart from the ones discussed here.

Before closing this section we would like to mention another phenomenological consequence of the D-particle foam model, namely a qualitative prediction on potential CPT violating modifications~\cite{omega} of the Einstein-Podolsky-Rosen (EPR) correlations of neutral meson pairs produced in meson factories~\cite{bernabeu}.
Specifically, if there is a non-trivial density of D-foam defects present during the decay of, say, the $\Phi $-Meson in Kaon factories, then in general one may have a modification of the standard EPR correlators between the long ($K_L$) and short-lived ($K_S$) kaon products of the decay, which is parametrized by a complex parameter $\omega$, whose amplitude
may be estimated to be given by~\cite{bernabeu}:
\begin{equation}\label{om1}
|\omega | \sim g_s^2 \frac{\xi^2 \, k^4 }{M^2_s \, (m_S -m_L)^2 },
\end{equation}
where $\xi^2$ is a fudge factor, depending on the density of the D-foam at red-shifts $z=0$ (i.e.\ in the current era) and  its fluctuations $\Delta^2$. However, the fact that neutral Kaons have quark and gluon substructure complicates the situation. Indeed, due to their electric charge, the quarks are not supposed to interact dominantly with the foam, only the gluons can do this~\cite{emnnewuncert,li}. This may introduce additional (and significant) suppression factors in $\xi$ coming from the strong interaction sector of the effective field theory describing the Kaon dynamics.

The current experimental bounds on $\omega$ may be summarized by the results of the
KLOE experiment at the Da$\Phi$NE $\Phi$-factory in Frascati (Italy)~\cite{kloe}:
\begin{equation}\label{om2}
|\omega |_{\rm exp} \, < \,  1.0 \times 10^{-3} \quad {\rm at~ 95\% ~C.L.}
\end{equation}
At least one order of magnitude improvement is expected on these bounds with the KLOE-2 experiment at the Upgraded DA$\Phi$NE~\cite{upgrade}. Taking into account that $|m_L - m_S| \sim 3.48 \times 10^{-15}$~GeV, and ignoring possible suppression of $\xi$ from the strong interactions, we observe from (\ref{om1}), that for low string scales $M_s/g_s$ of order of  10~TeV, for which the D-foam contribution to the DM relic abundances (\ref{dmcontr}) could be significant, the above bound (\ref{om2}) translates to $\xi < 10^{-12}$.
For high string scales, where $M_s/g_s$ is of the order of Planck mass, there are no significant constraints on $\xi$.

In the context of the microscopic models considered here, one may estimate~\cite{mavroomega} $\xi$ theoretically, for the case where there is \emph{with certainty } one D-particle defect present during the decay of a $\Phi$ meson, so that the initial entangled state of the Kaon products of the decay will be affected by the presence of the foam.
In this \emph{optimistic} scenario, one finds $|\xi |^2 \sim (m_L^2 + m_S^2)/k^2 $, where $k$ is the spatial momentum of the Kaon states in the detector. In general, though, there are fudge factors in front of this estimate, taking proper account of the probability of a D-particle being present during the decay of the initial $\Phi$ meson, which itself depends on the details of the foam (such as density \emph{etc}.), as well as the non-elementary quark and gluon structure of the Kaon particles, which plays an important suppressing r\^ole of foam effects, since the foam interactions between the electrically charged quarks and the neutral gluons have different strengths, as discussed above~\cite{emnnewuncert,li}. Thus, at present, lacking a microscopic non-perturbative detailed estimate of these complicated effects, we can only impose bounds on such parameters. Nevertheless, the above diverse tests indicate
how in general space-time foam models of the type considered here may be constrained in the near future.

\subsection{On the production of TeV-mass-scale D-particles at Colliders}

We would like now to comment briefly on the production of D-particle defects in particle collisions at high energy colliders, such as the LHC. In general, the production mechanisms will depend on the type of D-particles (and  on  the associated microscopic string theory model) considered. In this section we would like to make the discussion more general.

We will first distinguish our D-particle cosmological model from the generic D-matter analysed in \cite{D0matter}. In the latter work, D-particles are viewed as ordinary effectively point-like excitations of the vacuum. They arise from the appropriate compactification of Dp$'$ branes  around p$'$ cycles, with radii of the order of the string length $\ell_s = 1/M_s$. Since they  have non-trivial couplings with
Standard Model particles, they behave like standard dark matter candidates; hence their production mechanism at colliders will not differ from those of conventional dark matter. In fact in \cite{D0matter}, such D-particles have   been treated as providing a leading candidate for dark matter in our Universe. As in our case above, the mass of these D-particles is of order $M_s/g_s$, i.e. TeV scale defects can be present in low scale string theory models.

In general brane defects have spin structures; hence stable D-matter states can be either bosonic or fermionic, corresponding to the bosonic or fermionic zero modes of stable D-branes respectively.  In our cosmological model described in previous sections, we treated the D-particles as background defects, where their spin was not relevant for our induced effects; on the other hand, when we consider issues like the production of localised branes, then
their spin is relevant, since the corresponding interaction cross sections, for instance with Standard Model particles, such as nucleons,
do depend on spin. In this respect, the production of localised D-branes by the high energy collision  of Standard Model particles at colliders can incorporate even more complicated charged defects, consisting of various configurations of compactified Dp$'$-branes around p$'$-dimensional manifolds.

\begin{figure}[ht]%
\centering
\subfloat[]{\label{sfig:signal}\includegraphics[width=0.37\textwidth]{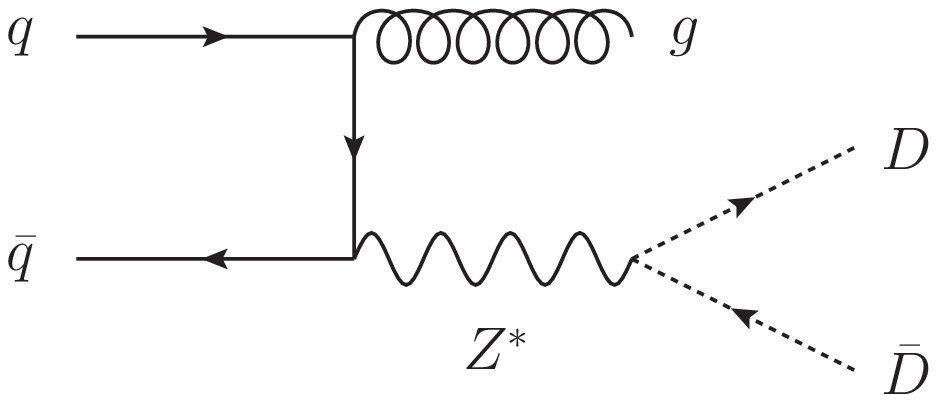}}
~ ~
\subfloat[]{\label{sfig:DM-prod}\includegraphics[width=0.32\textwidth]{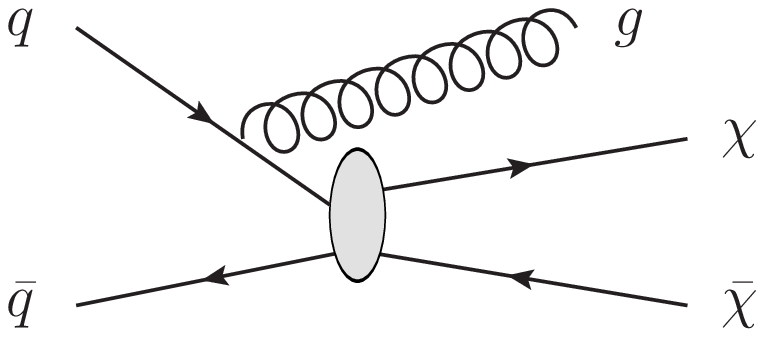}}\\
\subfloat[]{\label{sfig:bkg1}\includegraphics[width=0.37\textwidth]{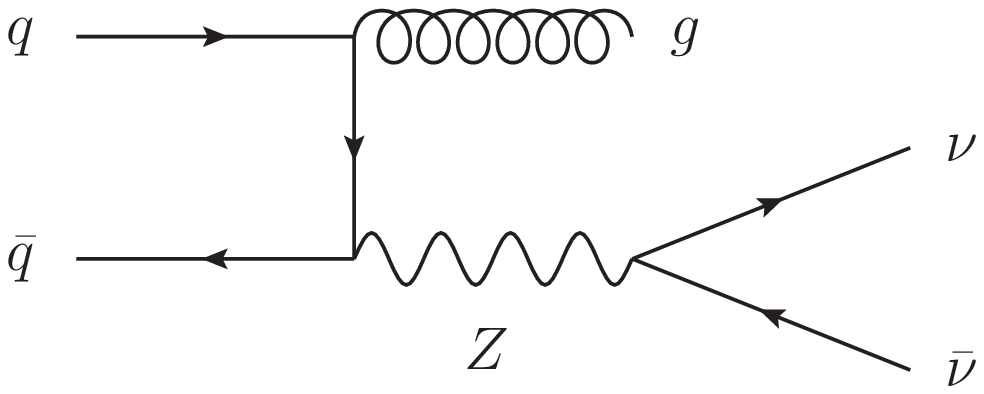}}
~ ~
\subfloat[]{\label{sfig:bkg2}\includegraphics[width=0.37\textwidth]{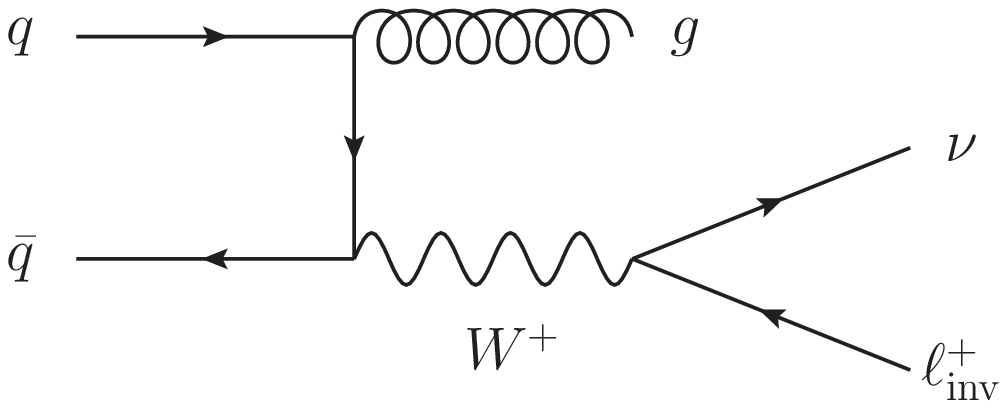}}%
\caption{(a) Feynman diagrams at the parton level for the production of D-particles by, say, q$\overline{{\rm q}}$ collisions in a generic D-matter low energy model of \cite{D0matter}, based on (\ref{waysD1}). This is only an example. There are many other processes for the production of D-matter from standard boson decays or gauge boson fusion, which we do not consider in our qualitative discussion here. (b) Production of conventional dark-matter particle-antiparticle ($\chi\bar{\chi}$) in effective field theories, assuming that dark matter, which may co-exist with D-matter, couples to quarks via higher-dimensional contact interactions~\cite{kolbmaverick,lhcmaverick}.  (c), (d) The dominant background processes within the Standard Model framework.}
\label{fig:dproduction1}
\end{figure}

In a low-energy, string-inspired, effective field-theory action,  the leading interactions of the D-particles with Standard Model matter are provided by terms with the generic structure (omitting Lorentz derivative or Dirac-matrix structures for brevity)~\cite{D0matter}
  \begin{equation}\label{waysD1}
   \propto g_{D} \, {\overline {\rm D}} \, {\rm D} \, {\rm (gauge~bosons})  ~.
   \end{equation}
The symbol $\propto $ in front of each type of interaction is included to denote form factors that arise from tree-level string amplitude calculations with appropriate Dirichlet and Neumann boundary conditions for the open strings~\cite{D0matter}. The relevant string amplitudes have three-vertex insertions on world-sheet discs, with the appropriate
boundary conditions: Dirichlet for the vertex operators describing an excitation of D-particles, and Neumann for ordinary open strings describing (ordinary) gauge bosons. The resulting effective three-point amplitude is expanded in powers of the $\sqrt{g_s}$  of the open string coupling $g_s < 1$, assumed perturbative,
$$ g_s^{1/2} \mathcal{F}(s,t, \alpha') + g_s \mathcal{G}(s,t, \alpha') + \dots $$
where $s,t $ are Mandelstam variables. In the context of string/brane theory, the Yang-Mills coupling $g_{\text{\it YM}} \sim g_s^{1/2} $.  Thus, the effective coupling $g_D$ is proportional to $g_{\text{\it YM}}$, but renormalised by appropriate kinematic factors
and higher-order corrections  in $g_s$. For momenta of the same order as the low-string scale the corrections may be important at high energy scattering. Unfortunately string corrections of D-branes are not understood very well at present to be able to give any quantitative prediction of the effective coupling $g_D$, especially for low scale D-particles; for our purposes we can only use generic effective field theory arguments in order to study the phenomenological features of the D-particles. These are in general  dependent on the microscopic-model. The terminology ``gauge bosons'' in (\ref{waysD1}) will represent not only gauge bosons but also other Standard Model particles.  As a result of (\ref{waysD1}), for instance, one may have the graphs of fig.~\ref{sfig:signal}, arising from quark/antiquark scattering. The D-matter/antimatter pairs can be produced by the decay of intermediate off-shell Z-bosons of the Standard Model, which is in agreement with (\ref{waysD1}).

The  D-matter pairs produced in a hadron collider will traverse the detector and exit undetected as they are weakly interacting only; however there will be large transverse missing energy ($E{\rm _T^{miss}}$). Hence analyses requiring high $E{\rm _T^{miss}}$ and one energetic jet would be of high relevance. The dominant background from Standard Model processes involves the decay of $Z$ to a neutrino pair and of $W^+$ to a lost (``invisible'' $\ell_{\rm inv}$) lepton and a neutrino (\emph{cf.} figs. \ref{sfig:bkg1} and~\ref{sfig:bkg2}, respectively). Such searches have been performed by CDF at the Tevatron~\cite{cdf-monojet} and by the ATLAS~\cite{atlas-monojet} and CMS~\cite{cms-monojet} experiments of the LHC at centre-of-mass energies of 7~TeV giving null results. The interpretation of the such analyses may be deployed to constrain D-matter parameters, as will be discussed below in the context of our specific D-foam model.

In the approach of \cite{D0matter}, the D-matter behaves as the dominant species of dark matter in the Universe;
they do not consider (as we do here) the effect of D-matter in the bulk region of the brane world. The cross section for the
above production of D${\overline{{\rm D}}}$-pairs,  $\sigma( q\, {\overline{q}} \rightarrow  {\rm D} \, \overline{{\rm D}} + \dots )$, is related to
the total Dark-Matter annihilation cross section $\sigma_{\rm total~annih} $ into Standard Model particles via the \emph{principle of detailed balance} (which one may assume to characterise string theory effective action models),  $ {\rm D} \, \overline{{\rm D}} \rightarrow f {\overline{f}} $, where $f$ denotes Standard Model particles (including quarks of course). The latter cross section can be constrained by cosmological data, if the D-matter is assumed to play the r\^ole of the dominant species in the Universe, since the total annihilation cross section is inversely proportional to the relic abundance of D-matter, $\Omega_D h^2 \propto \frac{1}{\langle \sigma_{\rm total~ann.}|\, v\rangle} $, in a standard notation. It is for this reason that the mass parameters and couplings of the D-matter
have to be constrained in such a way that the region of the cross section is near the `WIMP-miracle' region.
In this approach one also considers the scattering of the D-matter against nucleons, which provides, as usual, complementary constraints on the dark matter parameters (especially its mass and couplings) via direct searches. Needless to say that the relevant cross sections depend crucially on whether the D-matter is bosonic or fermionic (D-particles can have integer or half-integer spins, as a result of the corresponding bosonic or fermionic zero modes they characterise D-matter).

However, it is possible that, the D-matter particles being relatively heavy and thus slow moving,  deposit all their energy inside the detector. Provided the deposited energy from the D-${\rm \overline{\rm D}}$-particle pair is large enough, this may lead , for instance, to other observable effects in the MoEDAL detector of the LHC collider~\cite{Moedal}, which is a plastic nuclear tracks detector designed to search for magnetic monopoles and other highly ionising particles that may characterise several models of new physics, such as supersymmetry. Despite being electrically neutral, D-particles can operate in a similar way as a monopole, by destroying the chemical bonds in the plastic sheets that surround the MoEDAL detector, thereby leading literally to the formation of holes.
In fact D-particles are analogous to `t Hooft-Polyakov monopoles (solitons) in string theory. One difference is that the couplings of D-particles
may be perturbative, in contrast to the monopole case, which has a coupling $\propto 1/g_{\text{\it YM}}$.
From the shape of the various holes along the path inside the plastic sheets one can in principle determine the mass
of the heavy particle and its type.

Finally a third way of producing TeV-mass D-particles at particle colliders would be through the formation of TeV-scale black holes, which have short life times and then decay, \`{a} la Hawking, giving rise to pairs of particle-antiparticles that describe localised excitations of the
low-energy effective field theory along with stable D-particle-${\overline {\rm D}}$-antiparticle pairs.


As we have indicated, our model for D-foam has some essential differences from the generic D-matter models of \cite{D0matter}.
First, in our type IIA models of string foam~\cite{dfoam,westmuckett,emnnewuncert} the D-particles are \emph{truly point-like} and  the D-foam is considered as a \emph{stochastically fluctuating background}, with no (or suppressed) couplings between the D-particle defect and Standard Model fields. For us, the dominant interactions of D-foam with ordinary matter are gravitational in nature, provided by the distortion of space-time due to the recoil of the defect during the capture/string-splitting processes.
In this sense, D-particle in our models \emph{cannot} play the r\^ole of dark matter in the Universe. Their presence and gravitational interactions (through recoil) with ``conventional'' dark matter candidates, affect the relic abundance of the latter.

In these type of models, D-particle defects  can only be produced at colliders by first forming TeV-size black holes in low scale string models, which then undergo Hawking radiation, in which pairs of TeV-mass D-particle/anti-particle are produced along with all other Standard Model or other exotic particle
pairs, depending on the model. In this sense, for this type of string IIA D-foam~\cite{westmuckett,emnnewuncert}, the only collider signatures are the indirect ones via the increased relic abundances of dark matter, as discussed in the previous section.

In general, however, the D-particles (arising from compactification) in other types of D-foam models, such as  type IIB string theory models of \cite{li}, can couple
to the Standard Model fields, but with  couplings suppressed in comparison with the standard
model ones. In such models one may consider the formation of D-particles in some high energy collisions and in the early universe, along the lines of D-matter discussed previously~\cite{D0matter}.
Nevertheless in our D-foam cosmology, as we have emphasized, in contrast to the models of \cite{D0matter},  the D-particle contributions to the Universe vacuum energy density can be compensated by interactions  of D-particles in the bulk regions of our configurations (\emph{cf.} fig.~\ref{fig:recoil}). In section \ref{sec:Hubble}, it was noted that there are negative energy contributions from bulk D-particle defects that cancel out (on average) the positive energy contributions due to the D-particle masses and relic densities  (\emph{cf}. (\ref{popul})). Hence, in our models of D-foam, in contrast to the generic D-matter considerations of ref.~\cite{D0matter},  there is \emph{no} restriction in the relevant annihilation cross sections into Standard Model particle/antiparticle  pairs, $f\, \overline{f}$,
$\sigma_{\rm a} ({\rm D}\, {\rm D} \to f {\overline f} )$. Such restrictions apply only to the conventional dark matter annihilation cross sections that \emph{do} exist in our D-foam cosmology.

Let us give a concrete example of D-particle production at colliders in our type IIB string models of D-foam, considered in ref.~\cite{li}. Here D-particles are essentially D3-branes wrapped up around appropriately compactified three cycles (\emph{e.g}. tori)~\cite{li}. Such defects will look ``effectively'' point-like from a four-dimensional space time point of view, but the difference from the truly point-like defects of the models of \cite{dfoam,westmuckett,emnnewuncert} lies on the fact that in this case there are non-trivial (dimensionless) couplings $g_{37}^2$ between  the D-``particles''   and the Standard Model ones~\cite{li}
\begin{equation}\label{couplings}
g_{37}^2 = \frac{(1.55 )^4}{V_{A3} \, R'}  \, {{\tilde g}_7}^2 = \frac{(1.55 \ell_s)^4}{V_{A3} \, R'} {g}_7^2 \equiv \eta \, g_{7}^2~,
\end{equation}
where  $V_{A3}$ is the average three-spatial-dimensional volume which contains one D-particle (assuming uniform distribution of defects in the foam), $R'$ is the compactification radius of the dimension transverse to the wrapped-up-D3-brane (D-``particle'') and ${\tilde g}_7$ denotes the (dimensionful)  gauge couplings of open strings leaving in the D7-brane in the model of \cite{li}, which after appropriate compactification describes our observable world (the corresponding dimensionless couplings are defined as ${g}_7 = {\tilde g}_7/\ell_s^2$). Hence, ${g}_7$ is proportional to the Yang-Mills couplings ($g_{\text{\it YM}} \propto \sqrt{g_s}$, where $g_s$ is the (dimensionless) string coupling) of the various Standard Model gauge bosons, say. It is important to notice that the form (\ref{couplings}) characterises exclusively the D-foam model of \cite{li}, and should not be confused with the generic models of D-matter discussed in \cite{D0matter}.
As discussed in \cite{li}, in order to avoid observable Lorentz-violating effects in the dispersion relation of electrons that would affect the
synchrotron radiation spectrum from distant nebulae, such as Crab, by effects that are not observed experimentally, one must have
\begin{equation}\label{etamag}
\eta \equiv \frac{(1.55 \ell_s)^4}{V_{A3} \, R'} < \, 10^{-6}
\end{equation}
Such a bound, for instance, is easily satisfied in our D-foam string models with $ V_{A3} \sim (10 \ell_s)^3, \, R' \sim 3.4 \times 10^2 \, \ell_s$.

In view of (\ref{couplings}), and assuming, for definiteness that D-particles are \emph{fermionic}, one has trilinear interactions between the type IIB-D-``particles'' with ordinary Standard Model particles, such as gauge bosons, of the form (\ref{waysD1}) (omitting for brevity Lorentz derivative or Dirac-matrix structures for brevity)~\cite{D0matter,li}
  \begin{equation}\label{waysD}
   \propto g_{37} \, {\overline {\rm D}} \, {\rm D} \, {\rm (gauge~bosons})  ~,
   \end{equation}
where, as in (\ref{waysD1}),  the symbol $\propto $ in front of each type of interactions denote kimematical form factors that arise from  tree-level string amplitude calculations. In view of (\ref{etamag}), in this model, the trilinear coupling of D-particle/antiparticle pairs with the Standard Model gauge bosons would be much weaker than any typical Standard Model coupling. Indeed, the mass scale
of the effective four-fermion ($q\,{\overline q}\, {\rm D}\, {\overline {\rm D}}$) interaction, representing the mediating off-shell $Z$-boson
entering the D-particle production process in fig.~\ref{sfig:signal},  has the form
\begin{equation}\label{eq:mz}
M_{\rm eff} = \frac{M_Z}{\sqrt{g_{37} g_7}} \sim \frac{M_Z}{g_{7}}\, \eta^{-1/4}
\end{equation}
where $M_Z \sim 90 $~GeV is the mass of the Z-boson and $g_{7}$ represents the Standard Model weak interaction coupling on the brane world. Thus, on account of (\ref{etamag}), $M_{\rm eff}$
can be of order of a few TeV, making the D-production graph of fig.~\ref{sfig:signal} of similar order to the conventional dark-matter production graphs in the model (\emph{cf.} fig.~\ref{sfig:DM-prod}).
However, in our case, unlike standard scenarios involving only a single type of dark matter dominant in a given channel~\cite{kolbmaverick,lhcmaverick},  we have a mixture of dark-matter contributions, coming from D-matter and conventional (WIMP-type) dark matter ($\chi$). As a result, the detailed phenomenology is  more complicated, and depends on the relative masses of stable D-matter and conventional dark matter. In  the case of low-string scale (TeV) D- and dark- matter, one is in the border-line region for defining the low-energy effective low-energy string action, and stringy corrections are important in our model.

The null searches of, say, ATLAS detector in finding missing energy + jets, will thus impose, in the context of our model, a bound on the dark matter mass $m_\chi$ and D-particle couplings $g_{37}$ (and thus $\eta$, and therefore the D-foam density). With the available LHC data at $\sqrt{s}=7$~TeV and 1~fb$^{-1}$ integrated luminosity, the current lower bounds on the mass scale below which the effective field theory of dark matter is valid, is ${\cal O}(200$~GeV)~\cite{lhcmaverick}. Therefore we need the full LHC potential (1~fb$^{-1}$ at $\sqrt{s}=14$~TeV) to produce, and thus constrain the parameters of, our TeV-scale D-particles ({\emph c.f.} (\ref{eq:mz})).

It should be remarked at this stage that, since $\eta$ depends on the density of foam,
 the larger the  elementary (``unit'') 3-volume containing a D-particle or the radius $R'$,
i.e. the smaller the D-foam density,
the smaller the effective coupling, and hence the weaker the relevant interactions. In this case, the D-particle interactions with the Standard Model excitations can be purely perturbative. Unlike monopoles, which are non-perturbative for weakly coupled strings, the D-particles can be treated perturbatively. In fact,  this similarity of the D-particles  with the 't Hooft-Polyakov monopole, makes the MoEDAL detector of the LHC~\cite{Moedal} a natural place to look for their potential signatures through post-production distinctive signatures (``etching'') on the plastic sheets of the  detector.
In this way, the non observation of such D-particle pairs (with masses in the TeV range) at high energy collisions can lead to bounds on their couplings (\ref{couplings}), and  on the density of D-particles. 

It must be stressed that in our cosmological models of D-foam, the D-particle defects do not constitute the bulk of the dark matter. Thus, their annihilation cross sections, which are proportional to $(g_{37} g_7) /M_D^2 = (g_{37} g_7 g_s^2) /M_s^2 $, do not have to be in the `WIMP-miracle' regions, and can be much smaller for TeV-mass D-particles, $M_D \sim M_s/g_s$ (corresponding to compactification radii of order of the string length). Indeed, as we discussed in section \ref{sec:Hubble}, as a result of the existence of D-particles in the bulk regions of our configurations (\emph{cf.} fig.~\ref{fig:recoil}), there are negative contributions to the energy density of our brane world, induced by such bulk defects, which cancel out any positive energy contributions to the Universe's energy density coming from frozen D-particle relics (\emph{cf}. (\ref{popul})). Hence there is no restriction in the relevant annihilation cross sections
$\sigma_{\rm a} ({\rm D}\, {\rm D} \to f {\overline f} )$, where $f$ are Standard Model particles, say. Such restrictions apply only to the conventional dark matter annihilation.

\bigskip

\section{Conclusions and Outlook \label{sec:conclu}}

In the present article we have computed Dark Matter thermal relics (via the appropriately modified Boltzmann equation) in the presence of a particular type of  stringy space-time foam, including brany D-particle defects as the foam structures. We have seen that, as a result of the interactions of the DM particle(s) with the defects in the foam, there are induced Finsler-type metric distortions, depending ---in addition to the space-time coordinates--- also on the pertinent momentum transfer. These momentum-dependent metrics have profound effects on the geodesic equation of the DM particles, leading to modifications in the pertinent Boltzmann equation determining the (thermal) relic abundances.

The foam acts as a source of particle production~\cite{ariadneplb}, but also affects particle statistics in a way reminiscent of
non-extensive statistics of Tsallis type. In this latter respect, our D-particle foam models may be viewed as
providers of  microscopic models of non-extensive statistics in Cosmology. The net result, is that the DM relic abundance in the presence of the foam is larger than the corresponding one in Standard foam-less Cosmology ($\Lambda$CDM model). The effects of the foam in such DM relic populations depend on the value of the string scale and coupling.
For conventionally high string scales, close to $10^{18}$~GeV, the effects are negligible. For low string scales, on the other hand, the foam effects may be constrained by WMAP data on colliders, as they imply considerably less available room in the parameter space for supersymmetric particle physics models than in the $\Lambda$CDM scenarios.

In this latter respect, the D-foam acts as a competing effect to those of time dependent dilatons, characterizing certain non-equilibrium stringy cosmologies. Indeed, in such models, the time dependent dilaton fields result in negative sources (sinks) in the Boltzmann equation determining the DM relic abundances. In the absence of foam effects, this may lead to a dilution of DM abundances by as much as 1/10 of those in the Standard Cosmological Model. This would have profound effects on collider, and in particular LHC, phenomenology. The presence of low-string scale D-foam, on the other hand, acts in the opposite sense, since it tends to increase the relic abundance. Thus, in the presence of D-foam, the time dependent dilaton effects may be neutralized in some models.

Complementary constraints on the D-particle foam may be provided by high energy gamma-ray astronomy, via the induced vacuum refractive index that such models entail. Moreover, interesting tests can be made
at particle interferometer experiments, such as neutral Kaon or other neutral meson factories, as a result of the
modified entanglement properties of neutral mesons, due to their non-trivial interaction with the D-foam.

Production of TeV-mass D-particles at high-energy colliders, such as the LHC, has also been briefly discussed with the conclusion that such a production could be possible and lead to observable signatures characterised by missing transverse energy only when the collider operates at its full energy. On the other hand, such relatively light defects when produced inside the MoEDAL detector of the LHC may deposit sizable amounts of energy via collisions with the molecules of the nuclear track plastic arrays, thus being detectable through the induced tracks.

 Needless to say  there is still a lot to be done before definite conclusions are reached on the falsification of models dealing with the quantum structure of space-time. Nevertheless, certain models, such as the D-foam examined here, may be falsified or constrained significantly by a plethora of diverse cosmological or astroparticle physics tests in the foreseeable future.

\bigskip

\section*{Acknowledgements}

The work of N.E.M. and S.S. is partially supported by the European Union through the FP6 Marie
Curie Research and Training Network \emph{UniverseNet} (MRTN-CT-2006-035863).
V.A.M.\ acknowledges support by the Spanish Ministry of Science and Innovation (MICINN) under the project FPA2009-13234-C04-01 and by the Ram\'on y Cajal contract RYC-2007-00631 of MICINN and CSIC.
The work of A.V. is supported in part by a King's College London (UK) graduate quota studentship. Feynman diagrams were drawn using JaxoDraw~\cite{jaxodraw}.

\vspace{0.2cm}

\section*{APPENDIX A: Distribution Functions in non-Expanding Foam Universes}

The number density $n$ and energy density $\rho$ of species in our stochastic space-time foam, when one ignores the Universe's expansion, are given by the basic formulae:
\begin{equation}\label{no2}
\\n=\frac{g}{(2\pi)^3}\int{\langle\langle n \rangle\rangle d^3k}
\end{equation}
and
\begin{equation}\label{dens2}
\rho=\frac{g}{(2\pi)^3}\int{ \langle\langle n \omega_r \rangle\rangle d^3k },
\end{equation}
with $\langle\langle n \rangle\rangle $ given in (\ref{dist2}) in the text.

The evaluation of the momentum integrals involved is facilitated by going to spherical coordinates, i.e.\ setting  $k_1=k\sin\theta\cos\phi$, $k_2=k\sin\theta \sin\phi$ and $k_3=k \cos\theta$, where $k=|\vec{k}|=\sqrt{k_1^2+k_2^2+k_3^2}$ and $\theta$ and $\phi$ are the ordinary polar and azimuth angles running between $[0,\pi]$ and $[0,2\pi]$ respectively. Then,  we obtain $d^3k=dk\,k^2\sin \theta\, d\theta\, d\phi$ and therefore the angular part of the integration of the different powers of $k_j$ appearing in (\ref{dist}) yields:
\begin{equation}\label{ang1}
\int_{0}^{2\pi}d\phi\int_{0}^{\theta}d\theta\sin \theta=4\pi,
\end{equation}
\begin{equation}\label{ang2}
\int_{0}^{2\pi}d\phi\int_{0}^{\theta}d\theta
 k_j^2\sin \theta=k^2\left(\frac{4\pi}{3}\right),
\end{equation}
\begin{equation}\label{ang3}
\int_{0}^{2\pi}d\phi\int_{0}^{\theta}d\theta
 k_j^4\sin \theta=k^4\left(\frac{4\pi}{5}\right).
\end{equation}

 We commence our analysis with the number density calculation.  The integral in (\ref{no2}) can be written in the following compact form:
\begin{eqnarray}\label{int2}
&&\int{\langle\langle n \rangle\rangle d^3k}=(4\pi) \tilde{J}(0,1,m,\xi)\nonumber\\
&&+\overline{\hat{\sigma}_0^2}\left\{-\frac{2\pi}{3}\beta^3\left(\tilde{P}(1,1,m,\xi)-\xi \tilde{P}(1,2,m,\xi)\right)+\frac{8\pi}{5}\beta^4\left(\tilde{J}(2,1,m,\xi)-3\xi \tilde{J}(2,2,m,\xi)+2\xi^2\tilde{J}(2,3,m,\xi)\right)\right\},\nonumber\\
\end{eqnarray}

\noindent where we have set for brevity: $\bar{\hat{\sigma}}_0^2\equiv\left(\hat{\sigma}_{01}^2+\hat{\sigma}_{02}^2+\hat{\sigma}_{03}^2\right)$
 with $\hat{\sigma}_{0i}= \frac{\sigma_{0i}}{\beta}$ (we note that the final results will be expressed in terms of the parameters $\sigma_{0i}$) and defined:
\begin{eqnarray}\label{integrals}
&&\tilde{J}(l,n,m,\xi)=\int_{m}^{\infty} dE \frac{E\left(E^2-m^2\right)^{l+1/2}}{\left(\exp\left(\beta\left(E-\mu\right)\right)+\xi\right)^n},\nonumber\\
&&\tilde{P}(l,n,m,\xi)=\int_{m}^{\infty} dE \frac{E^2\left(E^2-m^2\right)^{l+1/2}}{\exp\left(\beta\left(E-\mu\right)\right)+\xi}.
\end{eqnarray}

The above formulae (\ref{int2},\ref{integrals}) are used to calculate the number densities of the various particle species, either relativistic or not, at different eras of the evolution of the Universe.

In the relativistic limit $\left(k_{B}T\gg m\right)$, we can consider
$m\rightarrow 0$, and therefore the integrals (\ref{integrals})
can be rewritten as: \begin{equation}
\int_{0}^{\infty}d\kappa\frac{\kappa^{s}}{\left(e^{\kappa-\mu\beta}+\xi\right)^{n}}\equiv f_{s}\left(n;\mu\beta,\xi\right).\label{general}\end{equation}
Hence:
\begin{eqnarray}
 \int{\langle\langle n\rangle\rangle d^{3}k} &=& (4\pi)\beta^{-3}f_{2}(1;\mu\beta,\xi)+\nonumber \\
 &  & \bar{\hat{\sigma}}_{0}^{2}\beta^{-3}\left\{ \frac{2\pi}{3}\left[-f_{5}(1;\mu\beta,\xi)+\xi f_{5}(2;\mu\beta,\xi)\right]+\frac{8\pi}{5}\left[f_{6}(1;\mu\beta,\xi)-3\xi f_{6}(2;\mu\beta,\xi)+2\xi^{2}f_{6}(3;\mu\beta,\xi)\right]\right\},\nonumber \\ \label{int3}\end{eqnarray}
where the functions $f_{s}\left(1;\mu\beta,\xi\right)$, $f_{s}\left(2;\mu\beta,\xi\right)$
and $f_{s}\left(3;\mu\beta,\xi\right)$ have the explicit form
\begin{eqnarray}\label{f1}
 &  & f_{s}(1;x,1)=-\Gamma(s+1)Li_{1+s}\left(-e^{x}\right),\nonumber\\
 &  & f_{s}(1;x,-1)=\Gamma(s+1)Li_{1+s}\left(e^{x}\right),\end{eqnarray}
with $Li_{n}(z)$ the polylogarithm function: $Li_{n}(z)\equiv\sum_{k=1}^{\infty}\frac{z^{k}}{k^{n}}$
defined for $\left|z\right|<1$. From the recursion formula \begin{equation}
f_{s}\left(n+1;x,\xi\right)=\frac{1}{\xi}f_{s}\left(n;x,\xi\right)-\frac{1}{n\xi}\frac{d}{dx}f_{s}\left(n;x,\xi\right),\end{equation}
we find \begin{equation}
f_{s}(2;x,\xi)=\xi\Gamma\left(s+1\right)\left(Li_{1+s}\left(-\xi e^{x}\right)-Li_{s}\left(-\xi e^{x}\right)\right)\label{f2}\end{equation}
 and \begin{equation}
f_{s}(3;x,\xi)=\xi\Gamma\left(s+1\right)\left(\frac{3}{2}Li_{s}\left(-\xi e^{x}\right)-Li_{1+s}\left(-\xi e^{x}\right)-\frac{1}{2}Li_{s-1}\left(-\xi e^{x}\right)\right).\label{f3}\end{equation}

In the ultra-relativistic limit $\mu\ll k_{B}T$, and in leading order
we can set $\mu\beta=0$ in (\ref{int3}), yielding finally the relativistic
number density for bosons and fermions:

\begin{equation}
n_{b}=\frac{gT^{3}}{2\pi^{2}}\Gamma(3)\zeta(3)+\frac{g}{8\pi^{3}}T^{5}\bar{\sigma}_{0}^{2}\zeta(5)\left(-\frac{2\pi}{3}\Gamma(6)+\frac{8\pi}{5}\Gamma(7)\right)\label{number1}\end{equation}
 and
\begin{equation}
n_{f}=\frac{3}{4}\frac{gT^{3}}{2\pi^{2}}\Gamma(3)\zeta(3)+\frac{g}{8\pi^{3}}T^{5}\bar{\sigma}_{0}^{2}\left\{ \frac{2\pi}{3}\Gamma(6)\left[-2\eta(6)+\eta(5)\right]+\frac{8\pi}{5}\Gamma(7)\left[6\eta(7)-6\eta(6)+\eta(5)\right]\right\}, \label{number2}\end{equation}
\noindent where $\zeta$ and $\eta$ are the Riemann zeta and Dirichlet
eta functions respectively and we have set: $\bar{\sigma}_{0}^{2}=\sum_i\sigma_{0i}^{2}$.

The first term in both (\ref{number1}) and (\ref{number2}) represents
the standard result whereas the second term gives the D-particles
foam correction. This correction, does not have the usual scaling
with temperature $T^{3}$ but scales as $T^{5}$ instead. However
this is not really in contradiction with Standard Cosmology since
the correction, weighted by the small parameters $\sigma_{0i}$, can
still be far below current observation limits.

In the non-relativistic, large-mass limit, the quantities $\xi$  can be neglected in the denominators of the integrands in (\ref{integrals}), that now reduce  to the integrals (upon integration over $x\equiv \beta E$):
\begin{eqnarray}
\tilde{g}(l,n,m,\beta)&=&\frac{1}{\beta^{2l+3}}\int_{m\beta}^{\infty}dx x\left(x^2-(m\beta)^2\right)^{l+1/2}\left(\exp\left(x-\mu\beta\right)\right)^{-n}\nonumber \\
&=&\frac{1}{\beta^{2l+3}}\frac{\Gamma\left(\frac{3}{2}+l\right)}{\sqrt{\pi}}m^{2+l}
\left(\frac{2}{n}\right)^{l+1}K_{2+l}(mn\beta)\beta^{2+l}\,e^{\beta\mu n},\nonumber\\
\tilde{h}(l,n,m,\beta)&=&\frac{1}{\beta^{2l+4}}\int_{m\beta}^{\infty}dx x^2\left(x^2-(m\beta)^2\right)^{l+1/2}\left(\exp\left(x-\mu\beta\right)\right)^{-n} \nonumber \\
&=&\frac{1}{\beta^{2l+4}}\left(\frac{2m\beta}{n}\right)^{1+l}K_{1+l}(mn\beta)
\frac{\Gamma\left(\frac{3}{2}+l\right)}{\sqrt{\pi}}\left(1+\beta^{2}m^{2}\right)\,e^{\beta\mu n},
\end{eqnarray}
where $K_{\nu}$ is the modified Bessel function of the second kind
of order $\nu$.

In terms of these functions we obtain for the heavy-species number density (\ref{no}):
\begin{eqnarray}\label{nonrel2b}
n&=&\frac{g}{8\pi^3}\int{\langle\langle n \rangle\rangle d^3k}\nonumber \\
&=&\frac{g}{8\pi^3}\,
\beta^{-3}\left\{(4\pi) g(0,1,m,\beta)e^{\beta\mu }+\frac{16\pi}{5}\bar{\hat{\sigma}}_0^2\left[g(2,1,m,\beta)e^{\beta\mu }-2\xi g(2,2,m,\beta)e^{2\beta\mu }+\xi^2g(2,3,m,\beta)e^{3\beta\mu }\right]\right.\nonumber\\
&&\left.-\frac{2\pi}{3}\bar{\hat{\sigma}}_0^2\left[h(1,1,m,\beta)e^{\beta\mu }-\xi h(1,2,m,\beta)e^{2\beta\mu }\right]
-\frac{8\pi}{5}\bar{\hat{\sigma}}_0^2\left[g(2,1,m,\beta)e^{\beta\mu }-\xi g(2,2,m,\beta)e^{2\beta\mu }\right]\right\}\nonumber\\
&=&g\left(\frac{m}{2\pi\beta}\right)^\frac{3}{2}e^{-(m-\mu)\beta}
+\frac{\sqrt{2}g}{5\pi^{3/2}}\left(m^{7/2}\beta^{-3/2}\right)\bar{\sigma}_0^2\left(\frac{15}{2}e^{-(m-\mu)\beta}-\frac{45}{16\sqrt{2}}\xi e^{-2(m-\mu)\beta}+\frac{15}{27\sqrt{3}}\xi^2e^{-3(m-\mu)\beta}\right)\nonumber\\
&&-\frac{g}{12\pi^{3/2}}\left(\beta^{-2}+m^2\right)\bar{\sigma}_0^2\left(\frac{m}{\beta}\right)^{3/2}\left(3e^{-(m-\mu)\beta}-\frac{3}{4\sqrt{2}}\xi e^{-2(m-\mu)\beta}\right),
\end{eqnarray}
where, in the subleading contributions, $\xi=+1$ applies to fermions and $\xi=-1$ to bosons. In deriving (\ref{nonrel2b}) we have used the asymptotic limit of the Bessel function for large $x$ values, $K_{\nu}\left(x\right)\approx \sqrt{\frac{\pi}{2x}}e^{-x}$, a consideration valid in our non-relativistic approach ($x\equiv m\beta\gg 1$).

In order to obtain the energy densities (\ref{dens}) in this model, one needs to calculate the average over the statistics of the \emph{product} of the distribution function $n_r$ as given in (\ref{dist}) and the energy $\omega_r$ (\ref{energy2}). By keeping again powers up to $r_i^2$ and dropping $r_ir_j$  terms for $i\neq j$, we find:
\begin{eqnarray}\label{ennum}
n_r\omega_r &\simeq& \frac{1}{\exp(\beta(E-\mu))+\xi}\left(E-\sum_{j}k_j^2r_j+\frac{E}{2}\sum_{j}k_j^2r_j^2\right)\nonumber\\
&&+\beta\frac{\exp(\beta(E-\mu))}{(\exp(\beta(E-\mu))+\xi)^2}\left(2E\sum_{j}k_j^2r_j-\frac{E^2}{2}\sum_{j}k_j^2r_j^2-2\beta E \sum_{j}k_j^4r_j^2-2\sum_{j}k_j^4r_j^2\right)\nonumber\\
&&+\frac{\exp(2\beta(E-\mu))}{(\exp(\beta(E-\mu))+\xi)^3}4\beta^2E\sum_{j}k_j^4r_j^2.
\end{eqnarray}

Averaging over $r_i$ and  integrating over the angular part (in spherical coordinates) of $d^3k$ ,
yields:
\begin{eqnarray}\label{int6}
\int{\langle\langle n_r\omega_r \rangle\rangle d^3k} &=& \int _{0}^{\infty} 8\pi dk k^2 \left\{\frac{1}{2}\frac{E}{\exp(\beta(E-\mu))+\xi}\right.\nonumber\\
&&\left.+\bar{\hat{\sigma}}_0^2\beta^2\left(\frac{1 }{12}\frac{k^2E}{\exp(\beta(E-\mu))+\xi}-\frac{\beta}{5}\frac{\exp(\beta(E-\mu))k^4}{(\exp(\beta(E-\mu))+\xi)^2}-\frac{\beta}{12}\frac{\exp(\beta(E-\mu))}{(\exp(\beta(E-\mu))+\xi)^2}k^2E^2\right.\right.\nonumber\\
&&\left.\left.-\frac{\beta^2}{5}\frac{\exp(\beta(E-\mu))}{(\exp(\beta(E-\mu))+\xi)^2}k^4E+\frac{2}{5}\beta^2\frac{\exp(2\beta(E-\mu))}{(\exp(\beta(E-\mu))+\xi)^3}Ek^4\right)\right\}~.
\end{eqnarray}
The reader should recall that  $\bar{\hat{\sigma}}_0^2\equiv\beta^{-2}\sum_i\sigma_{0i}^2$.

In terms of the parameter $E^2=k^2+m^2$, (\ref{int6}) takes the form:
\begin{eqnarray}\label{int8}
\int{\langle\langle n_r\omega_r \rangle\rangle d^3k} &=&
8\pi\beta^{-4}\left\{\frac{1}{2}P(0,1,m,\xi)\right.\nonumber\\
&&\left.+\bar{\hat{\sigma}}_0^2\left(\frac{1}{12}P(1,1,m,\xi)-\frac{1}{5}J(2,1,m,\xi)+\frac{1}{5}\xi J(2,2,m,\xi)-\frac{1}{12}Q(1,1,m,\xi)+\frac{1}{12}\xi Q(1,2,m,\xi)\right.\right.\nonumber\\
&&\left.\left.-\frac{1}{5}P(2,1,m,\xi)+\frac{1}{5}\xi P(2,2,m,\xi)+\frac{2}{5} P(2,1,m,\xi)-\frac{4}{5}\xi P(2,2,m,\xi)+\frac{2}{5}\xi^2P(2,3,m,\xi)\right)\right\},\nonumber\\
\end{eqnarray}
where we have set:
\begin{equation}\label{1}
J(l,n,m,\xi)=\int_{m\beta}^{\infty}dx\frac{x\left(x^2-m^2\beta^2\right)^{l+1/2}}{\left(\exp(x-\mu\beta))+\xi\right)^n},
\end{equation}
\begin{equation}\label{2}
P(l,n,m,\xi)=\int_{m\beta}^{\infty}dx\frac{x^2\left(x^2-m^2\beta^2\right)^{l+1/2}}{\left(\exp(x-\mu\beta))+\xi\right)^n}
\end{equation}
and
\begin{eqnarray}\label{3}
&&Q(l,n,m,\xi)=\int_{m\beta}^{\infty}dx\frac{x^3\left(x^2-m^2\beta^2\right)^{l+1/2}}{\left(\exp(x-\mu\beta)+\xi\right)^n}
=J(l+1,n,m,\xi)+m^2J(l,n,m,\xi).
\end{eqnarray}

The relations between $J(l,n,m,\xi)$, $P(l,n,m,\xi)$ and the corresponding quantities $\tilde{J}(l,n,m,\xi)$ and $\tilde{P}(l,n,m,\xi)$ defined in (\ref{integrals}) are:
\begin{eqnarray}
&&\tilde{J}(l,n,m,\xi)=\frac{1}{\beta^{2l+3}}J(l,n,m,\xi),\nonumber\\
&&\tilde{P}(l,n,m,\xi)=\frac{1}{\beta^{2l+4}}P(l,n,m,\xi).
\end{eqnarray}

Now from (\ref{int8}) one can calculate the pressure and energy density for any kind of species at any era of the evolution of the universe by calculating appropriately the integrals (\ref{1}), (\ref{2}) and (\ref{3}).

The non-relativistic pressure density $p$, for a heavy dark matter species of mass $m \gg |\vec{k}| \equiv k $, is obtained from the general expression for the pressure
\begin{equation}\label{pressuregeneral}
p= \frac{g}{(2\pi)^3} \, \int \langle \langle \frac{n}{3\,\omega_r} \rangle\rangle k^2 d^3 k
\end{equation}
upon taking the limit $m \gg k$, \emph{i.e.}\ $\omega_r \simeq m $ to leading order, and using as definition of temperature
\begin{equation}\label{minktemp}
\frac{g}{(2\pi)^3} \, \int \langle \langle n \rangle\rangle k^2 d^3 k \equiv n \, m\, T~.
\end{equation}
(the reader is invited to compare this with the definition in eq. (\ref{temperature}) in the text, for the case of an expanding Universe).
This implies for the non-relativistic pressure $p = \frac{n}{3}T + \dots $, where the dots denote foam corrections of order $\sigma_j^2$, suppressed also by $k_j/m \ll 1$ terms.

Similarly, the non-relativistic energy density is found to be: $\rho=m\,n + \mathcal{O}(\sigma_j^2)$. For isotropic foam situations, the (small) foam corrections can be easily seen to be cast in the form
$\rho = m \, n (1 + T m \bar{\sigma}_0^2)$, where the above-mentioned temperature definition (\ref{minktemp}) has been used.

From these considerations, for cold dark matter species $T \ll m$, the pertinent equation of (approximate) dust emerges: $w = p/\rho =
\frac{T}{3\,m}\left(1 + \mathcal{O}(\sigma_j^2)\right)  \ll 1 $ to leading order in the small foam corrections.

In the case of relativistic matter, the energy density is found to be:
\begin{eqnarray}\label{reldens}
\rho&=&\frac{g}{\pi^2}\left\{\beta^{-4}\frac{1}{2}f_3(1,\beta\mu,\xi)
+\bar{\sigma}_0^2\beta^{-6}\left(\frac{1}{12}f_5(1,\beta\mu,\xi)-\frac{1}{5}f_6(1,\beta\mu,\xi)+\frac{1}{5}\xi  f_6(2,\beta\mu,\xi)-\frac{1}{12}f_6(1,\beta\mu,\xi)\right.\right.\nonumber\\
&&\left.\left.+\frac{1}{12}\xi f_6(2,\beta\mu,\xi)-\frac{1}{5}f_7(1,\beta\mu,\xi)+\frac{1}{5}\xi f_7(2,\beta\mu,\xi)+\frac{2}{5} f_7(1,\beta\mu,\xi)-\frac{4}{5}\xi f_7(2,\beta\mu,\xi)+\frac{2}{5}\xi^2f_7(3,\beta\mu,\xi)\right)\right\}.\nonumber\\
\end{eqnarray}

Calculating the pressure of a relativistic gas, using the general formula (\ref{pressuregeneral}), and comparing it
with the expressions for the energy density (\ref{reldens}) above, it is straightforward to observe that the presence of the foam leads to deviations from the standard equation of state for radiation $w_{\rm rad} =\frac{1}{3}$, by terms proportional to the foam fluctuations $\mathcal{O}(\sigma_j^2)$. This is an important difference from the case of the
 non-extensive (Tsallis) statistics cosmology of \cite{pessah}, where the relativistic equation of state retains its conventional form $w_{\rm rad}^{\rm Tsallis} =1/3$.

Upon using relations (\ref{f1}), (\ref{f2}) and (\ref{f3}) and setting: $\mu=0$, we find from (\ref{reldens}):
\begin{equation}\label{reldens1b}
\rho_{b}=\frac{g_b}{2\pi^2}\Gamma(4)\zeta(4)T_b^4+\frac{g_b}{\pi^2}\bar{\sigma}_0^2T_b^6\left(\frac{1}{12}\Gamma(6)\zeta(6)-\frac{17}{60}\Gamma(7)\zeta(6)+\frac{1}{5}\Gamma(8)\zeta(6)\right)=\frac{\pi^2}{30}g_bT_b^4+g_b\frac{2\pi^4}{189}\bar{\sigma}_0^2T_b^6,
\end{equation}
\noindent for bosons and
\begin{eqnarray}\label{reldens2b}
\rho_{f}&=&\frac{g_f}{2\pi^2}\Gamma(4)\eta(4)T_f^4+\frac{g_f}{\pi^2}\bar{\sigma}_0^2T_f^6\left(\frac{1}{12}\Gamma(6)\eta(6)-\frac{17}{30}\Gamma(7)\eta(7)+\frac{17}{60}\Gamma(7)\eta(6)+\frac{6}{5}\Gamma(8)\eta(8)-\frac{6}{5}\Gamma(8)\eta(7)+\frac{1}{5}\Gamma(8)\eta(6)\right)\nonumber\\
&=&\left(\frac{7}{8}\right)\frac{\pi^2}{30}g_fT_f^4+\frac{g_f}{\pi^2}\bar{\sigma}_0^2T_f^6\left(\frac{18941\pi^6}{15120}-\frac{50841}{8}\zeta(7)+\frac{127\pi^8}{200}\right)=\left(\frac{7}{8}\right)\frac{\pi^2}{30}g_fT_f^4+g_f\frac{793.32}{\pi^2}\bar{\sigma}_0^2T_f^6,
\end{eqnarray}
\noindent for fermions. These equilibrium distributions will be important in
estimating the correction to the effective number of degrees of freedom.

 Changes to distribution functions can have cosmological
implications in principle and we considered them for the freeze
out of dark matter candidates.

\section*{APPENDIX B: Geodesics Equations in Stochastic Finsler Space-Times}

The geodesics are obtained by using the standard technique of the
Euler-Lagrange equations for the action $S$ given by $$
 S = \int_{\tau_{1}}^{\tau_{2}}\sqrt{L}d\tau \quad
{\rm  with} \quad  L=g_{\mu\nu}\left(x,\frac{dx^{\alpha}}{d\tau}\right)\frac{dx^{\mu}}{d\tau}\frac{dx^{\nu}}{d\tau},$$
leading to:
\begin{equation}
\frac{d}{d\tau}\frac{\partial L}{\partial\left(\frac{dx^{\mu}}{d\tau}\right)}-\frac{\partial L}{\partial x^{\mu}}=0,
\end{equation}
since $\tau$ is an affine parameter. This yields:

\begin{equation}
N_{\mu}^{\kappa}\frac{d^{2}x^{\mu}}{d\tau^{2}}=-\Gamma_{\mu\nu}^{\kappa}\frac{dx^{\mu}}{d\tau}\frac{dx^{\nu}}{d\tau}-\frac{1}{2}g^{\kappa\alpha}g_{\mu\nu:\alpha,\beta}\frac{dx^{\mu}}{d\tau}\frac{dx^{\nu}}{d\tau}\frac{dx^{\beta}}{d\tau}\label{modgeodesic}
\end{equation}
with the convention $g_{\mu\nu:\alpha}\equiv\frac{\partial}{\partial v^{\alpha}}g_{\mu\nu}$
and $g_{\mu\nu,\alpha}\equiv\frac{\partial}{\partial x^{\alpha}}g_{\mu\nu}$.
Here, \begin{equation}
N_{\mu}^{\kappa}=\delta_{\mu}^{\kappa}+g^{\kappa\alpha}\frac{dx^{\nu}}{d\tau}\left(g_{\alpha\nu:\mu}+g_{\mu\nu:\alpha}\right)+\frac{1}{2}g^{\kappa\alpha}g_{\beta\nu:\alpha:\mu}\frac{dx^{\beta}}{d\tau}\frac{dx^{\nu}}{d\tau}.\label{eq:2}\end{equation}
Equivalently, equation (\ref{modgeodesic}) is written as:
 \begin{equation}\label{geod} \frac{d^{2}x^{\lambda}}{d\tau^{2}}=-\left(N^{-1}\right)_{\rho}^{\lambda}\Gamma_{\mu\nu}^{\rho}\frac{dx^{\mu}}{d\tau}\frac{dx^{\nu}}{d\tau}-\frac{1}{2}\left(N^{-1}\right)_{\rho}^{\lambda} g^{\rho\alpha}g_{\mu\nu:\alpha,\beta}\frac{dx^{\mu}}{d\tau}\frac{dx^{\nu}}{d\tau}\frac{dx^{\beta}}{d\tau}.
 \end{equation}
We should stress that this is a general modification of the geodesic
equation for an arbitrary dependence of the metric on $v$. In our
case however, things simplify a lot since $g_{\mu\nu:\beta}$ is zero in all cases except for terms $g_{0i:i}=g_{i0:i}=mr_{i}a^4(t)$ and also $g_{\beta\nu:\alpha:\mu}=0$. The inverse metric $g^{\mu\nu}$
is readily calculated and is given by
 \[g^{00}=-\frac{a^{2}(t)}{q},\quad g^{0i}=\frac{k_{i}r_{i}a^{2}(t)}{q},\quad g^{ii}=\frac{1+a^{2}(t)\sum_{j\neq i}k_{j}^{2}r_{j}^{2}}{q},\quad g^{ij}=-\frac{k_{i}k_{j}r_{i}r_{j}a^{2}(t)}{q},\]
 with $q=a^{2}(t)\left[a^{2}(t)\sum_{i}k_{i}^{2}r_{i}^{2}+1\right]$.

We need $\left(N^{-1}\right)_{\mu}^{\kappa}$. For a general metric, this can be written as:
\begin{equation}
\left(N^{-1}\right)_{\mu}^{\kappa}=\delta_{\mu}^{\kappa}-U_{\mu}^{\kappa}+\left(U^{2}\right)_{\mu}^{\kappa},
\end{equation}
where
 \begin{equation} U_{\mu}^{\kappa}=g^{\kappa\alpha}\frac{dx^{\nu}}{d\tau}\left(g_{\mu\nu:\alpha}+g_{\alpha\nu:\mu}\right)+\frac{1}{2}g^{\kappa\alpha}g_{\beta\nu:\alpha:\mu}\frac{dx^{\beta}}{d\tau}\frac{dx^{\nu}}{d\tau}.
 \end{equation}
 This expression may seem complicated but we remember that in the end we always work up to (and include) terms
 ${\mathcal O}(r^{2})$.
The other ingredient in the calculation is the non-zero Christoffel
symbols which are found to be: \[
\Gamma_{00}^{0}=\frac{2a(t)\dot{a}(t)\left[\sum_ {i} r_{i}^2k_{i}^2\right]}{1+a^2(t)\left[\sum_{i} r_{i}^2k_{i}^2\right]},\quad
\Gamma_{0i}^{0}=\Gamma_{i0}^{0}=\frac{r_{i}k_{i}a(t)\dot{a}(t)}{1+a^2(t)\left[\sum_{i} r_{i}^2k_{i}^2\right]},\quad
\Gamma_{ii}^{0}=\frac{a(t)\dot{a}(t)}{1+a^2(t)\left[\sum_{i} r_{i}^2k_{i}^2\right]},\]

\[
\Gamma_{00}^{i}=\frac{2Hr_{i}k_{i}}{1+a^2(t)\left[\sum_{i} r_{i}^2k_{i}^2\right]},\quad\Gamma_{0i}^{i}=\Gamma_{i0}^{i}=H-\frac{a(t)\dot{a}(t)r_{i}^2k_{i}^2}{1+a^2(t)\left[\sum_{i} r_{i}^2k_{i}^2\right]},\quad
\Gamma_{0j}^{i}=-\frac{r_{i}k_{i}r_{j}k_{j}a(t)\dot{a}(t)}{1+a^2(t)\left[\sum_{i}r_{i}^2k_{i}^2\right]},\]

\[\Gamma_{ii}^{i}=-\frac{r_{i}k_{i}a(t)\dot{a}(t)}{1+a^2(t)\left[\sum_{i} r_{i}^2k_{i}^2\right]},\quad
\Gamma_{jj}^{i}=-\frac{r_{i}k_{i}a(t)\dot{a}(t)}{1+a^2(t)\left[\sum_{i}r_{i}^2k_{i}^2\right]},\quad
\Gamma_{ii}^{i}=-\frac{r_{j}k_{j}a(t)\dot{a}(t)}{1+a^2(t)\left[\sum_{i} r_{i}^2k_{i}^2\right]},
\]
\noindent where $H\equiv \frac{\dot{a}(t)}{a(t)}$ is the Hubble parameter.
We also note that in what follows, the use of summation convention is generally
eschewed and on the occasions that it is used, it will be clearly indicated.

After detailed calculations, where only terms up to order $r^{2}$ have been kept and  cross terms of the form $r_{i}r_{j}$ for $i\neq j$ have been dropped, equation (\ref{geod}), for $\mu=i=1,2,3,$ simplifies to:
\begin{eqnarray}\label{geod2}
 \frac{d^{2}x^{i}}{d\tau^{2}} &=& -\frac{2}{m^{2}} Hp^{i}p^{0}-\frac{2}{m^{2}}a(t)\dot{a}(t)r_{i}p^{i}\left(p^{0}\right)^{2}+\frac{8}{m^{2}}a^{5}(t)\dot{a}(t)r_{i}^{2}\left(p^{i}\right)^{3}p^{0}\nonumber \\
 &  & +\frac{2}{m^{2}}a^{3}(t)\dot{a}(t)r_{i}p^{i}
 \sum_{j}\left(p^{j}\right)^2+\frac{4}{m^{2}}a^{3}(t)\dot{a}(t)r_{i}^{2}p^{i}\left(p^{0}\right)^{3}-\frac{4}{m^{2}}a^{5}(t)\dot{a}(t)r_{i}^{2}p^{i}p^{0}
 \sum_{j}\left(p^{j}\right)^2.\end{eqnarray}
These equations are used in the modified Liouville equation that determines species abundances in the stochastic D-foam space-time.


\begin{thebibliography}{99}
\bibitem{kostelecky}
  A.~Kostelecky,
  Phys.\ Lett.\  {\bf B701}, 137-143 (2011).
  [arXiv:1104.5488 [hep-th]] and references therein.



\bibitem{finsler} See, for instance: D. Bao, S.~S.~Chern and Z.~Shen, \emph{An introduction
to Finsler Geometry} (Springer-Verlag (NY, 2000)).

\bibitem{finsler2} In the context of D-particle foam, such Finsler-type metrics have been first derived in:
J.~R.~Ellis, N.~E.~Mavromatos and D.~V.~Nanopoulos,
  Int.\ J.\ Mod.\ Phys.\  A {\bf 13}, 1059 (1998)
  [arXiv:hep-th/9609238];
  Gen.\ Rel.\ Grav.\  {\bf 32}, 127-144 (2000).
  [gr-qc/9904068].
For a short recent review on this topic see: N.~E.~Mavromatos,
  PoS {\bf QG-PH}, 027 (2007)
  [arXiv:0708.2250 [hep-th]] and references therein.

\bibitem{finsler2b} Also Finsler metrics in string theory, but in a different context that in ref. \cite{finsler2}, have been previously suggested in : S.~I.~Vacaru,
  arXiv:hep-th/0211068;
S.~I.~Vacaru,
  arXiv:hep-th/0310132;
In a field theory context, such metrics have been discussed, among other works, in:
 G.~Y.~Bogoslovsky,
  arXiv:0706.2621 [gr-qc];
  arXiv:0712.1718 [hep-th].
G.~W.~Gibbons, J.~Gomis and C.~N.~Pope,
  Phys.\ Rev.\  D {\bf 76}, 081701 (2007)
  [arXiv:0707.2174 [hep-th]];
  A.~P.~Kouretsis, M.~Stathakopoulos, P.~C.~Stavrinos,
  Phys.\ Rev.\  {\bf D79}, 104011 (2009).
  [arXiv:0810.3267 [gr-qc]].
 L.~Sindoni,
  Phys.\ Rev.\  D {\bf 77}, 124009 (2008)
  [arXiv:0712.3518 [gr-qc]];
M.~Anastasiei and S.~I.~Vacaru,
  J.\ Math.\ Phys.\  {\bf 50}, 013510 (2009)
  [arXiv:0710.3079 [math-ph]].
In the context of generic phenomenological models of non-standard dispersion relations in quantum gravity see:
F.~Girelli, S.~Liberati and L.~Sindoni,
  Phys.\ Rev.\  D {\bf 75}, 064015 (2007)
  [arXiv:gr-qc/0611024];
J.~Magueijo and L.~Smolin,
  Class.\ Quant.\ Grav.\  {\bf 21}, 1725 (2004)
  [arXiv:gr-qc/0305055].
J.~Skakala and M.~Visser,
  arXiv:0810.4376 [gr-qc].

\bibitem{finsler3}
  F.~Girelli, S.~Liberati, L.~Sindoni,
  Phys.\ Rev.\  {\bf D78}, 084013 (2008).
  [arXiv:0807.4910 [gr-qc]]. See also
  [arXiv:0909.3834 [gr-qc]] and references therein.

\bibitem{finslercosmo} A.~P.~Kouretsis, M.~Stathakopoulos, P.~C.~Stavrinos,
  Phys.\ Rev.\  {\bf D82}, 064035 (2010).
  [arXiv:1003.5640 [gr-qc]].


\bibitem{ariadneplb} N.~E.~Mavromatos, S.~Sarkar and A.~Vergou,
  Phys.\ Lett.\  B {\bf 696}, 300 (2011)
  [arXiv:1009.2880 [hep-th]].



  \bibitem{dfoam} J.~R.~Ellis, N.~E.~Mavromatos and D.~V.~Nanopoulos,
 Gen.\ Rel.\ Grav.\ \textbf{32}, 127 (2000); 
 Phys.\ Rev.\ D \textbf{61}, 027503 (2000); 
 Phys.\ Rev.\ D \textbf{62}, 084019 (2000);

 \bibitem{westmuckett} J.~R.~Ellis, N.~E.~Mavromatos
and M.~Westmuckett, 
Phys.\ Rev.\ D \textbf{70}, 044036 (2004); 
\emph{ibid.} \textbf{71}, 106006 (2005)~;
 J.~R.~Ellis, N.~E.~Mavromatos, D.~V.~Nanopoulos, M.~Westmuckett,
  Int.\ J.\ Mod.\ Phys.\  {\bf A21}, 1379-1444 (2006).
  [gr-qc/0508105].


\bibitem{emnnewuncert} J.~R.~Ellis, N.~E.~Mavromatos and D.~V.~Nanopoulos,
 Phys.\ Lett.\ B \textbf{665}, 412 (2008);
Int.\ J.\ Mod.\ Phys.\  A {\bf 26}, 2243 (2011)
  [arXiv:0912.3428 [astro-ph.CO]].



\bibitem{li} T.~Li, N.~E.~Mavromatos, D.~V.~Nanopoulos and D.~Xie,
 Phys.\ Lett.\ B \textbf{679}, 407 (2009). 


\bibitem{snIa} A.~G.~Riess \textit{et al.} [Supernova Search
Team Collaboration], 
 Astron.\ J.\ \textbf{116}, 1009 (1998);
S.~Perlmutter \textit{et al.} [Supernova Cosmology Project Collaboration],
 Astrophys.\ J.\ \textbf{517}, 565 (1999);
 R.~Amanullah {\it et al.},
  Astrophys.\ J.\  {\bf 716} (2010) 712
  [arXiv:1004.1711 [astro-ph.CO]].

\bibitem{cmb}
 D.~N.~Spergel {\it et al.}  [WMAP Collaboration],
  Astrophys.\ J.\ Suppl.\  {\bf 148} (2003) 175
  [arXiv:astro-ph/0302209].

\bibitem{7yrwmap} E.~Komatsu {\it et al.},
  ``\emph{Seven-Year Wilkinson Microwave Anisotropy Probe (WMAP) Observations:
  Cosmological Interpretation,}''
  arXiv:1001.4538 [astro-ph.CO] and references therein.

\bibitem{bao} D.~J.~Eisenstein \textit{et al.} {[}SDSS Collaboration{]},
 Astrophys.\ J.\ \textbf{633}, 560 (2005); 
 H.~J.~Seo and D.~J.~Eisenstein, 
Astrophys.\ J.\ \textbf{598}, 720 (2003); 
and references therein;
M.~Tegmark {\it et al.}  [SDSS Collaboration],
  Phys.\ Rev.\  D {\bf 69} (2004) 103501
  [arXiv:astro-ph/0310723].

\bibitem{lensing} L.~Fu \emph{et al.}, Astronomy \& Astrophysics
\textbf{479}, 9 (2008); L.~Guzzo \emph{et al.}, Nature \textbf{451},
541 (2008).

\bibitem{wheeler}J. A. Wheeler and K. Ford,\textit{\ Geons, Black
Holes and Quantum Foam}: \textit{A\ Life in Physics }(Norton, New
York, 1998).

\bibitem{ford} For a comprehensive review see: L.~H.~Ford, 
 Int.\ J.\ Theor.\ Phys.\ \textbf{44}, 1753 (2005) {[}arXiv:gr-qc/0501081{]}
and references therein. 


\bibitem{lorentz} For reviews see: T.~Jacobson, S.~Liberati and
D.~Mattingly, 
 Annals Phys.\ \textbf{321}, 150 (2006); 
N.~E.~Mavromatos, 
 J.\ Phys.\ Conf.\ Ser.\ \textbf{174}, 012016 (2009) {[}arXiv:0903.0318
{[}astro-ph.HE{]}{]}, and references therein. 






\bibitem{D0matter} G.~Shiu, L.~-T.~Wang,
  Phys.\ Rev.\  {\bf D69}, 126007 (2004).
  [hep-ph/0311228].


\bibitem{arrival} G.~Amelino-Camelia, J.~R.~Ellis, N.~E.~Mavromatos,
D.~V.~Nanopoulos and S.~Sarkar, 
 Nature \textbf{393}, 763 (1998); 
 J.~R.~Ellis, K.~Farakos, N.~E.~Mavromatos, V.~A.~Mitsou and
D.~V.~Nanopoulos, 
 Astrophys.\ J.\ \textbf{535}, 139 (2000); 
J.~R.~Ellis, N.~E.~Mavromatos, D.~V.~Nanopoulos, A.~S.~Sakharov
and E.~K.~G.~Sarkisyan, 
 Astropart.\ Phys.\ \textbf{25}, 402 (2006) {[}Astropart.\ Phys.\ \textbf{29},
158 (2008){]}. 



\bibitem{sussk1} N.~Seiberg and E.~Witten, 
 JHEP \textbf{9909}, 032 (1999); 
N.~Seiberg, L.~Susskind and N.~Toumbas, 
 JHEP \textbf{0006}, 044 (2000). 


\bibitem{msugra} P.~Nath, R.~L.~Arnowitt and A.~H.~Chamseddine,
 Nucl.\ Phys.\ B \textbf{227}, 121 (1983); 
H.~P.~Nilles, 
 Phys.\ Rept.\ \textbf{110}, 1 (1984). 


\bibitem{neutralino} J.~R.~Ellis, K.~A.~Olive, Y.~Santoso and
V.~C.~Spanos, 
 Phys.\ Lett.\ B \textbf{565}, 176 (2003) {[}arXiv:hep-ph/0303043{]}.
A.~B.~Lahanas and D.~V.~Nanopoulos, 
 Phys.\ Lett.\ B \textbf{568}, 55 (2003) {[}arXiv:hep-ph/0303130{]}.
For a review see: A.~B.~Lahanas, N.~E.~Mavromatos and D.~V.~Nanopoulos,
 Int.\ J.\ Mod.\ Phys.\ D \textbf{12}, 1529 (2003). 


\bibitem{elmn} 
 A.~B.~Lahanas, N.~E.~Mavromatos and D.~V.~Nanopoulos, 
 PMC Phys.\ A \textbf{1} (2007) 2; 
 Phys.\ Lett.\ B \textbf{649}, 83 (2007). 

\bibitem{cosmo} G.~A.~Diamandis, B.~C.~Georgalas, N.~E.~Mavromatos, E.~Papantonopoulos and I.~Pappa,
  Int.\ J.\ Mod.\ Phys.\  A {\bf 17}, 2241 (2002)
  [arXiv:hep-th/0107124];
G.~A.~Diamandis, B.~C.~Georgalas, N.~E.~Mavromatos and E.~Papantonopoulos,
  Int.\ J.\ Mod.\ Phys.\  A {\bf 17}, 4567 (2002)
  [arXiv:hep-th/0203241];
G.~A.~Diamandis, B.~C.~Georgalas, A.~B.~Lahanas, N.~E.~Mavromatos and D.~V.~Nanopoulos,
  Phys.\ Lett.\  B {\bf 642} (2006) 179
  [arXiv:hep-th/0605181].

\bibitem{dutta} B.~Dutta, A.~Gurrola, T.~Kamon, A.~Krislock,
A.~B.~Lahanas, N.~E.~Mavromatos and D.~V.~Nanopoulos, 
 Phys.\ Rev.\ D \textbf{79}, 055002 (2009). 



\bibitem{kolb} 
 E.~W.~Kolb and M.~S.~Turner, ``\emph{The Early universe},''
 Front.\ Phys.\ \textbf{69} 1 (1990). 

\bibitem{tsallissarkar} N.~E.~Mavromatos and S.~Sarkar,
  Phys.\ Rev.\  D {\bf 79}, 104015 (2009)
  [arXiv:0812.3952 [hep-th]].

\bibitem{pessah} D. F.~Torres, H.~Vucetich and A.~Plastino,
Phys.\ Rev.\ Lett. \textbf{79}, 1588 (1997);
M.~E.~Pessah, D.~F.~Torres and H.~Vucetich,
  Physica A {\bf 297}, 164 (2001)
  [arXiv:gr-qc/0105017];
M.E.~Pessah and D. F.~Torres,
Physica A \textbf{297}, 201 (2001).


\bibitem {tsallis}C. Tsallis, J. Stat. Phys.\textbf{ 52 }, 479 (1988).

\bibitem {polch2}J. Polchinski, Phys. Rev. Lett. \textbf{75} 4724 (1995).

\bibitem {coll} W. Fischler, S. Paban, and M. Rozali, Phys Lett \textbf{B381}, 62 (1996).

\bibitem {kogan} I.~I.~Kogan, N.~E.~Mavromatos and J.~F.~Wheater,
  Phys.\ Lett.\  B {\bf 387}, 483 (1996)
  [arXiv:hep-th/9606102];
I.~I.~Kogan and N.~E.~Mavromatos,
  Phys.\ Lett.\  B {\bf 375}, 111 (1996)
  [arXiv:hep-th/9512210].

\bibitem{szabo} N.~E.~Mavromatos and R.~J.~Szabo,
  Phys.\ Rev.\  D {\bf 59}, 104018 (1999)
  [arXiv:hep-th/9808124].

\bibitem {mav2}  N.~E.~Mavromatos,
\emph{Logarithmic conformal field theories and
strings in changing backgrounds},
in \emph{Shifman, M. (ed.) et al.: From
fields to strings, I. Kogan memorial Volume 2}, 1257-1364. (World Sci. 2005),
and references therein. [arXiv:hep-th/0407026].

\bibitem{gravanis} E.~Gravanis and N.~E.~Mavromatos,
  JHEP {\bf 0206}, 019 (2002); 
  arXiv:hep-th/0106146.

\bibitem {Dparticle} N.E.~Mavromatos and Sarben Sarkar,
  Physical\ Review\ D {\bf 72}, 065016 (2005)
  [arXiv:hep-th/0506242].

\bibitem{sen} A.~Sen,
  JHEP {\bf 9808}, 012 (1998)
  [arXiv:hep-th/9805170].



  \bibitem{eyras} E.~Eyras, S.~Panda,
  Nucl.\ Phys.\  {\bf B584}, 251-283 (2000).
  [hep-th/0003033]. See also:
O.~Bergman, M.~R.~Gaberdiel,
  Phys.\ Lett.\  {\bf B441}, 133-140 (1998).
  [hep-th/9806155].


\bibitem{loops} N.~D.~Lambert, I.~Sachs,
  JHEP {\bf 0102}, 018 (2001).
  [hep-th/0010045].
M.~Bertolini, A.~Lerda,
  Fortsch.\ Phys.\  {\bf 49}, 441-448 (2001).
  [hep-th/0012169].



 \bibitem{Gravanis} E.~Gravanis, N.~E.~Mavromatos,
  Phys.\ Lett.\  {\bf B547}, 117-127 (2002)
  [hep-th/0205298]; for a discussion on the role of no force condition for non-BPS D-particle see
  also the work of these authors in :
  [hep-ph/0104234];
A.~Campbell-Smith, N.~E.~Mavromatos,
  Phys.\ Lett.\  {\bf B488}, 199-206 (2000).
  [hep-th/0003262].


  \bibitem{witten} E.~Witten,
  Int.\ J.\ Mod.\ Phys.\  {\bf A10}, 1247-1248 (1995).
  [hep-th/9409111].



  \bibitem{bachas} C.~Bachas,
  [hep-th/9503030].





\bibitem{Dmatter} S.H.~Hansen, D.~Egli and C.~Salzmann,
New Astronomy, \textbf{10}, 379 (2005).

\bibitem{strongcoupl} See, for instance:
  M.~R.~Douglas, D.~N.~Kabat, P.~Pouliot and S.~H.~Shenker,
  Nucl.\ Phys.\  B {\bf 485}, 85 (1997)
  [arXiv:hep-th/9608024] and references therein.



\bibitem{emninfl} J.~R.~Ellis, N.~E.~Mavromatos, D.~V.~Nanopoulos, A.~Sakharov,
  New J.\ Phys.\  {\bf 6}, 171 (2004).
  [gr-qc/0407089].

\bibitem{MAGIC2}  J.~Albert {\it et al.}  [MAGIC Collaboration] and
J.~R.~Ellis, N.~E.~Mavromatos, D.~V.~Nanopoulos, A.~S.~Sakharov and E.~K.~G.~Sarkisyan,
  Phys.\ Lett.\  B {\bf 668}, 253 (2008).

\bibitem{fermi} A.~A.~Abdo {\it et al.}  [Fermi LAT and Fermi GBM Collaborations],
  Science {\bf 323} (2009) 1688;
FERMI GMB/LAT~Collaborations,
  arXiv:0908.1832 [astro-ph.HE].



\bibitem{mavroreview2010} N.~E.~Mavromatos,
  Int.\ J.\ Mod.\ Phys.\  A {\bf 25}, 5409 (2010)
  [arXiv:1010.5354 [hep-th]].



\bibitem{bernstein} J.~Bernstein, \emph{Kinetic theory in an Expanding Universe} (Cambridge Univ.\ Press, 1988).

\bibitem{Wu}
M.~Krook and T.T.~Wu,
The Physics of Fluids, {\bf 69} 1589 (1977).

\bibitem{spanos} For a recent review see:
  J.~L.~Feng,
  [arXiv:1003.0904 [astro-ph.CO]] and references therein;
See also: A.~B.~Lahanas, D.~V.~Nanopoulos and V.~C.~Spanos,
  Phys.\ Rev.\  D {\bf 62} (2000) 023515
  [arXiv:hep-ph/9909497];

\bibitem{triginer} J. Triginer, W. Zimdahl and D Pav\'{o}n, 
Class.\ Quantum \ Grav. \textbf{13}, 403 (1996).

\bibitem{pioline} I.~Antoniadis and B.~Pioline, 
 Nucl.\ Phys.\ B \textbf{550}, 41 (1999). 

\bibitem{riotto} D.~J.~H.~Chung, E.~W.~Kolb and A.~Riotto,
 Phys.\ Rev.\ D \textbf{60}, 063504 (1999). 

\bibitem{slepton} S.~Profumo and A.~Provenza, 
 JCAP \textbf{0612}, 019 (2006). 


\bibitem{nonequil-observ}
  J.~R.~Ellis, N.~E.~Mavromatos, V.~A.~Mitsou and D.~V.~Nanopoulos,
  Astropart.\ Phys.\  {\bf 27} (2007) 185
  [arXiv:astro-ph/0604272];
N.~E.~Mavromatos and V.~A.~Mitsou,
  Astropart.\ Phys.\  {\bf 29} (2008) 442
  [arXiv:0707.4671 [astro-ph]].


\bibitem{emnfits} J.~Ellis, N.~E.~Mavromatos and D.~V.~Nanopoulos,
  Phys.\ Lett.\  B {\bf 674}, 83 (2009)
  [arXiv:0901.4052 [astro-ph.HE]].


\bibitem{omega}  J.~Bernabeu, N.~E.~Mavromatos and J.~Papavassiliou,
  Phys.\ Rev.\ Lett.\  {\bf 92}, 131601 (2004)
  [arXiv:hep-ph/0310180].


\bibitem {bernabeu} J.~Bernabeu, N.~E.~Mavromatos and S.~Sarkar,
  Phys.\ Rev.\  D {\bf 74}, 045014 (2006)
  [arXiv:hep-th/0606137].

\bibitem{kloe} F.~Ambrosino {\it et al.}  [KLOE Collaboration],
  Phys.\ Lett.\  B {\bf 642}, 315 (2006)
  [arXiv:hep-ex/0607027];
A.~Di Domenico,
  arXiv:0904.1976 [hep-ex].

\bibitem{upgrade} See, e.g.: G.~Amelino-Camelia {\it et al.},
  Eur.\ Phys.\ J.\  C {\bf 68}, 619 (2010)
  [arXiv:1003.3868 [hep-ex]] and references therein.

\bibitem{mavroomega} N.~E.~Mavromatos,
  J.\ Phys.\ Conf.\ Ser.\  {\bf 171}, 012007 (2009)
  [arXiv:0904.0606 [hep-ph]].


\bibitem{kolbmaverick} M.~Beltran, D.~Hooper, E.~W.~Kolb, Z.~A.~C.~Krusberg and T.~M.~P.~Tait,
  JHEP {\bf 1009}, 037 (2010)
  [arXiv:1002.4137 [hep-ph]].

\bibitem{lhcmaverick}  A.~Rajaraman, W.~Shepherd, T.~M.~P.~Tait and A.~M.~Wijangco,
  Phys.\ Rev.\ D {\bf 84}, 095013 (2011)
  [arXiv:1108.1196 [hep-ph]].



\bibitem{cdf-monojet}   T.~Aaltonen {\it et al.}  [CDF Collaboration],
  Phys.\ Rev.\ Lett.\  {\bf 101} (2008) 181602
  [arXiv:0807.3132 [hep-ex]];
  T.~Aaltonen {\it et al.} [CDF Collaboration], arXiv:1203.0742 [hep-ex] (2012).

\bibitem{atlas-monojet}
  G.~Aad {\it et al.}  [ATLAS Collaboration],
  Phys.\ Lett.\ B {\bf 705} (2011) 294
  [arXiv:1106.5327 [hep-ex]];
ATLAS Collaboration, Tech.\ Rep.\ ATLAS-CONF-2011-096, CERN, Geneva (2011).

\bibitem{cms-monojet}
  S.~Chatrchyan {\it et al.}  [CMS Collaboration],
  Phys.\ Rev.\ Lett.\  {\bf 107} (2011) 201804
  [arXiv:1106.4775 [hep-ex]].

\bibitem{Moedal} J. L.~Pinfold, "The MoEDAL Technical Design Report", CERN-LHCC-2009-006, 2009
(http://moedal.web.cern.ch/content/moedal-technical-design-report-tdr) and references therein ;
see also J.~Pinfold [MOEDAL Collaboration],
  CERN Cour.\  {\bf 50N4}, 19 (2010) .





\bibitem{jaxodraw}
  D.~Binosi and L.~Theussl,
  Comput.\ Phys.\ Commun.\  {\bf 161} (2004) 76
  [hep-ph/0309015];
  D.~Binosi, J.~Collins, C.~Kaufhold and L.~Theussl,
  Comput.\ Phys.\ Commun.\  {\bf 180} (2009) 1709
  [arXiv:0811.4113 [hep-ph]].

\end{thebibliography}
\end{document}